\documentclass[12 pt]{article}

\usepackage{fullpage}
\usepackage{graphics}
\usepackage{graphicx}
\usepackage{citesort}

\newcommand{\be}{\begin{equation}}
\newcommand{\ee}{\end{equation}}
\def\bea {\begin{eqnarray}}
\def\eea {\end{eqnarray}}  
\def \bk{{\bf k}}
\def \bkp{{\bf k^\prime}}

\def \bq {{\bf q}}

\def \iwm {i\omega_m}
\def \iwmp {i\omega_{m^\prime}}
\def \kiwm {\bk,i\omega_m}
\def \kiwmp {\bk^\prime,i\omega_{m^\prime}}
\def\wtilde{\tilde {\omega }}
\def\YBCO{YBaCu$_3$O$_{7-x}$}
\def\BKBO{Ba$_{1-x}$K$_x$BiO$_3$}   

\def\taup{\tau^\prime}

\begin{document}


\title{ Electron - Phonon Superconductivity}


\author{F. Marsiglio and J.P. Carbotte}

\date{\today}



\maketitle


\tableofcontents

\section{Introduction}

A fairly sophisticated description of electron-phonon superconductivity
has existed since the early 1960's, following the work of Eliashberg
\cite{eliashberg60}, Nambu \cite{nambu60}, Morel and Anderson \cite{morel62}, 
and Schrieffer {\it et al.} \cite{schrieffer63}. All of this work
extended the original ideas of Bardeen, Cooper, and Schrieffer
\cite{bardeen57} on superconductivity, to include dynamical phonon exchange
as the root cause of the effective attractive interaction between electrons
in a metal. 
For certain superconducting materials, Eliashberg theory (as this
description is generally called) provides a very accurate description of
the superconducting state. Nonetheless, as B.T. Matthias was fond of
iterating \cite{matthias69}, this description 
was never considered (by him and others)
particularly helpful for discovering new, high temperature superconductors
\cite{remark0}. Part of the problem remains that a truly accurate
description of the normal state has not been forthcoming. Part of
{\em that} problem is the `curse' of Fermi Liquid Theory. To the
extent that the electron-phonon coupling causes relatively innocuous
corrections to most normal state properties, its underlying characteristics
remain undetectable (indeed, as will be reviewed here, the characteristics
of the electron-phonon interaction are made more apparent in the
superconducting state). An exception may be the A15 compounds, whose
anomalous normal state properties might help us achieve further understanding
of the electron-phonon interaction in these materials \cite{anderson83}.

This review will barely touch upon normal state properties influenced by
the electron-phonon interaction.  A considerable literature continues
to develop on this topic, including a more microscopic treatment of
model systems with simple electon-ion interactions. There have been
many theoretical developments in the last two decades, many of which
have been directed towards understanding the high temperature oxides.
Some references will be provided in the Appendix, but, for the bulk of
the chapter, we will
focus primarily on the superconducting state in `conventional'
superconductors. In the past, many reviews have been written on the role
of the electron-phonon interaction in superconductors. The reader is
directed in particular to the reviews by Carbotte \cite{carbotte90},
Rainer \cite{rainer86}, Allen and Mitrovi\'c \cite{allen82}, and
Scalapino \cite{scalapino69} (they are listed here in inverse chronological
order). While we have repeated much of what already exists in these reviews,
we felt it was important for completeness in the present volume, and
because the material is presented with a slightly different outlook than
has been done in the past.

The first section provides an overview of the subject as we see it, with
some details relegated to the Appendix. This is followed by a discussion 
of our knowledge of the electron-phonon interaction in metals, including
an update on old ideas to use the optical conductivity to extract this
information. The next two sections provide a very brief review of the
impact of the electron phonon interaction on the superconducting critical
temperature, the energy gap, the specific heat, and critical magnetic fields.
The next section examines dynamical response functions. Again, largely
because of the discovery of the high temperature superconductors, workers
were prompted to re-examine in more detail the effect of stronger
electron phonon coupling on various response functions. For example, as
will be discussed in the pertinent subsection, the lack of a coherence
peak in the NMR relaxation time was observed. Does this (on its own) indicate 
an exotic mechanism, or can it be explained by damping effects due to
a substantial electron phonon coupling ? Answers to such questions are
reviewed in this section. Finally, we end with a summary, including some
remarks on various non-cuprate but non-conventional superconductors. The
Appendix will sketch some derivations and provide references to more recent
literature. 

\section{The Electron-Phonon Interaction: Overview}

\subsection{Historical Developments}

The history of superconductivity is an immense and fascinating subject
\cite{dahl92}. While the discovery of superconductivity occurred in
1911 \cite{onnes11}, from a theoretical point of view, a first breakthrough
occurred with the discovery of the Meissner-Ochsenfeld effect
\cite{meissner33}, and the
understanding that this implied that the superconducting state was a
thermodynamic phase \cite{gorter33}. During this time a few attempts
were made at proposing a mechanism for superconductivity \cite{thomas90},
but, by 1950, when London's book \cite{london50} appeared, nothing
concerning mechanism was really known \cite{remark1}. 

In 1950 several important developments took place \cite{hoddeson92}; first, 
two independent isotope effect measurements were performed on Hg 
\cite{maxwell50,reynolds50}, which indicated that the superconducting
transition was intimately related to the lattice, probably through the
electron-phonon interaction. These experiments were all the more
remarkable because in 1922 Onnes and Tuyn had looked for an isotope
effect in superconducting Pb, and, within the experimental accuracy
of the time, had found no effect \cite{onnes23}.

Secondly, Fr\"ohlich \cite{frohlich50} adopted, for the first time,
a field-theoretical approach to problems in condensed matter. In
particular, he studied the electron-phonon interaction in metals,
and demonstrated, through second order perturbation theory, that
electrons exhibit an effective attractive interaction through
the phonons. Although the theory as formulated was incomplete, it
did lay the foundations for subsequent work. In fact one of the essential
features of this mechanism was summarized in his introduction 
\cite{frohlich50}: ``Nor is it accidental that very good conductors
do not become superconductors, for the required relatively strong
interaction between electrons and lattice vibrations gives rise
to large normal resistivity.'' His theory correctly produced an
isotope effect (recognized in a {\it Note Added in Proof}), and,
moreover, foreshadowed the discovery of the perovskite superconductors, by
suggesting that the number of free electrons per atom should be
reduced.

After hearing about the isotope effect measurements, Bardeen also
formulated a theory of superconductivity based on the electron-phonon
interaction, wherein he determined the ground state energy variationally
\cite{bardeen50}.
Both of these theories failed to properly explain superconductivity,
essentially because they focussed on the single-electron self-energies,
rather than the two-electron instability \cite{hoddeson92}. Another
breakthrough occurred a little later when Fr\"ohlich \cite{frohlich54}
used a self-consistent method to determine an energy lowering proportional
to $\exp{(-1/\lambda)}$, where $\lambda$ is the dimensionless 
electron-phonon coupling constant. This showed how essential singularities
could enter the problem, and why no perturbation expansion in $\lambda$
would succeed in this problem (although in fact the energy
lowering is due to a Peierls instability, not superconductivity). 

A parallel development meanwhile had been taking place in the problem
of electron propagation in polar crystals, i.e. the study of polarons.
In fact, this problem dates back to at least 1933 \cite{landau33}, when
Landau first introduced the idea of a ``polarization'' cloud due
to the ions surrounding an electron, which, among other things,
renormalized its properties. Fr\"ohlich also addressed this problem, first 
in 1937 \cite{frohlich37}, and then again in 1950 \cite{frohlich50b}. Lee, 
Low and Pines \cite{lee52} subsequently took up the problem, also using
field-theoretic techniques, to provide a solution to the intermediate
coupling polaron problem. This problem was taken on later by Feynman
\cite{feynman55}, then by Holstein and others \cite{holstein59}, along
with many others to the present day. In fact, as described in the Appendix,
a small group of physicists continues to emphasize 
polaron physics as being critical to high temperature superconductivity
in the perovskites.
 
Pines, having worked with Bohm on electron-electron interactions, and
having just used field-theoretic techniques in the polaron problem,
now combined with Bardeen to derive an effective electron-electron
interaction, taking into account both electron-electron interactions
{\it and} lattice degrees of freedom 
\cite{bardeen55}. The result was the effective interaction 
Hamiltonian between two electrons with wave vectors ${\bf k}$
and ${\bf k^\prime}$ and energies $\epsilon_{\bf k}$ and
$\epsilon_{\bf k^\prime}$ \cite{ashcroft76}:
\begin{equation}
V_{\bf k,k^\prime}^{\rm eff} = {4 \pi e^2 \over ({\bf k-k^\prime})^2 +
k_\circ^2} 
\biggl[ 
1 + {\hbar^2 \omega^2({\bf k-k^\prime}) \over
({\epsilon_{\bf k} - \epsilon_{\bf k^\prime}})^2  - 
\hbar^2 \omega^2({\bf k-k^\prime}) }
\biggr],
\label{bardeen_pines}
\end{equation}
where $k_\circ$ is the Thomas-Fermi wave vector, and $\omega({\bf q})$
is the dressed phonon frequency. Eq. (\ref{bardeen_pines}) is an
{\it effective} interaction; a more formal and general approach,
utilizing Green functions, will be given later. Nonetheless, it
is clear that this effective interaction captures the essence of
``overscreening'', i.e. for electronic energy differences less than
the phonon energy, the phonon contribution to the screened interaction
has the opposite sign from the electronically screened interaction,
and exceeds it in magnitude. Physically \cite{bardeen61}, one 
electron makes a transition, which excites a phonon, accompanied 
by an ionic charge density fluctuation. A second electron undergoes 
a transition caused by this induced charge density fluctuation. If 
the differences in the electron energies is small compared to the
phonon excitation energy, the second electron is {\it actually}
attracted to the first. This is shown pictorially in Fig. 1.

Eq. (\ref{bardeen_pines}) represents the starting point for the
two-electron interaction in metals. It was further simplified for
both the Cooper pair calculation \cite{cooper56} and the
Bardeen-Cooper-Schrieffer (BCS) \cite{bardeen57} calculation. 
The progression of events that ultimately led to a successful
theory for BCS has been well documented \cite{hoddeson92}. Most
of this part of the story had little to do with the details of
the attractive mechanism, but rather with the pairing theory itself. Thus,
one can divide the theory of superconductivity into two separate
conquests: first the establishment of a pairing formalism, which
leads to a superconducting condensate, given some attractive
particle-particle interaction, and secondly, a mechanism by which
two electrons might attract one another. BCS, by simplifying the 
interaction, succeeded in establishing the pairing formalism. They
were able to explain quite a number of experiments, previously
performed, in progress at the time of the formulation of the theory,
and many that were to follow. However, one might well ask to what 
extent the experiments support the electron-phonon {\it mechanism}
as being responsible for superconductivity \cite{remark2}. Indeed, one of the
elegant outcomes of the BCS pairing formalism is the universality
of various properties; at the same time this universality means
that the theory really doesn't distinguish one superconductor 
from another, and, more seriously, one mechanism from another.
Fortunately, while many superconductors do display universality,
some do not, and these, as it turns out, provided very strong support
for the electron-phonon mechanism, as initially motivated by
Fr\"ohlich \cite{frohlich50} and by Bardeen and Pines \cite{bardeen55}. 
Much of this chapter will be concerned with these deviations from
universality.

After the BCS paper appeared, several workers rederived their results
using alternative formalisms. For example, Anderson used an RPA treatment
of the reduced BCS Hamiltonian in terms of pseudospin operators
\cite{anderson58}, and Bogoliubov and others \cite{bogoliubov58,valatin58}
developed more general methods, later to be adapted to inhomogeneous
superconductivity by de Gennes \cite{degennes66}. Finally, Gor'kov
\cite{gorkov58} developed a Green function method, from which both
the BCS results, and the Ginzburg-Landau phenomenology \cite{ginzburg50}
could be derived, near the transition temperature, $T_c$. 

The Gor'kov formalism proved to be the most useful, for the purposes
of generalizing BCS theory (with its model effective interaction) to
the case where the electron-phonon interaction is properly taken into
account in the superconducting state. This was done by Eliashberg 
\cite{eliashberg60}, as well as
Nambu \cite{nambu60}, and later partially by Morel and Anderson 
\cite{morel62} and more completely by Schrieffer and coworkers 
\cite{schrieffer63,schrieffer64,scalapino66}. Around the same time
tunneling became a very useful spectroscopic probe of the superconducting
state \cite{giaever60}; besides providing an excellent measure of the
gap in a superconductor, it also revealed the fine detail of the
electron-phonon interaction \cite{rowell63}, to such an extent that
tunneling data could be ``inverted'' to tell us about the underlying
electron-phonon interactions \cite{mcmillan65}. These developments
have been well documented in the Parks treatise \cite{parks69}. In
particular retardation effects are covered in the articles by
Scalapino \cite{scalapino69} and McMillan and Rowell \cite{mcmillan69}.
An interesting historical perspective is provided in the article by
Anderson \cite{anderson69}.

In the meantime, developments in our understanding of the polaron
were occurring in parallel. The problem of phonon-mediated superconductivity
and the problem of the impact of electron-phonon interactions on a single
electron are obviously related, but, after the initial work by
Fr\"ohlich and Pines and coworkers, the two fields seem to have parted
ways. Indeed, an excellent summary of the status of polarons at that time
is Ref. \cite{kuper63}, where, however, there is essentially no
``cross-talk'' with the theory of superconductivity. Similarly, in the
treatise by Parks \cite{parks69} there is essentially no discussion
of polarons \cite{schrieffer69}, in spite of the fact that 
the `polaron' really is the
essential building block of the BCS theory of superconductivity. So,
for example, a perusal of the index of the classic texts on 
superconductivity, by Schrieffer \cite{schrieffer64}, Blatt 
\cite{blatt64}, Rickayzen \cite{rickayzen65}, de Gennes 
\cite{degennes66}, and Tinkham \cite{tinkham75} reveals not a single entry
\cite{remark3}.
The reason for this is that the electron-phonon coupling strength
in all known superconductors was deemed to be sufficiently weak that
the only effect on normal state properties was a slightly increased 
electron effective mass. Thus, the electronic state is presumed to
be well described by Fermi Liquid Theory, upon which the BCS theory
(and its modifications) is based. It is important to keep this in
mind; for this reason we will refrain from referring to Eliashberg
theory as a strong coupling theory (we ourselves have used this term
in the past). Eliashberg theory goes beyond BCS theory because it
includes retardation effects; however, it is still a weak coupling
theory, in the sense that the Fermi energy is the dominant energy,
and the quasiparticle picture remains intact.

We make this distinction because in recent years polaron theory 
has experienced a renaissance, and some attempts to explain high
temperature superconductivity have utilized polaron and bipolaron
concepts. The bipolaron is simply a bound state of two polarons,
analogous to the Cooper pair, except that the latter requires
a Fermi sea to exist (at least in three dimensions) whereas the
former exists as a tightly bound pair in the absence of a Fermi
sea. In this respect bipolaron theories resemble the quasichemical
theory advocated by Schafroth and coworkers \cite{schafroth54,blatt64}
in the 1950's. Tightly bound electron pairs are now recognized as the
strong coupling limit of the BCS ground state; the transition to the
normal state is, however, governed by very different (and as yet
undetermined) excitations compared to BCS theory. We will refer to
some of this work in the course of this chapter.

To complete this brief historical tour, we should add that in 1964,
with the suggestion of a theorist \cite{cohen64}, what
has emerged as a new class of superconductors was discovered 
\cite{schooley64}. The actual superconducting compound was doped
Strontium Titanate (SrTiO$_3$), a perovskite with low carrier
density. This compound, along with BaPb$_{0.75}$Bi$_{0.25}$O$_3$, another
doped perovskite discovered in 1975 \cite{sleight75} with a transition
temperature of 12 K, were the precursors to the modern high temperature
superconductors discovered by Bednorz and M\"uller \cite{bednorz86}.
In fact, with fortuitous foresight, Schooley {\it et al.} \cite{schooley64a}
remarked, ``If SrTiO$_3$ had magnetic properties, a complete study
of this material would require a thorough knowledge of all of solid state
physics.'' Little did they know that in 1986 perovskites would be
discovered, that not only had high superconducting transition
temperatures, but also exhibited a plethora of magnetic phenomena.
We should also note that the so-called cuprates, which presently exhibit 
superconducting transition temperatures up to 160 K (under pressure),
all contain CuO$_2$ layers, whereas the cubic oxides (such as
SrTiO$_3$, BaPb$_{0.75}$Bi$_{0.25}$O$_3$, and Ba$_{1-x}$K$_x$BiO$_3$ \,\,
\cite{mattheiss88} (with $T_c \approx 30$ K)) do not. For this reason
many workers have come to regard the layered cuprates and the cubic
oxides as belonging to two completely separate (and unconventional)
classes, even though they are both essentially low carrier density
perovskites. 

\subsection{Electron-Ion Interaction}
\subsubsection{Overview}

A useful ab initio theory has to begin from some fundamental starting
point. In condensed matter systems the starting point is usually taken
to be electrons and ions (with their charges, and masses, etc.) along
with the chemical composition of the material \cite{rainer86}.
Given these ingredients, the prescription for calculation is, in
principle, straightforward. One has to solve the many-body Schrodinger
equation, with a Hamiltonian consisting of one-body kinetic energy
terms and the two-body Coulomb interaction. The form of these terms,
along with all the constants involved, are known, so all that is required
to solve the problem is perhaps some ingenuity along with unlimited 
computer resources. This has been referred to by Laughlin as the
Condensed Matter version of ``The Theory of Everything'' \cite{laughlin00}.

Of course the difficulty is that, even if one could solve this problem,
one would not recognize what the solution represented. The notion
of ionic collective modes (i.e. phonons), for example, would not
be very transparent in such an approach. More obscure still would be
the distinction between a superconducting state versus a metallic state.

Instead, an approach which separates the complex many-body problem into
smaller, more tractable pieces, has traditionally been adopted in
condensed matter, and in particular in the problem of superconductivity
\cite{bardeen57,scalapino69,rainer86}. The most systematic approach
has been discussed by Rainer \cite{rainer86}. The premise in this
approach is the observation that many metals (amongst which many undergo
a transition to a superconducting state) are well described by Landau
Fermi Liquid Theory. This allows for an asymptotic expansion in small
parameters like $k_BT_c/E_F$, $\hbar \omega_{\rm phon}/E_F$ and 
$1/k_F\ell$, where $E_F$ ($k_F$) is the Fermi energy (wavevector),
$\omega_{\rm phon}$ is a typical phonon frequency, and $\ell$ is the
electron mean free path. He separates the problem into the ``{\it high
energy problem}'' (effect of Coulomb interactions amongst the electrons
themselves as well as between the electrons and the fixed nuclear potentials),
and the ``{\it low energy problem}'' (the dressing of conduction electrons
with phonons), and the eventual formation of the superconducting state.
Most of this review will concern the low energy problem. In our opinion the
high energy problem is not at all solved at present, from a truly
``ab initio'' approach. For example, strictly speaking, one cannot rely 
on any of the expansion parameters mentioned above, because one does
not know, in principle, whether one has a metal with a well-defined
Fermi surface, to begin with. Nonetheless, by appealing to experimental
observation, one can use for many cases the fact that nature has
already solved the high energy problem, and proceed from there to
solve the low energy part. This has been the dominant philosophy
throughout most of the last four decades towards understanding
superconductivity.

The difficulty with this approach was exemplified by the discovery
of superconductivity in the layered perovskites; band structure
calculations for the parent compound (La$_2$CuO$_4$) demonstrated
that it was a metal, when in fact the real material was an 
antiferromagnetic insulator. This problem was later repaired 
\cite{anisimov91}, but it remains the case that band structure calculations
fail to properly take into account strong Coulomb correlations, and
remain somewhat powerless to reliably predict a breakdown of the Fermi Liquid
picture.

With these caveats, the ``ab initio'' approach of Ref. \cite{rainer86}
has experienced excellent success in cases where a metallic state is
known to exist, and experimental input has been used in the theory.
We will comment in particular on the ``low energy'' part of the theory
later in this chapter. A thorough discussion is available in
Ref. \cite{rainer86}. 

\subsubsection{Models}

The net result of a proper handling of the ``high energy'' problem in the case
of a well-behaved metal is a set of input parameters for the low energy
problem that are simple enough to make the remaining part of the problem
appear to have arisen from a non-interacting model. The distinction is that
the input parameters (band structure, phonon spectrum, etc.) come not
directly from specified model parameters, but rather from previous
calculation and/or experiment. For this reason, we now discuss possible
models for the electron-phonon interaction, which, for the moment, we
view as fundamental models in their own right, and {\bf not} as models
which somehow parameterize (and disguise) the ``high energy problem''.

The reason for this is that we hope to accomplish several tasks
simultaneously. First, we will in effect work through the ``low energy
problem'' discussed in the previous subsection. Secondly, we will touch
upon some of the more recent work on electron-phonon Hamiltonians, which
are characterized not so much by comparison with experiment as comparison
with some ``exact'' solution, as attained, for example, by Quantum Monte
Carlo methods \cite{scalapino88,vonderlinden92}. Thirdly,
we will also be able to make contact with recent ongoing work on the
polaron (and bipolaron). These latter two topics are presented here more
by way of a digression. Some further detail is presented in an Appendix,
but for a more thorough discussion the cited literature will have to be
consulted.

It is always tempting to immediately compare the results of a calculation 
with experiment; agreement justifies the starting model (in this context
this would mean the Hamiltonian, with associated parameters), 
whereas disagreement would tend to rule
out the starting model as a candidate. In the many-body problem, however,
life is not so simple. For one thing, we {\it know} the starting
Hamiltonian, as emphasized in the previous subsection. We {\it will}
get agreement with experiment if we were only able to routinely calculate
any observable. However, in our endeavour to understand many-body
systems, we have grown to utilize {\it effective} Hamiltonians, which
would capture the {\it essence} of the phenomenon under investigation.
The purpose of this strategy is twofold; we make sense of the many-body
system in terms we can understand, and we make the calculation itself
more tractable in practice. 

There are many Hamiltonians in condensed matter physics, which were
derived as effective Hamiltonians for some particular problem, but, which
have since taken on a life of their own. This is true because (a) they
have withstood solution in spite of their simplicity, and (b) they
epitomize some qualitative aspect of the more general problem.
Famous examples are the Heisenberg/Ising model for spins, and the
Hubbard model for fermions with spin degrees of freedom. In the
electron-phonon problem several effective models have arisen over the years, 
the three most prominent of which have been the Fr\"ohlich Hamiltonian
\cite{frohlich50}, the Holstein model \cite{holstein59}, and 
the BLF (Bari\v si\'c-Labb\'e-Friedel) model \cite{barisic70} (also known
as the SSH (Su-Schrieffer-Heeger) model \cite{su79}).
The Fr\"ohlich Hamiltonian was derived in a continuum approximation
(see Ref. \cite{frohlich62} or \cite{feynman72} for a derivation),
and results in a coupling between the electron density and the ionic
momentum (a canonical transformation changes this to the
ionic displacement) which diverges as the momentum transfer between
electron and ions goes to zero. This Hamiltonian has been the subject
of many investigations of the polaron. Holstein proposed his model
as a simplification in which the interaction between electron and ion
is more local; in fact in some ways the simplification Hubbard
\cite{hubbard63} invoked to replace the long-range Coulomb interaction
is analogous to the simplification that the Holstein model represents
compared to the Fr\"ohlich Hamiltonian. Both the Fr\"ohlich  and
Holstein models represent couplings of the electron to an {\it optical}
phonon mode. We will focus on the Holstein model since it is particularly
amenable to numerical simulations. In contrast, the BLF (SSH) model couples  
the electron to the relative displacement of nearby ions, i.e. an 
{\it acoustic} phonon mode. The physics is simple; in the Holstein model
ionic distortions affect the electron energy level at a particular site,
while in the BLF model ionic displacements affect the electron hopping
amplitude. These are represented pictorially in Fig. 2, although of
course the coupling is dynamic.

The BLF model gained prominence in the 1980's \cite{heeger88} when
it was used to describe
solitons in conducting polymers; otherwise comparatively little effort
has been expended towards an understanding of its properties, particularly
in two or three dimensions. 
The BLF Hamiltonian is
\bea
H & = & \sum_i {{\bf p}^2_i \over 2M}\phantom{aa} + \phantom{aa}
\sum_{<ij>} {1 \over 2} K\bigl({\bf u}_i - {\bf u}_j\bigr)^2
\nonumber \\
& & -\sum_{<ij> \atop \sigma} \bigl(t_{ij} - {\bf \alpha} \cdot
({\bf u}_i - {\bf u}_j) \bigl(c^\dagger_{i\sigma}c_{j\sigma}
\phantom{a} + \phantom{a} h.c.\bigr),
\label{ham_BLF}
\eea                    
where the first line refers to the ions, with mass $M$ and spring constant
$K$. The ionic degrees of freedom are described by the ion momentum, 
${\bf p}_i$, and displacement, ${\bf u}_i$, at site $i$. The electrons are
described by creation (annihilation) operators $c^\dagger_{i\sigma}$
($c_{i\sigma}$) for an electron with spin $\sigma$ at site $i$. The
electron hopping amplitude is given by $t_{ij}$; this in turn is
modulated by ionic vibrations, and therefore results in the electron-ion
coupling with strength $|{\bf \alpha}|$. The coupling constant $|{\bf \alpha}|$
is proportional to the gradient of the hopping overlap integral between
electron orbitals on two neighbouring sites. 

Equation \ref{ham_BLF} gives rise to the standard electron-phonon
Hamiltonian, as written in momentum space:
\be
H  =  \sum_{\bk \sigma}\epsilon_\bk c^\dagger_{\bk \sigma} c_{\bk \sigma}
\phantom{a}  + \phantom{a} \sum_\bq \hbar \omega_\bq a^\dagger_\bq a_\bq
\phantom{a} + \phantom{a} {1 \over \sqrt{N}} \sum_{\bk \bk^\prime \atop \sigma}
g(\bk,\bk^\prime) \bigl(a_{\bk - \bk^\prime} + a^\dagger_{-(\bk - \bk^\prime)}
\bigr) c^\dagger_{\bk^\prime \sigma} c_{\bk \sigma}.
\label{ham_BKF_mom}
\ee             
We have used the conventional oscillator operators, $a_\bq =
{M\omega_\bq \over 2\hbar}\bigl( {\bf u}_\bq + i{\bf p}_\bq \bigr)$ and
the standard Fourier expansions, $c^\dagger_{i\sigma} = {1 \over \sqrt{N}}
\sum_\bk e^{i\bk \cdot {\bf R}_i} c^\dagger_{\bk \sigma}$, etc. The phonon
dispersion is given by $\omega_\bq$, where, in principle, $\bq$ includes
branch indices as well as momenta within the first Brillouin zone, and
$g(\bk,\bk^\prime)$ is the coupling function. For the BLF Hamiltonian,
this coupling function has a very specific form (involving sine functions).
A more general consideration of the electron-ion interaction yields
a Hamiltonian of essentially the same form \cite{scalapino69,allen82}, but
where the parameters involved are understood to already contain the
``high energy'' effects alluded to earlier. State-of-the-art 
computations of the electron-ion coupling strength, are given, for
example, in Ref. \cite{krakauer93} (for La$_{2-x}$Sr$_x$CuO$_4$) and
in Ref. \cite{gunnarson97} (and references therein, for A$_3$C$_{60}$).
 
The Holstein Hamiltonian is
\be
H = -t \sum_{{<ij>}\atop \sigma} (c^\dagger_{i \sigma} c_{j \sigma} +
H.c.) + \sum_i
[{p^2_i \over 2M} + {1 \over 2} K{x}^2_i] - \alpha \sum_{i \sigma} x_i
n_{i \sigma},
\label{ham_Hol}
\ee
where the parameters are as before except that the displacement variable
$x_i$ represents the (one-dimensional) displacement of some optical mode 
(say a breathing mode) associated with the $i$th site, and the electron-ion
coupling $\alpha$ represents the change in site energy (per unit displacement)
associated with this mode. 
In momentum space this Hamiltonian is particularly simple:
\be
H  =  \sum_{\bk \sigma}\epsilon_\bk c^\dagger_{\bk \sigma} c_{\bk \sigma}
\phantom{a}  + \phantom{a} \sum_\bq \hbar \omega_E a^\dagger_\bq a_\bq
\phantom{a} + \phantom{a} {g \over \sqrt{N}} \sum_{\bk {\bf q} \atop \sigma}
\bigl(a_{\bf q} + a^\dagger_{-(\bf q)}
\bigr) c^\dagger_{{\bf k+q} \sigma} c_{\bk \sigma},
\label{ham_Hol_mom}
\ee             
where $\omega_E$ is the Einstein mode frequency and $g \equiv 
\sqrt{ {\alpha^2 \hbar \omega_E \over 2K} }$. This model has been
studied extensively in the last twenty years, at least partly
due to its simplicity. Some of this work is reviewed in the Appendix.

\subsection{Migdal Theory}

The primary language of many-body systems is the Green function,
or propagator. Many books have been written (see for example Refs.
\cite{kadanoff62,abrikosov63,mattuck67,fetter71,rickayzen80,mahan81}) 
about the Green
function formalism, so we will bypass a thorough discussion here. 
A sketch of the derivation of the Migdal \cite{migdal58} equation
for the electron self-energy is given in the Appendix.
Migdal argued that all vertex corrections are
$O(m/M)^{1/2}$ compared to the bare vertex, and therefore can be ignored.
Here $m$ ($M$) is the electron (ion) mass. This represents a tremendous
simplification, and allows one to solve a theory which should work for
arbitrary coupling strength (this is, in fact, not the case, for reasons
that will become apparent in the next section).

An ``exact'' formulation of the electron-phonon problem can be 
summarized \cite{migdal58,englesberg63,nakajima80} in terms of the 
Dyson equations (written in momentum and imaginary frequency space):
\be
G({\bf k},i\omega_m) = \bigl[ G_\circ({\bf k},i\omega_m)^{-1}
- \Sigma({\bf k},i\omega_m)
\bigr]^{-1}
\label{dyson_electron}
\ee
for the electron, and
\be
D({\bf q},i\nu_n) = \bigl[ D_\circ({\bf q},i\nu_n)^{-1} - \Pi({\bf q},i\nu_n)
\bigr]^{-1}
\label{dyson_phonon}
\ee
for the phonon, where $G({\bf k},i\omega_m)$ is the 
one-electron Green function, $D({\bf q},i\nu_n)$ is the phonon
propagator, and
$\Sigma({\bf k},i\omega_m)$ is the electron 
and $\Pi({\bf q},i\nu_n)$  the phonon self energy.
Then, 
\be 
\Sigma({\bf k},i\omega_m) = - { 1 \over N\beta} \sum_{{\bf k^\prime}, m^\prime}
g_{\bf k,k^\prime} D({\bf k - k^\prime}, i \omega_m - i\omega_{m^\prime})
G({\bf k^\prime},i\omega_{m^\prime}) \Gamma({\bf k^\prime},i\omega_{m^\prime};
{\bf k},i\omega_m;{\bf k - k^\prime},i \omega_m - i\omega_{m^\prime}),
\label{self_ele_exact}
\ee           
and
\be
\Pi({\bf q},i\nu_n) = {2 \over N\beta}\sum_{{\bf k}, m} g_{\bf k,k+q}
G({\bf k+q},i\omega_m + i\nu_n) G({\bf k},i\omega_m) 
\Gamma({\bf k+q},i\omega_m + i\nu_n;{\bf k},i\omega_m;{\bf q},i\nu_n),
\label{self_pho_exact}
\ee  
where the vertex function $\Gamma$ can only be defined in terms of
an infinite set of diagrams (i.e. not in closed form).

The non-interacting propagators are
\be
G_\circ({\bf k},i\omega_m) = \bigl[ i\omega_m - (\epsilon_{\bf k} - \mu)
 \bigr]^{-1}
\label{g_0}
\ee
for the electron and
\be
D_\circ({\bf q},i\nu_n) = \bigl[ -M(\omega^2({\bf q}) + \nu_n^2) \bigr]^{-1}
\label{d_0}
\ee
for the phonon, where $\epsilon_{\bf k}$ is the single electron dispersion
(band indices are implicit here and in the following), $\mu$ is the chemical
potential, and $\omega({\bf q})$ is the phonon dispersion.
In writing these relations we have adopted the finite temperature
Matsubara formalism, with Fermion ($i\omega_m \equiv i\pi T(2m-1)$)
and Boson ($i\nu_n \equiv i2\pi T n$) Matsubara frequencies, where
$m$ and $n$ are integers and $T$ is the temperature ($k_B \equiv 1$).
The Matsubara sums in Eqs. (\ref{self_ele_exact},\ref{self_pho_exact}) 
extend over all integers, and the
momentum sums extend over the first Brillouin zone. This convention
will be maintained unless noted otherwise.

Migdal's approximation was to set the vertex function $\Gamma$ equal
to the bare vertex, $g$. Then, the electron self-energy can be written:
\be
\Sigma({\bf k},i\omega_m) = - { 1 \over N\beta} \sum_{{\bf k^\prime}, m^\prime}
|g_{\bf k,k^\prime}|^2 D({\bf k - k^\prime}, i \omega_m - i\omega_{m^\prime})
G({\bf k^\prime},i\omega_{m^\prime}).
\label{self1}
\ee
Migdal \cite{migdal58} also included renormalization effects in the phonon
propagator. With an application to real materials in mind, however, 
the electron dispersion relations will have been obtained from
a band structure calculation, and the phonon properties will generally
have been taken from experiment.
In this case the phonon self energy is omitted entirely
(to avoid double counting). In addition electron-electron effects have
been omitted, as they have been presumed to be included already in 
the band structure and phonon calculations (to the best extent possible).

Alternatively, Eq. (\ref{self1}) can be viewed as having
been derived from some microscopic electron-ion Hamiltonian. For example,
in the case of the Holstein Hamiltonian, Eq. (\ref{ham_Hol}), 
$g_{\bf k,k^\prime} \rightarrow g$, the constant appearing in 
Eq. (\ref{ham_Hol_mom}), and the electron band structure is given
by $\epsilon_{\bf k} = -2t \cos{(k_x)}$ (in one dimension, and for 
nearest-neighbour hopping only). In addition, the phonon frequency
becomes dispersionless ($\omega({\bf q}) \rightarrow \omega_E$) and 
the phonon self energy is given
by some appropriate approximation. Such an identification is useful
for comparison to exact results (usually done numerically - see the
Appendix for references).

In the classical literature 
\cite{migdal58,englesberg63,shimojima70,grimvall81,allen82}, 
Eq. (\ref{self1}) is simplified in the following
way. First, very often the phonon propagator is provided separately,
usually by inelastic neutron scattering measurements
\cite{brockhouse62,stedman67}. To see how, one first writes the
phonon propagator in terms of its spectral representation \cite{allen82}:
\be
D({\bf q},i\nu_n) = \int_0^\infty d\nu B({\bf q},\nu) {2 \nu \over (i\nu_n)^2
- \nu^2}
\label{pho_spec}
\ee
where $B({\bf q},\nu)$ is the phonon spectral function
\be
B({\bf q},\nu) \equiv -{1 \over \pi} {\rm Im} D({\bf q}, \nu + i\delta).
\label{pho_spec1}
\ee
The spectral function is positive definite, and obeys a sum rule; it
is the quantity that is constructed with fits to high-symmetry phonon
dispersion curves measured by inelastic neutron scattering \cite{brockhouse62}.
Following this tact a calculation of the phonon self energy is no longer
required. Another simplification was recognized in Ref. \cite{englesberg63};
this is the use of the non-interacting electron Green function
$G_\circ({\bf k},i\omega_m)$ in the right hand side of Eq. (\ref{self1})
instead of the full self-consistent choice, $G({\bf k},i\omega_m)$.
This approximation is valid when particle-hole symmetry is present
{\it and} the infinite bandwidth approximation is invoked. This latter
approximation is used extensively in the early literature on
metals and superconductors; a systematic explanation of the logic is
provided in Ref. \cite{allen82}, and requires the usual hierarchy of
energy scales, $\omega_{\rm phon} << E_F$ ($\hbar \equiv 1$). The result
is
\be
\Sigma({\bf k},i\omega_m) = {1 \over N \beta} \sum_{{\bf k^\prime}, m^\prime}
\int_0^\infty d\nu |g_{\bf k,k^\prime}|^2 B({\bf k - k^\prime},\nu) 
{2 \nu \over (\omega_m - \omega_{m^\prime})^2 + \nu^2} 
G_\circ({\bf k^\prime},i\omega_{m^\prime}).
\label{unk}
\ee
The form of Eq. (\ref{unk}) allows one to introduce the electron-phonon
spectral function, 
\be
\alpha^2F({\bf k,k^\prime},\nu) \equiv N(\mu) 
|g_{\bf k,k^\prime}|^2 B({\bf k-k^\prime},\nu),
\label{alpha1}
\ee
where $N(\mu)$ is the electron density of states at the
chemical potential.
At this point one can introduce `Fermi surface Harmonics' 
\cite{allen76,allen82}, and define an electron self-energy with Fermi
momentum which depends on Matsubara frequency, and on the angle around
the Fermi surface. Elastic impurities would act to homogenize the
self-energy (as well as other properties), so a more useful function
for dirty superconductors is the Fermi-surface-averaged spectral function,
\be
\alpha^2F(\nu) \equiv {1 \over N(\mu)^2} \sum_{\bf k,k^\prime}
\alpha^2F({\bf k,k^\prime},\nu) 
\delta(\epsilon_{\bf k} - \mu) 
\delta(\epsilon_{\bf k^\prime} - \mu). 
\label{a2f}
\ee

To gain an understanding of electron-phonon effects, Englesberg and
Schrieffer \cite{englesberg63} solved this model for two simple phonon
models, the Einstein and Debye models. Here we summarize their results
for the Einstein model, with unmodified phonon spectrum, a simpler case 
since both the phonon spectrum and
the bare vertex function are independent of momentum.
In this case $g_{\bf k,k^\prime} \equiv g$ and $B({\bf q},\nu) \equiv
\delta(\nu - \omega_E)$. Using, in addition, the prescription
\be
{ 1 \over N} \sum_{\bf k} \rightarrow \int d\epsilon N(\epsilon)
\label{dos}
\ee
along with a constant density of states approximation, extended
over an infinite bandwidth, one obtains for the electron self energy
\be
\Sigma(i\omega_m) = \lambda \omega_E^2 \int_{-\infty}^\infty d\epsilon
\, {1 \over \beta} \sum_{m^\prime} {1 \over \omega_E^2 + (\omega_{m^\prime}
-\omega_m)^2} {1 \over i\omega_{m^\prime} - (\epsilon - \mu)},
\label{self2}
\ee
where we have used the standard definition for the electron-phonon
mass enhancement parameter, $\lambda$:
\be
\lambda \equiv 2 \int_0^\infty d\nu {\alpha^2F(\nu) \over \nu},
\label{lambda}
\ee
which, for the Einstein spectrum used here, reduces to
\be
\lambda = 2N(\epsilon_F)g^2/\omega_E.
\label{lambda_ein}
\ee
Performing the Matsubara sum yields
\be
\Sigma(i\omega_m) = {\lambda \omega_E \over 2} \int_{-\infty}^\infty d\epsilon
\biggl(
\, \,
{n(\omega_E) + 1 - f(\epsilon - \mu) \over i\omega_m - \omega_E - (\epsilon
- \mu)} +
{n(\omega_E) + f(\epsilon - \mu) \over i\omega_m + \omega_E - (\epsilon - \mu)}
\, \,
\biggr)
\label{self3}
\ee
where $f(\epsilon - \mu)$ is the Fermi function and $n(\omega_E)$ is the
Bose distribution function.
The remaining integral can also be performed \cite{allen82}
\be
\Sigma(z) = {\lambda \omega_E \over 2}
\biggl[ \, \,
-2\pi i (n(\omega_E) + 1/2 ) + 
\psi \bigl( {1 \over 2} + i{\omega_E - z \over 2\pi T} \bigr) - 
\psi \bigl( {1 \over 2} - i{\omega_E + z \over 2\pi T} \bigr)
\, \, \biggr]
\label{self4}
\ee
where $\psi (x)$ is the digamma function \cite{abramowitz64,allen82}
and the entire expression has been analytically continued to a general
complex frequency $z$. Because
we performed the Matsubara sum first, before replacing $i \omega_m$
with $z$, this is the physically correct analytic continuation
\cite{baym61}.

At zero temperature one can use well-documented properties of the
digamma function, or, more simply, refer to the analytic continuation
of Eq. (\ref{self3}), since the Bose and Fermi functions may be more
familiar. Since $n(\omega_E) \rightarrow 0$ and
$f(\epsilon - \mu) \rightarrow \theta(\mu - \epsilon)$ as $T \rightarrow 0$
($\theta(x)$ is the Heaviside step function), 
the self energy at $T = 0$ is
\be
\Sigma(z) = {\lambda \omega_E \over 2} \ln \biggl( {\omega_E - z \over
\omega_E + z} \biggr).
\label{self5}
\ee
Spectroscopic measurements yield properties as a function of real frequency;
because of the analytic properties of the Green function, this corresponds
to a frequency either slightly above or below the real axis. We will use
frequencies slightly above, and designate the infinitesmal positive
imaginary part by `$i\delta$'. Thus,
\be
\Sigma(\omega+i\delta) = {\lambda \omega_E \over 2} \Biggl[ \ln \mid 
{\omega_E - \omega \over \omega_E + \omega} \mid - i\pi \theta(
\mid \omega \mid - \omega_E) \Biggr].
\label{self6}
\ee
The real and imaginary parts of this self energy are shown in Fig. 3,
along with the non-interacting inverse Green function ($\omega -
(\epsilon_{\bf k} - \mu)$) to determine the poles of the electron
Green function (see Eq. (\ref{dyson_electron})) graphically. A
quantity often measured in single particle spectroscopies
is the spectral function, $A({\bf k},\omega)$ defined by
\be
A({\bf k},\omega) \equiv -{1 \over \pi} {\rm Im} G({\bf k},\omega + i\delta).
\label{spect_ele}
\ee
With this definition, we obtain, through Eq. (\ref{dyson_electron}) and 
(\ref{self6}),
\bea
A({\bf k},\omega) & = & \delta \biggl(
\omega - (\epsilon_{\bf k} - \mu ) - {\lambda \omega_E \over 2} \ln
\mid {\omega_E - \omega \over \omega_E + \omega} \mid
\biggr) \phantom{aaaaaaaaaa} {\rm if} \mid \omega \mid < \omega_E, 
\nonumber \\
 & & \phantom{a} 
\nonumber \\
 & = & {\lambda \omega_E / 2 \over  \biggl(
\omega - (\epsilon_{\bf k} - \mu ) - {\lambda \omega_E \over 2} \ln
\mid {\omega_E - \omega \over \omega_E + \omega} \mid
\biggr)^2 \, + \, \biggl( {\pi \lambda \omega_E \over 2} \biggr)^2 }
\phantom{aaa} {\rm if} \mid \omega \mid > \omega_E.
\label{spect_ele2}
\eea
Plots are shown in Fig. 4. Each spectral function displays a
quasiparticle peak, whose strength $a_{\bf k}$ and frequency
$\omega_{\bf k}$ is implicitly dependent on wavevector 
\be
a_{\bf k} = \biggl( 1 + {\lambda \over 1-(\omega_{\bf k}/\omega_E)^2 } 
\biggr)^{-1},
\label{residue}
\ee
where $\omega_{\bf k}$ is the solution (between $-\omega_E$ and $\omega_E$)
to the zero of the delta-function argument in Eq. (\ref{spect_ele2}).
For all momenta (or equivalently all $\epsilon_{\bf k} - \mu$) there is
a solution, whose frequency approaches $\omega_E$ asymtotically as
$\epsilon_{\bf k} - \mu \rightarrow \infty$. The weight of this peak
starts at the Fermi surface ($\epsilon_{\bf k} = \mu$) as $1/(1 + \lambda)$
and quickly goes to zero according to Eq. (\ref{residue}) as 
$\omega_{\bf k} \rightarrow \omega_E$, which occurs for $\epsilon_{\bf k} 
{> \atop \sim} 2\omega_E$. For larger $\epsilon_{\bf k}$ a quasiparticle
peak forms once again, albeit with non-zero width, at approximately the
non-interacting electron energy, $\epsilon_{\bf k} = \mu$. At intermediate
$\epsilon_{\bf k} \approx \omega_E$, the quasiparticle picture has broken
down, and a description as described here is required for a complete
picture. 

How well the Migdal approximation works in specific circumstances is
the subject of ongoing research (see, for example, Refs. 
\cite{scalettar89,marsiglio90,vonderlinden95,alexandrov95,botti00}, and the
Appendix. For example, Alexandrov {\it et al.} \cite{alexandrov87} found
an apparent breakdown (for coupling strengths greater than 1, within the
Holstein model) to the approximation when a finite electronic bandwidth
was taken into account. 

We have focussed on the modifications to the electron spectral function
due to the electron-phonon interaction. For excitations at the Fermi
level ($\epsilon_{\bf k} = \mu$), the quasiparticle pole remains there
($\omega_{\bf k_F} = 0$), remains infinitely long-lived (it is a
delta-function), but has a reduced weight, by a factor of $1 + \lambda$. 
This same factor enhances the effective mass, and alters
various normal state properties in a similar way \cite{prange64,grimvall81}.
For example, the low temperature electronic specific heat is linear in
temperature with coefficient usually denoted by $\gamma$, which is
proportional to the electron density of states. The electron-phonon
interaction enhances this coefficient by the same factor, $1 + \lambda$.
Other renormalizations are reviewed in Ref. \cite{grimvall81}.

\subsection{Eliashberg Theory}

Eliashberg theory is the natural development of BCS theory to include
retardation effects due to the `sluggishness' of the phonon response.
In fact, insofar as BCS introduced an energy cutoff, $\omega_D$ (the
Debye frequency), they included, in the most minimal way, retardation
effects. However, Eliashberg theory goes well beyond this approximation,
and handles momentum cutoffs and frequency cutoffs separately. We
begin this section with a very
brief review of BCS theory, followed by a more detailed discussion of
Eliashberg theory.

\subsubsection{BCS Theory}

Before one establishes a theory of superconductivity, one requires a
satisfactory theory of the normal state. In conventional superconductors,
Fermi Liquid Theory appears to work very well, so that, while we cannot
solve the problem of electrons interacting through the Coulomb interaction,
experiment tells us that Coulomb interactions give rise to well-defined
quasiparticles, i.e. a set of excitations which are in one-to-one
correspondence with those of the free-electron gas. The net result is that
one begins the problem with a `reduced' Hamiltonian,
\be 
H_{\rm red}  =  \sum_{\bk \sigma}\epsilon_\bk c^\dagger_{\bk \sigma} 
c_{\bk \sigma}
\phantom{a}  + \phantom{a} 
\sum_{\bk \bk^\prime}
V_{\bk,\bk^\prime} c^\dagger_{\bk \uparrow} c^\dagger_{-\bk \downarrow}
c_{-\bk^\prime \downarrow} c_{\bk \uparrow},
\label{ham_red}
\ee                
where, for example, the electron energy dispersion $\epsilon_\bk$
already contains much of the effect due to Coulomb interactions.
The important point is that well-defined quasiparticles with a
well-defined energy dispersion near the Fermi surface are assumed
to exist, and are summarized by the dispersion $\epsilon_\bk$.
The pairing interaction $V(\bk,\bk^\prime)$ is assumed to be
`left-over' from the main part of the Coulomb interaction, and this is
the part that BCS simply modelled, based on earlier work by Fr\"ohlich
\cite{frohlich50} and Bardeen and Pines \cite{bardeen55}. 

Complete derivations of BCS theory have been provided elsewhere in this
volume; here we state the final result \cite{schrieffer64}:
\be
\Delta_{\bf k} = - {1 \over N} \sum_{\bf k^\prime} 
V_{\bk,\bk^\prime} { \Delta_{\bf k^\prime}
\over 2 E_{\bf k^\prime} } \tanh{\beta E_{\bf k^\prime} \over 2},
\label{bcs1}
\ee
where 
\be
E_{\bf k} = \sqrt{(\epsilon_{\bf k} - \mu)^2 + \Delta_{\bf k}^2}
\label{eek}
\ee
is the quasiparticle energy in the superconducting state, and 
$\Delta_{\bf k}$ is the variational parameter used by BCS. 
An additional equation which must be considered alongside the gap
equation (\ref{bcs1}) is the number equation,
\be
n = 1 - {1 \over N} \sum_{\bk} {\epsilon_{\bk} - \mu \over E_{\bk}} 
\tanh{\beta E_{\bf k} \over 2}.
\label{bcs2}
\ee                
Given a pair potential and an electron density, one has to `invert'
these equations to determine the variational parameter $\Delta_{\bf k}$
and the chemical potential. For conventional superconductors the chemical
potential hardly changes on going from the normal to the superconducting
state, and the variational parameter is much smaller than the chemical
potential, with the result that the second equation was usually ignored.

BCS then modelled the pairing interaction as a negative (and therefore
attractive) constant with a sharp cutoff in momentum
space:
\be
V_{\bk,\bk^\prime} \approx - V \theta(\omega_D - \mid (\epsilon_{\bf k} 
-\mu) \mid) \theta(\omega_D - \mid (\epsilon_{\bf k^\prime} -\mu) \mid).
\label{bcs_pot}
\ee
Using this potential in Eq. (\ref{bcs1}), along with a constant 
density of states assumption over the entire range of integration,
we obtain
\be
{1 \over \lambda} = \int_0^{\omega_D} {d\epsilon \over E} \, 
\tanh{\beta E \over 2},
\label{bcs3}
\ee
where $\lambda \equiv N(\mu)V$. At $T = 0$, the integral can be done 
analytically to give
\be
\Delta = 2 \omega_D {\exp{(-1/\lambda)} \over 1 - \exp{(-1/\lambda)} }.
\label{bcs_zero}
\ee
In weak coupling this becomes the more familiar
\be
\Delta = 2 \omega_D \exp{(-1/\lambda)},
\label{bcs_zero_weak}
\ee    
while in strong coupling we obtain
\be
\Delta = 2 \omega_D \lambda.
\label{bcs_zero_strong}
\ee  
Both of these results are within the realm of BCS theory (at zero
temperature) \cite{leggett80,nozieres85}, although the latter 
generally requires
a self-consistent solution with the number equation, Eq. (\ref{bcs2}).

Close to the critical temperature, $T_c$, the BCS equation becomes
\be
{1 \over \lambda} = \int_0^{\beta \omega_D/2 } dx \,
{\tanh{x} \over x},
\label{bcs_tc}
\ee     
which can't be solved in terms of elementary functions for arbitrary 
coupling strength. Nonetheless, in weak coupling, one obtains
\be
T_c = 1.13\omega_D \exp{(-1/\lambda)},
\label{bcs_tc_weak}
\ee
and in strong coupling
\be
T_c = \omega_D \lambda /2.
\label{bcs_tc_strong}
\ee
It is clear that $T_c$ or the zero temperature variational parameter
$\Delta$ depend on material properties such as the phonon spectrum
($\omega_D$), the electronic structure ($N(\mu)$) and the electron-ion
coupling strength ($V$). However, it is possible to form various
thermodynamic ratios, which
turn out to be independent of material parameters. The obvious example
from the preceding equations is the ratio 
${2\Delta \over k_B T_c}$.
In weak coupling (most relevant for conventional superconductors), for
example, we obtain
\be
{2\Delta \over k_B T_c} = 3.53,
\label{gap_ratio}
\ee
a universal result, independent of the material involved.
Many other such ratios can be determined within BCS theory, and the 
observed deviations from these universal values contributed to
the need for an improved formulation of BCS theory. For example,
the observed value of this ratio in superconducting Pb was closer
to 4.5, a result that is readily understood with Eliashberg theory.
It is worth noting that simply extending BCS theory to the strong
coupling limit (see Eqs. (\ref{bcs_zero_strong},\ref{bcs_tc_strong})
above) results again in a universal constant, ${2\Delta \over k_B T_c} = 4$,
which is the maximum value attainable within BCS theory with a constant
interaction \cite{swihart62}, and is still clearly too low.

Other aspects of BCS theory, particularly those which prove to inadequately
account for the superconducting properties of some materials (notably Pb and
Hg) will not be reviewed here. Instead, we will make reference to the
BCS limit as we encounter various properties within the experimental or
Eliashberg context.

\subsubsection{Eliashberg Equations}

In most reviews and texts that derive the Eliashberg equations, the starting
point is the Nambu formalism \cite{nambu60}. While this formalism simplifies
the actual derivation, it also provides a roadblock to further understanding
for the uninitiated. For this reason we have followed the conceptually
much more straightforward approach (provided by Rickayzen \cite{rickayzen65},
for example) in the derivation outlined in the Appendix. The result can be
summarized by the following set of equations: 
\bea 
\Sigma(\kiwm) & \equiv & {1 \over N\beta} \sum_{\bkp,m^\prime}
{\lambda_{\bk \bkp}(\iwm - \iwmp) \over N(\mu)} G(\kiwmp)
\label{g1}
\\
\phi(\kiwm) & \equiv & {1 \over N\beta} \sum_{\bkp,m^\prime}
\Biggl[{\lambda_{\bk \bkp}(\iwm - \iwmp) \over N(\mu)}  - V_{\bf k k^\prime}
 \Biggr]F(\kiwmp),
\label{g2}
\\
G(\kiwm) & = & {G^{-1}_n(\kiwm) \over G^{-1}_n(\kiwm)  G^{-1}_n(-\bk,-\iwm)
+\phi(\kiwm) \bar{\phi}(\kiwm) }
\label{g3}
\\
F(\kiwm) & = & {\phi(\kiwm) \over G^{-1}_n(\kiwm)  G^{-1}_n(-\bk,-\iwm)
+\phi(-\bk,-\iwm) \bar{\phi}(-\bk,-\iwm) }
\label{g4}
\\
G^{-1}_n(\kiwm) & = & G^{-1}_\circ(\kiwm) - \Sigma(\kiwm).
\label{g5}
\eea
Another couple of equations identical to Eqs. (\ref{g2}) and (\ref{g4}),
except with $\bar{\phi}$ and $\bar{F}$ instead of $\phi$ and $F$, have
been omitted; they indicate that some choice of phase is possible, which
will be important for Josephson effects \cite{josephson62} but not for
what will be considered in the remainder of this chapter. Therefore, we
use $\bar{\phi} = \phi$ \cite{remark4}. 

Note that $G^{-1}_\circ(\kiwm)$ is
the inverse of the non-interacting Green function, in which Hartree-Fock
contributions from both the electron-ion and electron-electron interactions
are assumed to be contained.

Following the standard practice we have used a kernel given by
\be
\lambda_{\bk \bkp} (z) \equiv \int_0^\infty {2 \nu \alpha_{\bk \bkp}^2
F(\nu) \over \nu^2 - z^2} d \nu
\label{lambda_z}
\ee
\noindent where $\alpha_{\bk \bkp}^2 F(\nu)$ is given by Eq. (\ref{alpha1}).
Eqs. (\ref{g1}-\ref{lambda_z}) have been written in a fairly general way;
in this way they can be viewed as having arisen from a microscopic Hamiltonian
as in Eqs. (\ref{ham_BLF}-\ref{ham_Hol}) (although electron-electron
interactions have been included in the pairing channel only, and not in
the single electron self energy), or, alternatively, from a treatment
of real metals, where, as mentioned earlier, the electron
and phonon structure come from previous calculations and/or experiments.
These equations emphasize the electron-ion interaction; attempts to 
explain superconductivity through the electron-electron interactions
have been proposed in the past, mainly through collective modes
\cite{little64,ginzburg64,berk66,grabowski84,hirsch85,bickers87,takada88,millis90}; 
some of these attempts will be treated elsewhere
in this volume in the context of high temperature superconductivity.

Assuming the electron and phonon structure is given, Eqs.
(\ref{g1}-\ref{lambda_z}) must be solved for the two functions,
$\Sigma(\kiwm)$ and $\phi(\kiwm)$.
The procedure is as
follows: it is standard practice to
separate the self energy, $\Sigma(\kiwm)$, into its even and odd components
\cite{allen82}:
\bea
i\omega_m \bigl[ 1 - Z({\bf k},i\omega_m) \bigr]  & \equiv &
{1 \over 2} \bigl[ \Sigma(\kiwm) - \Sigma({\bf k},-i\omega_m) \bigr]
\nonumber \\
\chi(\kiwm) & \equiv & 
{1 \over 2} \bigl[ \Sigma(\kiwm) + \Sigma({\bf k},-i\omega_m) \bigr]
\label{even_odd}
\eea
where $Z$ and $\chi$ are both even functions of $i\omega_m$ (and, as
we've assumed all along, ${\bf k}$). Then, Eq. (\ref{g1}) becomes two
equations,
\bea
Z({\bf k},i\omega_m) = 1 + {1 \over N\beta}
\sum_{\bkp,m^\prime}
{\lambda_{\bk \bkp}(\iwm - \iwmp) \over N(\mu)} 
{ \bigl(\omega_{m^\prime} / \omega_m \bigr)
Z({\bf k^\prime},i\omega_{m^\prime}) \over 
\omega_{m^\prime}^2 Z^2({\bf k^\prime},i\omega_{m^\prime}) + 
\bigl( \epsilon_{\bf k^\prime} - \mu + \chi({\bf k^\prime},i\omega_{m^\prime})
\bigr)^2  + \phi^2({\bf k^\prime},i\omega_{m^\prime})}
\label{ga1}
\\                                                          
\chi({\bf k},i\omega_m) = - {1 \over N\beta}
\sum_{\bkp,m^\prime}
{\lambda_{\bk \bkp}(\iwm - \iwmp) \over N(\mu)}
{ \epsilon_{\bf k^\prime} - \mu + \chi({\bf k^\prime},i\omega_{m^\prime}) 
\over
\omega_{m^\prime}^2 Z^2({\bf k^\prime},i\omega_{m^\prime}) +
\bigl( \epsilon_{\bf k^\prime} - \mu + \chi({\bf k^\prime},i\omega_{m^\prime})
\bigr)^2  + \phi^2({\bf k^\prime},i\omega_{m^\prime})}
\label{ga2}   
\eea
along with the gap equation (Eq. (\ref{g2})):
\be
\phi({\bf k},i\omega_m) = {1 \over N\beta}
\sum_{\bkp,m^\prime}
\biggl(
{\lambda_{\bk \bkp}(\iwm - \iwmp) \over N(\mu)} - V_{\bf k k^\prime} \biggr) 
{ \phi({\bf k^\prime},i\omega_{m^\prime})
\over
\omega_{m^\prime}^2 Z^2({\bf k^\prime},i\omega_{m^\prime}) +
\bigl( \epsilon_{\bf k^\prime} - \mu + \chi({\bf k^\prime},i\omega_{m^\prime})
\bigr)^2 + \phi^2({\bf k^\prime},i\omega_{m^\prime})}.
\label{ga3}     
\ee
These are supplemented with the electron number equation, which determines
the chemical potential, $\mu$:
\bea
n & = & {2 \over N\beta} \sum_{\bkp,m^\prime} 
G({\bf k^\prime},i\omega_{m^\prime}) e^{i\omega_{m^\prime}0^+} 
\label{number_exact}
\\
& = & 1 - {2 \over N\beta} \sum_{\bkp,m^\prime} 
{ \epsilon_{\bf k^\prime} - \mu + \chi({\bf k^\prime},i\omega_{m^\prime})
\over
\omega_{m^\prime}^2 Z^2({\bf k^\prime},i\omega_{m^\prime}) +
\bigl( \epsilon_{\bf k^\prime} - \mu + \chi(({\bf k^\prime},i\omega_{m^\prime})
\bigr)^2  + \phi^2({\bf k^\prime},i\omega_{m^\prime})}.
\label{ga4}
\eea
These constitute general Eliashberg equations for the electron-phonon
interaction, in which electron-electron interactions enter explicitly
only in the pairing equation. 
Very complete calculations of these functions (linearized, for the 
calculation of $T_c$) were carried out for
Nb by Peter {\it et al.} \cite{peter77}, and for Pb by Daams \cite{daams79}.

The more standard practice
is to essentially confine all electronic properties to the Fermi surface;
then only the anisotropy of the various functions need be considered. Often
these are simply averaged over (due to impurities, for example), or the
anisotropy may be very weak and therefore neglected.
In this case the equations (\ref{ga1}-\ref{ga4}) can be written 
\bea
Z_m & = & 1 + \pi T \sum_{m^\prime} \lambda(i\omega_m - i\omega_{m^\prime})
{( \omega_{m^\prime} / \omega_m ) Z_{m^\prime} \over 
\sqrt{ \omega_{m^\prime}^2 Z^2_{m^\prime} + \phi^2_{m^\prime} }} A_0(m^\prime)
\label{gb1}
\\
\chi_m & = & - \pi T \sum_{m^\prime} \lambda(i\omega_m - i\omega_{m^\prime})
A_1(m^\prime)
\label{gb2}
\\        
\phi_m & = & \pi T \sum_{m^\prime} 
\biggl( \lambda(i\omega_m - i\omega_{m^\prime}) - N(\mu) V_{\rm coul} \biggr)
{\phi_{m^\prime} \over
\sqrt{ \omega_{m^\prime}^2 Z^2_{m^\prime} + \phi^2_{m^\prime} }} A_0(m^\prime)
\label{gb3}
\\    
n & = & 1 - 2 \pi T N(\mu) \sum_{m^\prime} A_1(m^\prime)
\label{gb4}
\eea
where we have adopted the shorthand $Z(i\omega_m) = Z_m$, etc, $\lambda(z)$
and $V_{\rm coul}$ represent appropriate Fermi 
surface averages of the quantities
involved, and the functions $A_0(m^\prime)$ and $A_1(m^\prime)$ are given
by integrals over appropriate density of states, using the prescription
(\ref{dos}) to convert from Eqs. (\ref{ga1}-\ref{ga4}) to 
Eqs. (\ref{gb1}-\ref{gb4}). If the electron density of states is assumed
to be constant, then, with the additional approximation of infinite
bandwidth, $A_0(m^\prime) \equiv 1$ (actually a cutoff,
$\theta(\omega_c - \mid \omega_{m^\prime} \mid)$, is required in
Eq. (\ref{gb3})), and $A_1(m^\prime) \equiv 0$. This last result
effectively removes $\chi_m$ (and Eqs. (\ref{gb2},\ref{gb4}) ) from
further consideration. An earlier review by one of us \cite{carbotte90}
covered the consequences of the remaining two coupled equations in
great detail. 

Nonetheless, a considerable effort has been devoted to examining gap
anisotropy, as well as variations in the electronic density of states near the
Fermi surface. We describe some of this work in the following few paragraphs.

Referring back to Eqs. (\ref{ga1}-\ref{ga4}), one can rewrite the summation
over ${\bf k^\prime}$ on the right-hand-side of these equations as an
integral over energy plus an integral over angle (for a given constant energy
surface). In carrying out the energy integration the energy dependent
electron density of states (EDOS), $N(\epsilon)$, introduces a new
weighting factor if $N(\epsilon)$ exhibits variations over the energy
scale of the phonon frequencies. On the other hand, the integration over
angle will account for variations of the gap and other quantities in the
integrands with momentum direction. There is a large literature on each of
these complicating effects, starting with anisotropy effects 
\cite{clem66,weber77}, and more recently with EDOS energy dependence 
\cite{horsch77,nettel77,lie78,allen82}. 

Concerning anisotropy, the observed universal decrease in $T_c$ with
increasing impurity concentration (i.e. so-called `normal' impurities,
deemed to be innocuous by Anderson's argument \cite{anderson59}) can be 
attributed to the washing out of gap anisotropy. To see why this decreases
$T_c$ (we omit here effects due to valence changes) we note that the
impurity potential scattering has a tendency to homogenize the gap 
on the Fermi surface. This tends to reduce the gap in some directions,
and it is these directions that make the maximum contribution to $T_c$, and
so $T_c$ is reduced.
A simple BCS calculation can demonstrate this analytically. One makes a separable
approximation for the pairing potential, Eq. (\ref{bcs_pot}), to be used
in the BCS equation (\ref{bcs1}):
\be
V_{\bk,\bk^\prime} = - V(1 + a_{\bf k})(1 + a_{\bf k^\prime}),
\label{separable}
\ee
where the same energy cutoffs are assumed, and $a_{\bf k}$ is a function
of momentum direction only. Assuming $a_{\bf k}$ to be small with
a Fermi surface average equal to zero (i.e. $<a_{\bf k}> = 0$) and
$a^2_{\bf k} = a^2$, with $<>$ denoting an angular average over the
Fermi surface, then clearly $\Delta_{\bf k} = \Delta_\circ (1 + a_{\bf k})$.
Solving the resulting equation yields
\be
<\Delta_{\bf k}> = \Delta_\circ = 2 \omega_D 
\exp{(- {1 \over \lambda (1+a^2)})} \biggl(1 -{3 \over 2} a^2 \biggr)
\label{bcs_ani1}
\ee
in the weak coupling approximation. Similarly, one can solve the $T_c$
equation, to obtain
\be
T_c = 1.13 \omega_D \exp{(- {1 \over \lambda (1+a^2)})}.
\label{bcs_ani2}
\ee
This last equation demonstrates that $T_c$ is increased by anisotropy.
Hence, increased scattering due to impurities will decrease $T_c$, as the
anisotropy is washed out. Finally, the gap ratio,
\be
{2 <\Delta_{\bf k}> \over k_BT_c} = 3.53 \biggl(1 - {3 \over 2} a^2 \biggr),
\label{bcs_ani3}
\ee
showing that anisotropy reduces this quantity.

How big can the anisotropy be in pure conventional superconductors~?
Microscopically the anisotropy is related to band structure anisotropy plus
anisotropy in the electron-phonon spectral function from Eq. (\ref{alpha1}),
$\alpha^2F({\bf k,k^\prime},\nu)$. In Fig. 5 we show the results of
a calculation of the gap anisotropy in Pb as a function of position
on the Fermi surface \cite{tomlinson76}. These calculations include
multiple-plane-wave effects for the electronic wave functions, and the 
corresponding distortions of the Fermi surface from a sphere, as well as 
anisotropy effects due to the phonons and umklapp processes in the electron
phonon interactions. The Figure illustrates the gap $\Delta_\circ(\theta,\phi)$
at zero temperature, as a function of $\theta$ for three constant $\phi$
arcs. Solid angle regions where the Fermi surface of Pb does not exist
are indicated by vertical  solid lines. It is clear that the pure Pb crystal
gap is highly anisotropic, varying by about 20\% over the Fermi
surface. As described above, impurities will wash out this anisotropy.
Nevertheless, such anisotropies can be observed in some low temperature
properties, like the specific heat. For more details the reader is referred
to Ref. \cite{weber77}.

The other complication we have mentioned is an energy variation in the EDOS,
as seems to exist in some A15 compounds. If this energy dependence occurs
on a scale comparable to $\omega_D$, then $N(\epsilon)$ cannot be 
assumed to be constant, and cannot be  taken outside of the integrals in
Eqs. (\ref{ga1}-\ref{ga4}). Such EDOS energy dependence is thought to be 
responsible for some of the anomalous properties seen in A15 compounds ---
their magnetic susceptibility and Knight shift \cite{clogston61}, and the
structural transformation from cubic to tetragonal 
\cite{testardi73,weger73,izymov74}. Several electronic band structure 
calculations \cite{klein78,ho78,pickett80,klein80} also find 
sharp structure in $N(\epsilon)$ at the Fermi level. An accurate description
of the superconducting state thus requires a proper treatment of this
structure. This was first undertaken to understand $T_c$ by Horsch and
Reitschel \cite{horsch77} and independently by Nettel and Thomas
\cite{nettel77}. A more general approach to understanding the effect of
energy dependence in $N(\epsilon)$ on $T_c$ was given by Lie and Carbotte
\cite{lie78}, who formulated the functional derivative $\delta T_c/\delta
N(\epsilon)$; they found that only values of $N(\epsilon)$ within 5 to
10 times $T_c$ around the chemical potential have an appreciable effect
on the value of $T_c$. More specifically they found that $\delta T_c/\delta
N(\epsilon)$ is approximately a Lorentzian with center at the chemical
potential; the function becomes negative only at energies $|\epsilon - \mu|
{> \atop \sim} 50 T_c$.

Irradiation damage experiments illustrate some of this dependency. 
For example, irradiation of Mo$_3$Ge causes an increase in $T_c$
\cite{gurvitch78}. Washing out gap anisotropy with the irradiation
cannot possibly account for an increase in $T_c$; instead, this result
finds a natural explanation in the fact that the chemical potential
for Mo$_3$Ge falls in a valley \cite{ghosh80} of the EDOS, and irradiation
smears the EDOS, thus increasing $N(\mu)$, and hence $T_c$.

For details on the formulation of Eliashberg theory with an energy
dependent $N(\epsilon)$ the reader is referred to the work of
Pickett \cite{pickett80b} and Mitrovi\'c and Carbotte \cite{mitrovic83},
and references therein. The energy dependent EDOS affects many properties.
To illustrate a typical result we show in Fig. 6 the effect of 
an energy dependent EDOS on the current (I)-voltage (V) characteristics 
of a tunneling junction \cite{mitrovic81,mitrovic83}. A detailed
discussion of tunneling appears in Section 3.3.2.
The tunneling conductance is proportional to the electron
density of states, and is denoted by $\sigma(\omega) \equiv {\rm Re}
\Biggl( {\omega \over \sqrt{\omega^2 - \Delta^2(\omega)}} \Biggr)$. Fig.
6 shows the difference with the BCS conductance, $\sigma(\omega)/
\sigma_{\rm BCS}(\omega) - 1$ vs. $\omega - \Delta_\circ$ 
\cite{mitrovic83,mitrovic81}. Fig. 6a (b) is for a peak (valley) in the
EDOS at the Fermi level. The solid curves include the effect of an energy 
dependent EDOS, while the dashed curves do not (the EDOS is approximated
by a constant value, $N(\mu)$). In these examples the electron phonon 
spectral density obtained for Nb$_3$Sn \cite{shen72} is used.

These differences can be highlighted in another way, shown in Fig. 7
\cite{mitrovic81,mitrovic83}.
Here, the ``effective'' electron phonon spectral density, 
$\alpha^2F(\Omega)_{\rm eff}$, is obtained
by inverting the solid curves in Fig. 6 under the assumption that
the EDOS is constant and equal to $N(\mu)$. The dashed curves give
Shen's original $\alpha^2F(\Omega)$ while the solid curves are the
result of (incorrectly) inverting the result obtained with an energy
dependent EDOS, but not accounting for it in the inversion process
itself. The actual EDOS used to generate the I-V characteristic 
is shown in the inset for each figure. It contains a peak in Fig. 7a
and a valley in Fig. 7b. Clearly a peak introduces a negative tail into
$\alpha^2F(\Omega)_{\rm eff}$, which of course is not present in the
actual $\alpha^2F(\Omega)$. For other important modifications the reader
is referred to the references. The rest of this chapter will focus
primarily on the `standard' theory, using Eqs. (\ref{gb1}-\ref{gb4})
with $A_0(m) \equiv 1$ and $A_1(m) \equiv 0$.

All of the equations discussed so far have been developed on the
imaginary frequency axis. Because practitioners in the field at the
time were interested in tunneling spectroscopy measurements \cite{rowell63},
the
theory was first developed on the real frequency axis 
\cite{schrieffer63,scalapino66}. The resulting
equations are complicated, even for numerical solution. It
wasn't until quite a number of years later that numerical work
returned to the imaginary axis \cite{owen71}, where, for 
thermodynamic properties, the numerical solution was very efficient 
\cite{bergmann73,rainer74,allen75,daams81}. The difficulty, however,
was that imaginary axis solutions are not suitable for dynamical
properties. We will return to the interplay between imaginary and real
frequency axis solutions as we ecounter them throughout the chapter. 

\section{The Phonons}

\subsection{Neutron Scattering}

When dealing with model Hamiltonians, the phonon dispersion relations
(before interaction with the electrons) are generally given, and simple:
they are Einstein modes, or Debye-like modes, for example. A noteable
exception is the case where the model contains anharmonic forces, in which
case even the `non-interacting' phonon spectrum is unknown.

In the case of real solids, and in particular metals, the situation
is much worse. In this case the electrons cannot be ignored, though
they can be treated in the Born-Oppenheimer approximation. Nonetheless
the results require parametrization (with input from other experiments)
and are generally not reliable. Pseudopotential methods 
\cite{phillips59,harrison66} can be applied to this problem,
again, with limited success. In contrast, the spectacular success of
inelastic neutron  scattering techniques \cite{brockhouse62,stedman67}
to simply measure the phonon dispersion curves in real metals effectively
eliminates the need to calculate them quantitatively. Various qualitative
effects, like the impact of electronic screening to the long wavelength
ionic plasma mode \cite{bohm50}, as well as the existence of Kohn 
anomalies \cite{kohn59}, all due to the presence of electrons, are
understood theoretically. For detailed results, however, Born-von Karman
fits to high symmetry phonon dispersions suffice for an excellent
description of the low temperature phonon properties. At temperatures
of order $10$ K, the phonons in most conventional superconductors are
completely determined, and no longer changing with temperature.   
Hence, as far as understanding (low temperature) superconductivity is
concerned, these higher temperature measurements are sufficient.

The measured dispersion curves, $\omega_{\bf q}$ (again, branch indices
are suppressed), are summarized in the frequency distribution
\be
F(\nu) = {1 \over N} \sum_{\bf q} \delta(\nu - \omega_{\bf q} ),
\label{dense_pho}
\ee
where $N$ is the number of ions in the system, and ${\bf q}$ is
a wavevector which ranges over the entire First Brillouin Zone (FBZ),
(and implicitly contains the branch index). It should be stressed that
this procedure is an idealization; in actual fact a set of `constant ${\bf q}$'
scans are performed (usually along high symmetry directions). A typical 
result \cite{brockhouse62} is shown in Fig. 8 for Pb, for a set of wavevectors
along the diagonal in reciprocal space.
Note that the neutron counts tend to form a peak as a function of energy
transfer (to the neutron), $\hbar \nu$. In general these peaks have a finite
width, i.e. broader than the spectrometer resolution; these are due to
a variety of effects, for example, anharmonic effects. Nonetheless,
because the peaks are relatively sharp compared to the centroid
energy, (i.e. the phonon inverse lifetimes are small compared to their
energies), these data are usually presented in the form of Fig. 9,
as a set of dispersion curves. Fig. 9 does obscure, however, the
lifetimes of the various phonons, and hence the validity of Eq.
(\ref{dense_pho}), where infinitely long-lived phonons are assumed
throughout the Brillouin zone, is called into question.

Nonetheless, for most of the Brillouin zone the approximation of
infinitely long-lived excitations is a good one (hence, the name,
phonon), and so the spectrum of excitations can be constructed
according to Eq. (\ref{dense_pho}). Such a procedure relies on coherent
neutron scattering. An alternative is to use {\it incoherent}
neutron scattering, whereby one measures the spectrum more or less directly.
This latter procedure has advantages over the former, but also includes
multiphonon scattering processes, and for non-elemental materials,
weighs the contribution from each element differently, according to
their varying scattering lengths. The result is often denoted the
`generalized density of states' (GDOS). A comparison for a Thallium-Lead
alloy is shown in Fig. 10 \cite{brockhouse68,roy70}. Also shown is the
result from tunneling, to be discussed in the next subsection. There
is clearly good agreement between the various methods. Amongst the two
neutron scattering techniques, inelastic
coherent neutron scattering produces the sharpest features, but requires
a model (i.e. a Born-von Karman fit) to extract the spectrum $F(\nu)$ from
the dispersion curves measured along high symmetry directions.

\subsection{The Eliashberg Function, $\alpha^2F(\nu)$: Calculations}

First-principle calculations of the electron-phonon spectral function,
$\alpha^2F(\nu)$ require a knowledge of the electronic wave functions,
the phonon spectrum, and the electron-phonon matrix elements between two
single-electron Bloch states. A fairly comprehensive review is given
in Ref. \cite{grimvall81}. For our purposes, we note that, since
the phonon spectrum will come from experiment, Eq. (\ref{alpha1}) requires
calculation of $g_{\bf k,k^\prime}$. It is \cite{carbotte90,grimvall81}
\be
g_{{\bf k,k^\prime} j} = <\psi_{\bf k} \mid {\bf \epsilon}^j
({\bf k-k^\prime}) \cdot {\bf \nabla V} \mid \psi_{\bf k^\prime} >
\left[  {\hbar \over 2 M \omega_j({\bf k
-k^\prime}) } \right]^{1/2} 
\label{gkk}
\ee
where, for this equation we have included the phonon branch index $j$
explicitly. The Bloch state is denoted $\mid \psi_{\bf k} > $, and
${\bf \epsilon}^j({\bf k})$ is the polarization vector for the
($j{\bf k}$)th phonon mode. The crystal potential is denoted $V$, and
as one might expect, the electron-phonon coupling depends on its
gradient. 

Tomlinson and Carbotte \cite{tomlinson77} used pseudopotential methods
\cite{appapillai73,anderson63} to compute $g_{{\bf k,k^\prime} j}$
and, from Eq. (\ref{alpha1}), $\alpha^2F(\nu)$, for Pb. The phonons were
taken from experiment \cite{brockhouse62,stedman67,kotov68,cowley74}
through Born - von K\'arm\'an fits. The result is plotted in Fig. 11,
along with results from tunneling experiments (to be described below).
The agreement is qualitatively very good; this provides very strong 
confirmation of the electron phonon mechanism of superconductivity.

Further details of more modern calculations of electron-phonon coupling
constants can be found in, for example, Refs. \cite{krakauer93} and 
\cite{gunnarson97} and references therein. Their reliability appears to
remain an issue, both with the high temperature cuprates, and perhaps less
so with the fulleride and more conventional superconductors.
The spirit of these calculations is somewhat different than the older ones,
in that coupling constants are extracted from the phonon linewidths, where
it is assumed that the phonon broadening is entirely due to the electron-ion
interaction (and not, say, anharmonic effects). Allen \cite{allen72,allen74}
derived a formula (Fermi's Golden Rule) for the inverse
lifetime, $\gamma_{\bf q}(\nu)$, of a phonon with momentum 
(and branch index) ${\bf q}$:
\be
\gamma_{\bf q} =  2 \pi \omega_{\bf q} \sum_{\bf k} 
|g_{{\bf k,k^\prime}}|^2 \left[
{f({\epsilon_{\bf k + q}} - \mu) - f({\epsilon_{\bf k}} - \mu) \over
\hbar \omega_{\bf q}} \right] 
\delta (\epsilon_{\bf k + q} + \hbar \omega_{\bf q} - \epsilon_{\bf k}),
\label{allen1}
\ee
where again we have suppressed both phonon branch indices and electron
band labels. Using this equation, in the approximation that the expression
$\left[ f(\epsilon_{\bf k + q} - \mu) - f(\epsilon_{\bf k} - \mu) \right]/
(\hbar \omega_{\bf q}) $ is replaced by $\delta ({\epsilon_{\bf k}} - \mu)$
makes it resemble Eq. (\ref{a2f}), so that one can write
\bea
\alpha^2F(\nu) & = & {1 \over \pi N(\mu)} {1 \over N} 
\sum_{\bf q} {1 \over 2} {\gamma_{\bf q} \over \hbar \omega_{\bf q}}
\delta (\nu - \omega_{\bf q})
\nonumber
\\
& = & {1 \over 3N} \sum_{\bf q} {1 \over 2} \omega_{\bf q} \lambda_{\bf q} 
\delta (\nu - \omega_{\bf q})
\label{allen2}
\eea
where the second line serves to define a ${\bf q}$-dependent coupling 
parameter:
\be
\lambda_{\bf q} \equiv {3 \over \pi N(\mu)} {\gamma_{\bf q} \over \hbar 
\omega^2_{\bf q}}.
\label{allen3}
\ee
It is through these relations that coupling parameters are often determined.

It is worth noting at this point that several moments of the function
$\alpha^2F(\nu)$ have played an important role in characterizing retardation
(and strong coupling) effects in superconductivity. Foremost amongst these
is the mass enhancement parameter, $\lambda$, already defined in
Eq. (\ref{lambda}); in addition,
the characteristic phonon frequency, $\omega_{\ln}$ is given by
\be
\omega_{\ln} \equiv \exp{\left[ {2 \over \lambda}
\int_0^\infty d\nu \, \ln{(\nu)} {\alpha^2F(\nu) \over \nu}  \right]}.
\label{wln}
\ee
Further discussion of these calculations can be found in Refs.
\cite{grimvall81,carbotte90}.

\subsection{Extraction from Experiment}

Experiments which probe dynamical properties do so as a function of
frequency, which is a real quantity. However, the Eliashberg equations
as formulated in the previous section are written on the imaginary
frequency axis. To extract information from these equations relevant to
spectroscopic experiments, one must analytically continue these
equations to the real frequency axis. Mathematically speaking, this is not
a unique procedure; one can often imagine several functions whose values
on the imaginary axis are equal, and yet differ elsewhere in the complex
plane (and in particular on the real axis). For example, replacing unity
by $-\exp{(\beta i\omega_m)}$, in any number of places in the equations
does not affect the imaginary axis equations, or their solutions, and
yet on the real axis the corresponding number of factors 
$-\exp{(\beta \omega)}$ will appear.

Physically speaking, however, the Green functions involved have to
satisfy certain conditions; complying with these conditions determines
the function uniquely \cite{baym61}. This allows a unique determination
of the analytic continuation of the Eliashberg equations on the real
axis. This procedure will be discussed in the following subsection,
followed by subsections on experimental spectroscopies, and how they
can be used to extract the Eliashberg function, $\alpha^2F(\nu)$.

\subsubsection{The Real-Axis Eliashberg Equations}

We begin with Eqs. (\ref{g1} - \ref{g5}). To analytically continue
Eqs. (\ref{g3} - \ref{g5}) is trivial; one simply replaces the imaginary
frequency $i\omega_m$ wherever it appears with $\omega + i\delta$. The
$i\delta$ remains to remind us that we are analytically continuing the
function to just above the real axis; it is important to specify this
since there is a discontinuity in the Green function as one crosses the
real axis. A simple replacement of $i\omega_m$ with $\omega + i\delta$ in
Eqs. (\ref{g1},\ref{g2}) (leaving the summations over $m^\prime$)
would in general be incorrect. The correct procedure is to first perform
the Matsubara sum, and then make the replacement. To perform the 
Matsubara sum, however, one has to introduce the spectral representation
for the Green functions, $G$ and $F$. These are given by
\bea
G({\bf k},i\omega_m) & = & \int_{-\infty}^{\infty} d\omega \,
{ A({\bf k},\omega) \over i\omega_m - \omega}
\label{speca}
\\
F({\bf k},i\omega_m) & = & \int_{-\infty}^{\infty} d\omega \,
{ C({\bf k},\omega) \over i\omega_m - \omega},
\label{specc} 
\eea
where $A({\bf k},\omega)$ is given by Eq. (\ref{spect_ele}) and 
$C({\bf k},\omega)$ is given by a similar relation:
\be 
C({\bf k},\omega) \equiv -{1 \over \pi} {\rm Im} F({\bf k},\omega + i\delta).
\label{spectc_ele}
\ee
The spectral representation for the phonons is already present in Eqs.
(\ref{g1},\ref{g2}). Therefore the Matsubara sum can be performed
straightforwardly (see, for example, Refs. \cite{mahan81,allen82}), and
the analytical continuation can be done. Upon integrating over momentum
(using, as in Eqs. (\ref{gb1}-\ref{gb4}) electron-hole symmetry and
a constant (and infinite in extent) density of electron states), one arrives
at the standard real-axis Eliashberg equations \cite{schrieffer63,allen82}.
These equations are much more difficult to solve than the imaginary axis
counterparts. They require numerical integration of principal value
integrals and square-root singularities, and the various Green function
components are complex. In contrast the imaginary axis equations are
amenable to computers (the sums are discrete) and the quantities involved
are real. Moreover a considerable number of thermodynamic and magnetic
properties can be obtained directly from the imaginary axis solutions. 

The discrepancy in computational ease between the two formulations led to 
an alternative path to dynamical
information, namely the direct analytic continuation of {\it the solutions}
of the imaginary axis equations to the real axis by a fitting procedure
with Pad\'e approximants \cite{vidberg77}. This method is in general
very sensitive to the input data, and has (surmountable
\cite{blaschke82,leavens85}) difficulties at
high temperatures and frequencies.

More recently yet another procedure was formulated \cite{marsiglio88},
which first requires a numerical solution of the imaginary axis equations,
followed by a numerical solution of analytic continuation equations. This
latter set is formally exact (i.e. no fitting required) and yet avoids the
complications of the real-axis equations. These equations are
\bea
\Sigma(\bk,z) ={1 \over N\beta}
\sum_{\bkp m^\prime=-\infty}^\infty
{\lambda_{\bk \bkp}(z - \iwmp) \over N(\mu)} G(\bkp,\iwmp)
\phantom{a} -
\phantom{ {1 \over N\beta}
\sum_{\bkp m^\prime=-\infty}^\infty
{\lambda_{\bk \bkp}(z - \iwmp) \over N(\mu)} G(\bkp,\iwmp)}
\nonumber \\
{1 \over N} \sum_{\bkp} \int_0^\infty d\nu
{\alpha_{\bk \bkp}^2F(\nu) \over N(\mu)}
\Biggl\{
\Bigl[ f(\nu - z) + N(\nu) \Bigr] G(\bkp,z-\nu) +
\Bigl[ f(\nu + z) + N(\nu) \Bigr] G(\bkp,z+\nu)
\Biggr\}
\label{anal_sig}
\\
\phantom{aaaaa}
\nonumber \\
\phi(\bk,z) = {1 \over N\beta}
\sum_{\bkp m^\prime=-\infty}^\infty
\Biggl[ {\lambda_{\bk \bkp}(z - \iwmp) \over N(\mu)} - V_{\bk \bkp}
\Biggr] F(\bkp,\iwmp)
\phantom{a} -               
\phantom{\Biggl[ {\lambda_{\bk \bkp}(z - \iwmp) \over N(\mu)} - U \Biggr]
F(\bkp,\iwmp)}
\nonumber \\
{1 \over N} \sum_{\bkp} \int_0^\infty d\nu
{\alpha_{\bk \bkp}^2F(\nu) \over N(\mu)}
\Biggl\{
\Bigl[ f(\nu - z) + N(\nu) \Bigr] F(\bkp,z-\nu) +
\Bigl[ f(\nu + z) + N(\nu) \Bigr] F(\bkp,z+\nu)
\Biggr\},
\label{anal_phi}
\eea              
where $z$ can actually be anywhere in the upper half-plane. Thus, for
example, Eqs. (\ref{g1},\ref{g2}) can be recovered by substituting
$z=i\omega_m$. On the other hand, once these equations have been solved,
one can substitute $z=\omega + i \delta$, and iterate the resulting
equations to convergence. When the ``standard'' approximations for
the momentum dependence are made (i.e. Fermi surface averaging, constant
density of states, particle-hole symmetry, etc.) the result is
\bea
Z(\omega + i\delta) & = & 1 + {i\pi T \over \omega}\sum_{m=-\infty}^{\infty}
\lambda(\omega-i\omega_m)
{\omega_m Z(\iwm) \over \sqrt{\omega_m^2 Z^2(\iwm) + \phi^2(\iwm)} }
\nonumber \\
& & +{i\pi\over \omega}
\int_0^{\infty}d\nu\,\alpha^2F(\nu)\Biggl\{\bigl[N(\nu)+f(\nu-\omega)
\bigr] {(\omega-\nu)Z(\omega-\nu + i \delta) \over
\sqrt{(\omega-\nu)^2 Z^2(\omega-\nu + i \delta) - \phi^2(\omega-\nu+i \delta)}}
\nonumber \\
& & +\bigl[N(\nu)+f(\nu+\omega)\bigr]
{(\omega+\nu)Z(\omega+\nu + i \delta) \over \sqrt{(\omega+\nu)^2 
Z^2(\omega+\nu + i \delta) -
\phi^2(\omega+\nu + i \delta)}}\Biggr\}
\label{anal_zz} \\
\phi(\omega + i \delta) & = & \pi T\sum_{m=-\infty}^{\infty}
\bigl[
\lambda(\omega-i\omega_m)-\mu^*(\omega_c)\theta(\omega_c-|\omega_m|)
\bigr]
{\phi(\iwm) \over \sqrt{\omega_m^2 Z^2(\iwm) + \phi^2(\iwm)} }
\nonumber \\
& & +i\pi\int_0^{\infty}d\nu\,\alpha^2F(\nu)\Biggl\{\bigl[N(\nu)+f(\nu-\omega)
\bigr] {\phi(\omega-\nu + i \delta) \over \sqrt{(\omega-\nu)^2 
Z^2(\omega-\nu + i \delta) -
\phi^2(\omega-\nu + i \delta)}}       
\nonumber \\
& & +\bigl[N(\nu)+f(\nu+\omega)\bigr]
{\phi(\omega+\nu + i \delta) \over \sqrt{(\omega+\nu)^2 
Z^2(\omega+\nu + i \delta) -
\phi^2(\omega+\nu + i \delta)}}\Biggr\}.
\label{anal_phi2}
\eea 
Note that in cases where the square-root is complex, the branch
with positive imaginary part is to be chosen. 

One important point has been glossed over in these derivations. Because of the
infinite bandwidth approximation, an unphysical divergence occurs in the
term involving the direct Coulomb repulsion, $V_{\bf k,k^\prime}$, both in
the imaginary axis formulation, Eq. (\ref{gb3}), and in the real-axis
formulation, Eq. (\ref{anal_phi2}). The solution to this difficulty is
to introduce a cutoff in frequency space (even though the original premise
was that the Coulomb repulsion was frequency {\it independent}), as is
apparent in the two equations. In fact, this cutoff should be of order the
Fermi energy, or bandwidth. However, this requires a summation (or integration)
out to huge frequency scales. In fact one can use a scaling argument 
\cite{bogoliubov59,morel62,marsiglio89} to
replace this summation (or integration) by one which spans a small
multiple ($\approx 6$) of the phonon frequency range. Hence the magnitude
of the Coulomb repulsion is scaled down, and becomes \cite{bogoliubov59}
\be
\mu^\ast(\omega_c) \approx {N(\mu) U  \over 1 + N(\mu) U \ln{\epsilon_F \over
\omega_c} },
\label{pseudo}
\ee
where $U$ is a double Fermi surface average of the direct Coulomb repulsion. 
This reduction is correct physically, in that the retardation due to the
phonons should reduce the effectiveness of the direct Coulomb repulsion
towards breaking up a Cooper pair. It does appear to overestimate this
reduction, however \cite{bonca00}. The analytic continuation of this
part of the equations has been treated in detail in Ref. \cite{leavens80}.

In the zero temperature limit, Eqs. (\ref{anal_zz},\ref{anal_phi2}) are
particularly simple. Then the Bose function is identically zero and
the Fermi function becomes a step function: $f(\nu - \omega) \rightarrow
\theta (\omega - \nu)$. Once the imaginary axis equations have been 
solved, solution of Eqs. (\ref{anal_zz},\ref{anal_phi2}) no longer requires
iteration. One can simply build up the solution by construction from
$\omega = 0$ (assuming $\alpha^2F(\nu)$ has no weight at $\nu = 0$); in fact,
if the phonon spectrum has no weight below a frequency, $\nu_{\rm min}$, then
only the first lines in Eqs. (\ref{anal_zz},\ref{anal_phi2})
need be evaluated. In particular, if the gap (still 
to be defined) happens to occur below this minimum frequency (often a good
approximation for a conventional superconductor) then the gap can be
obtained in this manner \cite{remark5}.

In the following two sections we explore the possibility of using Eqs.
(\ref{anal_zz},\ref{anal_phi2}) to obtain information about the microscopic
parameters of Eliashberg theory. 

\subsubsection{Tunneling}

Perhaps the simplest, most direct probe of the excitations of a solid 
is through
single particle tunneling. In this experiment electrons are injected into
(or extracted from) a sample, as a function of bias voltage, $V$. The
resulting current is proportional to the superconducting density of states
\cite{giaever60,meservey69,duke69,wolf85}:
\be
I_S(V) \propto \int d \omega {\rm Re} \left[ {| \omega | \over 
\sqrt{ \omega^2 - \Delta^2(\omega)}} \right] 
\left[ f(\omega) - f(\omega +V) \right],
\label{current}
\ee
where we have used the gap function, $\Delta(\omega)$, defined as
\be
\Delta(\omega) \equiv \phi(\omega + i \delta)/Z(\omega + i \delta).
\label{gapw}
\ee
The proportionality constant contains information about the density
of states in the electron supplier (or acceptor), and the tunneling
matrix element. These are usually assumed to be constant. If one
takes the zero temperature limit, then the derivative of the current
with respect to the voltage is simply proportional to the
superconducting density of states,
\be
\left( {dI \over dV}\right)_S / \left( {dI \over dV}\right)_N =
{\rm Re} \left(  |V| \over \sqrt{V^2 -\Delta^2(V)}\right) ,
\label{didv}
\ee
where $S$ and $N$ denote ``superconducting'' and ``normal'' state, 
respectively. The right hand side of Eq. (\ref{didv}) is simply the
density of states, computed within the Eliashberg framework (see, for
example, Ref. \cite{mcmillan69}). It is not at all apparent what the
structure of the density of states is from Eq. (\ref{didv}), until one
has solved for the gap function from Eqs. (\ref{anal_zz},\ref{anal_phi2})
and Eq. (\ref{gapw}).
At zero temperature the gap function $\Delta(\omega)$ is real and
roughly constant up to a frequency roughly equal to that constant. This
implies that the density of states will have a gap, as in BCS theory.
At finite temperature the gap function has a small imaginary part starting
from zero frequency (and, in fact the real part approaches zero at
zero frequency \cite{karakozov75}) so that
in principle there is no gap, even for an s-wave order parameter. In
practice, a very well-defined gap still occurs for moderate coupling, and
disappears at finite temperature only when the coupling strength is increased
significantly \cite{allen91,marsiglio91}. 

In Fig. 12 and 13 we show the current-voltage and conductance plots for
superconducting Pb, taken from McMillan and Rowell \cite{mcmillan69}.
These data were obtained from a superconductor-insulator-superconductor (SIS)
junction, with Pb being the superconductor on both sides of the insulating
barrier, so that, rather than directly using Eq. (\ref{didv}), the current
is given by a convolution of the two superconducting densities of states. 
Two features immediately stand out in these plots. First, a gap is clearly
present in Fig. 12, given by $2 \Delta_\circ$, where $\Delta_\circ$ is the
single electron gap defined by
\be
\Delta_\circ \equiv {\rm Re} \Delta (\omega = \Delta_\circ),
\label{gapedge}
\ee
a definition one can use for all temperatures. Secondly, a significant
amount of structure occurs beyond the gap region, as is illustrated in Fig. 13.

McMillan and Rowell were able to deconvolve their measurement, to produce
the single electron density of states shown in Fig. 14.
Since the superconducting density of states is given by the right hand
side of Eq. (\ref{didv}), the structure in the data must be a reflection
of the structure present in the gap function, $\Delta (\omega)$. The
structure in the gap function is in turn a reflection of the structure
in the input function, $\alpha^2F(\nu)$. In other words, Eqs.
(\ref{anal_zz},\ref{anal_phi2}) can be viewed as as a highly nonlinear
transform of $\alpha^2F(\nu)$. Thus the structure present in Fig. 14
contains important information (in coded form) concerning the electron-phonon
interaction. One has only to ``invert'' the ``transform'' to determine
$\alpha^2F(\nu)$ from the tunneling data. This is precisely what McMillan
and Rowell \cite{mcmillan65,mcmillan69} accomplished, first in the case of
Pb. 

The procedure to do this is as follows. First a ``guess'' is made for the
entire function, $\alpha^2F(\nu)$, and the Coulomb pseudopotential parameter,
$\mu^\ast$. Then the real axis Eliashberg equations ((\ref{anal_phi}) and
(\ref{anal_zz})) are solved, and the superconducting density of states
(Eq. (\ref{didv})) is calculated. The result attained will in general
differ from the experimentally measured function (represented, for example,
by Fig. 14); a Newton-Raphson procedure (using functional derivatives rather
than normal derivatives) is used to determine the correction to the initial
guess for $\alpha^2F(\nu)$ that will lead to better agreement. Very often
another parameter (for example, the measured energy gap value) is used to
fit $\mu^\ast$. This process is iterated until convergence is achieved. The
result for Pb is illustrated by the dotted curve in Fig. 11.

Once $\alpha^2F(\nu)$ (and $\mu^\ast$) has been acquired in this way
one can use the Eliashberg equations to calculate other properties, for
example, $T_c$. These can then be compared to experiment, and the agreement 
in general tends to be fairly good. One may suspect, however, a circular
argument, since the theory was used to produce the spectrum (from experiment),
and now the theory is used as a predictive tool, with the same spectrum.
There are a number of reasons, however, for believing that this procedure
has produced meaningful information. First, the spectrum attained has come
out to be positive definite, as is required physically. Second, the spectrum
is non-zero precisely in the phonon region, as it should be. Moreover, it
agrees very well with the calculated spectrum. Thirdly, as 
already mentioned, various thermodynamic properties are calculated with this
spectrum, with good agreement with experiment. Finally, the density of states
itself can be calculated in a frequency regime beyond the phonon region,
as is shown in Fig. 15.
The agreement with experiment is spectacular.

None of these indicators of success can be taken as definitive proof of
the electron-phonon interaction. For example, even the excellent agreement
with the density of states could be understood as a mathematical
property of analytic functions \cite{remark6}. Also, we have focussed
on Pb; in other superconductors this procedure has not been so straightforward.
For example, in Nb a proximity layer is explicitly accounted for in the
inversion \cite{wolf74,wolf85}, thus introducing extra parameters.
In the so-called A15 compounds (eg. Nb$_3$Sn, V$_3$Si, etc.), although
the measured tunneling results have been inverted \cite{shen72b}, several
experiments do not fit the overall electron-phonon framework \cite{anderson83}.

More details are provided in Ref. \cite{carbotte90}. An alternate inversion 
procedure is also provided there \cite{galkin74}, which utilizes a
Kramers-Kronig relation to extract $\Delta(\omega)$ from the tunneling
result. An inversion of ${\rm Im} \phi(\omega + i \delta)$ then removes
$\mu^\ast$ from the procedure. A variant of this, where the imaginary
axis quantity $\Delta(i\omega_m)$ is extracted directly from the tunneling
I-V characteristic, and then the imaginary axis equations are inverted for
$\alpha^2F(\nu)$, also works \cite{remark7}, but the accuracy requirements
for a unique inversion are very debilitating.

\subsubsection{Optical Conductivity}

In principle, any spectroscopic measurement will contain a signature of
$\alpha^2F(\nu)$. In particular, several attempts have been made to infer
$\alpha^2F(\nu)$ from optical conductivity measurements in the superconducting
state \cite{joyce70,allen71,farnworth74}. In this section we describe a 
procedure for extracting $\alpha^2F(\nu)$ from the normal state
\cite{marsiglio98}.

A common method to determine the optical conductivity is to measure the
reflectance \cite{timusk89} as a function of frequency, usually at normal
incidence. The reflectance, $R(\nu)$, is defined as the absolute ratio
squared of reflected over incident electromagnetic wave amplitude.
The complex reflectivity is defined by
\be
r(\nu) \equiv R^{1/2}(\nu)\exp{(i\theta(\nu))},
\label{reflectivity}
\ee
where $\theta(\nu)$ is the phase, and is obtained through a Kramers-Kronig
relation from the reflectance \cite{timusk89}                              
\be
\theta(\nu) = {\nu \over \pi} \int_0^\infty {\ln{R(\nu^\prime)} - \ln{R(\nu)}
\over \nu^2 - {\nu^\prime}^2 } d\nu^\prime.
\label{kk}
\ee
The complex reflectivity is related to the complex index of refraction,
$n(\nu)$,
\be
r(\nu) \equiv {1 - n(\nu) \over 1 + n(\nu)},
\label{index}
\ee
which, finally, is related to the complex conductivity, $\sigma(\nu)$ (using
the dielectric function, $\epsilon(\nu)$):
\be
\epsilon(\nu) \equiv n^2(\nu) = \epsilon_\infty + {4 \pi i \sigma(\nu) \over
\nu},
\label{dielectric}
\ee
where $\epsilon_\infty$ is the dielectric function at high frequency 
(in principle, for infinite frequency this would be unity). 
It is through such transformations that the `data' is often 
presented in `raw' form. Nonetheless, assumptions are 
required to proceed through these steps; for example, Eq. (\ref{kk}) 
indicates quite
clearly that the reflectance is required over all positive frequencies. Thus
extrapolation procedures are required at low and high frequencies; a more
thorough discussion can be found in \cite{wooten72}; see also 
\cite{marsiglio97}.

For this review, we will consider both static impurities and phonons as
sources of electron scattering. Both contribute to the optical conductivity,
and can be treated theoretically either with the Kubo formalism or with a
Boltzmann approach \cite{mahan81}. In the Born approximation the result for
the conductivity, in the normal state, at zero temperature, is \cite{allen71}:
\be
\sigma(\nu) = {\omega^2_P \over 4\pi} {i \over \nu}
\int_0^\nu d\omega {1 \over \nu + i/\tau - \Sigma(\omega)
- \Sigma(\nu - \omega) }
\label{sigma}
\ee
\noindent where
\be
\Sigma(\omega) = \int_0^\infty d\Omega \alpha^2F(\Omega)
\ln | {\Omega - \omega \over \Omega + \omega} | - i\pi
\int_0^{|\omega|} d \Omega \ \alpha^2F(\Omega)
\label{selfenergy}
\ee
\noindent is the effective electron self-energy due to the electron-phonon
interaction. The spectral function that appears in Eq. (\ref{selfenergy})
is really a closely related function, as has been discussed by Allen 
\cite{allen71} and Scher \cite{scher70}. For our purposes we will treat 
them identically.  The other two parameters that enter these expressions
are the electron plasma frequency, $\omega_P$, and the (elastic) 
electron-impurity scattering rate, $1/\tau$.

Equation (\ref{sigma}) has been written to closely resemble the Drude
form,
\be
\sigma_{\rm Dr}(\nu) = {\omega_P^2 \over 4 \pi} { i \over \nu + i/\tau};
\label{sigma_drude}
\ee 
the equation could well be recast in this form, with a frequency-dependent
scattering rate and effective mass (in the plasma frequency) \cite{dolgov91}.
Eqs. (\ref{sigma}) and (\ref{selfenergy}) make clear that the optical 
conductivity is given by two integrations over the electron-phonon spectral
function. One would like to ``unravel'' this
information as much as possible before attempting an inversion, so that,
in effect, the signal is ``enhanced''. To this end one can attempt
various manipulations \cite{marsiglio95,dolgov95,marsiglio96}. 

As a first step one can make a weak coupling
type of approximation to obtain \cite{marsiglio98} the {\it explicit} result:
\be
\alpha^2F(\nu) = {1 \over 2\pi} {\omega_P^2 \over 4\pi}
{d^2 \over d\nu^2} \biggl\{ \nu Re{1 \over \sigma(\nu)} \biggr\}.
\label{explicit}
\ee
\noindent  Note that the conductivity data, including a measurement of the
plasma frequency, provides us with both the shape {\it and magnitude} of
$\alpha^2F(\nu)$. Eq. (\ref{explicit}) works extremely well, as Fig. 16 shows,
in the case of Pb. It tells us that, with a judicious
manipulation of the conductivity data, the underlying electron-phonon spectral
function emerges in closed form. The very simple formula, Eq. (\ref{explicit})
introduces some errors --- it was derived with some approximations --- as
can be seen in Fig. 16. In fact, a full numerical inversion will also
succeed \cite{marsiglio99,shulga01}; the first reference requires a
Newton-Raphson iteration technique, while the second uses an adaptive method
(in the superconducting state).

Eq. (\ref{explicit}) was first applied to K$_3$C$_{60}$ \cite{marsiglio98}
to help determine whether or not this class of superconductor was driven
by the electron-phonon interaction. The result is shown in Fig. 17
and provides convincing evidence that the alkali-doped fullerene
superconductors are driven by the electron-phonon mechanism. 
We will return to these superconductors in a later section, and further
examine the optical conductivity in the superconducting state in another
section.

\section{The Critical Temperature and the Energy Gap}

Perhaps the most important property of a superconductor is the critical
temperature, $T_c$. For this reason a considerable amount of effort has been
devoted both towards new materials with higher superconducting $T_c$, and,
on the theoretical side, 
towards an analytical solution of the linearized Eliashberg
equations (set $\phi{_m{^\prime}}$ to zero, where it appears in the
denominator in Eqs. (\ref{gb1} - \ref{gb4}) ) for $T_c$ 
(see \cite{allen82,carbotte90} for reviews); the experimental `holy grail'
has enjoyed some success, particularly in the last 15 years; the theoretical
goal has had limited success. In fact numerical solutions are
so readily available at present, that the absence of an analytical solution
is not really debilitating to understanding $T_c$. 

In the conventional theory there are two input "parameters": a function
of frequency, $\alpha^2F(\nu)$, about which we have already said much, and
$\mu^\ast(\omega_c)$, a number which summarizes the (reduced) Coulomb
repulsion experienced by a Cooper electron pair. The focus of this chapter
will be the effect of size and functional form of $\alpha^2F(\nu)$ on $T_c$.

\subsection{Approximate Solution: The BCS Limit}

The first insight into $T_c$ comes from reducing the Eliashberg theory to
a BCS-like theory. This is accomplished by approximating the kernel
\be
\lambda(i\omega_m - i\omega_{m^\prime}) \equiv \int_0^\infty 
{2\nu \alpha^2F(\nu) \over \nu^2 + (\omega_m - \omega_{m^\prime})^2}
\label{lambda_m} 
\ee
by a constant as long as the magnitude of the two Matsubara frequencies
are within a frequency rim of the Fermi surface \cite{allen75}, 
taken for convenience
to be $\omega_c$, the cutoff used for the Coulomb repulsion, $\mu^\ast$.
That is, 
\be
\lambda(i\omega_m - i\omega_{m^\prime}) = \cases{ \lambda  \phantom{aaaa}
{\rm for \phantom{aa} both} \phantom{aaaa}
|\omega_m|, |\omega_{m^\prime}| < \omega_c,\cr
0 \phantom{aaaa} {\rm otherwise,}}
\label{squarewell}
\ee
where $\lambda \equiv \lambda(0)$ has already been defined in 
Eq.~(\ref{lambda}).
Then, the linearized version of Eq. (\ref{gb1}) (with $A_0(m^\prime) = 1$),
for the renormalization function, $Z(i\omega_m)$, reduces to
\be
Z(i\omega_m) \approx 1 + \lambda .
\label{gb1_approx}
\ee
Using this and solving the linearized version of Eq. (\ref{gb3}) for
the pairing function yields
\be
{1 + \lambda \over \lambda - \mu^\ast} = \psi \biggl( {\omega_c \over 2\pi T_c}
+ {1 \over 2} \biggr) - \psi \bigl( {1 \over 2} \bigr),
\label{tc_psi}
\ee
where $\psi(x)$ is the digamma function. The cutoff in these 
equations is along the Matsubara frequency axis; this procedure 
is to be contrasted with the BCS procedure, which introduced a cutoff in 
momentum space. The former is more physical, insofar as
the true electron-phonon interaction comes from retardation effects, which
occur in the temporal domain; hence the cutoff should occur in the
frequency (either real, or imaginary) domain. In practice, the two 
procedures are connected, so they produce the same physical equation 
in the weak coupling limit.

Returning to Eq. (\ref{tc_psi}), for large $x$, $\psi(x) \approx
\log{(x)}$, so, in the weak coupling limit ($T_c << \omega_c$), we obtain a
BCS-like equation,
\be
T_c=1.13 \omega_c \exp{\biggl(-{1 + \lambda \over \lambda - \mu^\ast} \biggr)}. 
\label{tc_weak}
\ee
This equation has essentially summarized all the detailed information contained
in the electron-phonon spectral function $\alpha^2F(\nu)$ into two parameters,
$\lambda$ and $\omega_c$.  The mass enhancement parameter, $\lambda$, is 
a simple moment of $\alpha^2F(\nu)$ (see Eq. (\ref{lambda})), while the
parameter $\omega_c$ physically is meant to represent some typical phonon
frequency. In more refined treatments \cite{mcmillan68,allen75}, 
$\omega_c$ is given by some moment of $\alpha^2F(\nu)$ as well. 
For example, in Ref. \cite{allen75}, the logarithmic average is used to
define $\omega_{\ln}$ (see Eq. (\ref{wln})), a quantity we shall use 
extensively in the following sections. They modified the McMillan equation
\cite{mcmillan68} to read
\be
k_B T_c = {\hbar \omega_{\ln} \over 1.2} \exp{\biggl(- {1.04(1 + \lambda)
\over \lambda - \mu^\ast (1 + 0.62 \lambda) } \biggr)}.
\label{mcmill}
\ee
A derivation of this equation is given in Refs. \cite{mcmillan68,allen75}.
        
\subsection{Maximum $T_c$, Asymptotic Limits, and Optimal Phonon Spectra}

Eq. (\ref{tc_weak}) (or Eq. (\ref{mcmill})) describes the weak 
coupling limit of
Eliashberg theory reasonably well. It errs in the strong coupling 
limit; for example, it predicts that $T_c$ saturates as $\lambda$
increases, whereas the Eliashberg equations themselves predict
that $T_c$ grows indefinitely with $\lambda$ \cite{remark8}.
Asymptotic results from Eliashberg theory can be
obtained correctly and analytically \cite{allen75,rainer73,carbotte90}
through a variety of arguments. The methodology based on scaling theorems
is particularly powerful, and has been applied to other thermodynamic
properties as well \cite{carbotte90}.

The correct asymptotic result for $T_c$ is \cite{allen75,rainer73}:
\be
T_c = 0.183\sqrt{\lambda} \omega_E,
\label{tc_asy}
\ee
as $\lambda \rightarrow \infty$. In obtaining Eq. (\ref{tc_asy}), an
Einstein spectrum has been assumed (this is not required), which in 
turn is characterized by
two parameters: the weight, $A \equiv \lambda \omega_E /2$, and the
frequency, $\omega_E$. In writing Eq. (\ref{tc_asy}), one tacitly has
assumed that the parameter $\lambda \equiv 2 A/\omega_E$ is increased
while keeping the frequency $\omega_E$ fixed. In reality, the two parameters
are not independent --- this is the main point of the article by
Cohen and Anderson \cite{cohen72}. For example, often phonon softening
occurs {\em because} the coupling strength increases. In fact, this is
made explicit in McMillan's definition \cite{mcmillan68} of $\lambda$:
\be
\lambda \equiv {N(\epsilon_F) \alpha^2 \over M \omega_E^2},
\label{lam_mcm}
\ee
where $N(\epsilon_F)$ is the electron density of states at the Fermi
energy, $M$ is the ionic mass, and $\alpha^2$ is the electron-ion
coupling referred to in the Holstein Hamiltonian, Eq. (\ref{ham_Hol}) (in
a more realistic electron-phonon Hamiltonian, $\alpha^2$ would be
given by a Fermi surface average of the electronic matrix element of
the change in crystal potential as one atom is moved \cite{mcmillan68}).

To determine what the optimal phonon frequencies actually are, 
functional derivatives were introduced \cite{bergmann73}. These had
already been utilized extensively as an iterative aid in inverting
tunneling data with the Eliashberg equations \cite{mcmillan69}. The most
commonly used functional derivative is that of $T_c$ with respect to
infinitesmal changes in $\alpha^2F(\nu)$, {\em with fixed area}, $A \equiv
\int_0^\infty d\nu \, \alpha^2F(\nu)$, first computed by Bergmann and
Rainer \cite{bergmann73}. An approximate result, derived in Ref.
\cite{mitrovic81b}, is given by the expression
\be
{\delta T_c \over \delta \alpha^2F(\Omega) } = {1 \over 1 + \lambda}
\sum_{n = 1}^\infty { 4 \bar{\Omega} \over {\bar{\Omega}}^2 + 4 \pi^2 n^2},
\label{tcfun}
\ee
where $\bar{\Omega} \equiv \Omega/k_BT_c$, and the $B_n$ are numbers given
by
\be
B_n = \sum_{m=1}^n \biggl({1 \over n} {2 \over 2m-1} + {2 \over (2m-1)^2}
\biggr) - {\pi^2 \over 4}.
\label{numbers}
\ee
This function (which is universal) is shown in Fig. 18, and reflects well
the generic behaviour of the more precise calculation. It illustrates that
the optimal phonon frequency lies at some finite frequency (i.e. non-zero,
{\em and} non-infinite), which is a factor of order 10 times the critical
temperature. Thus if one could imagine shifting small amounts of
weight in $\alpha^2F(\nu)$ around then $T_c$ would increase if spectral
weight is shifted either from very high or from very low frequencies towards
frequencies near the maximum of the curve shown in Fig. 18.

The reasoning above leads naturally to the concept of an optimum spectrum,
first determined by Leavens \cite{leavens75}, and elaborated upon in Refs.
\cite{blezius86,carbotte86,marsiglio87,schossmann87,akis88}.
In an optimum spectrum calculation, one imagines having a fixed area of 
$\alpha^2F(\nu)$, and asks at what frequency it would best be situated 
in order to optimize some particular property. An appropriate scaling
of the linearized Eliashberg equations for an Einstein phonon spectrum
with frequency $\omega_E$ leads to the result
\be
T_c/A = f({\bar{\omega}}_E,\mu^\ast),
\label{scaling}
\ee
where $A$ is the area, ${\bar{\omega}}_E \equiv \omega_E/A$, and $f$ is
a universal function of ${\bar{\omega}}_E$, to be determined numerically 
for each choice of $\mu^\ast$ (a very weak $A$ dependence in the cutoff
associated with $\mu^\ast$ has been neglected). The result is a curve with
a maximum at ${\bar{\omega}}_E \approx 1$; placing a spectral function
at this frequency will yield the maximum $T_c$. This procedure yields a
result,
\be
T_c \le A c(\mu^\ast),
\label{optimal}
\ee
where $c(\mu^\ast)$ is a function of $\mu^\ast$ shown in Fig. 19.
Also shown are data from many superconductors for which $\alpha^2F(\nu)$
is known from tunneling spectroscopy, all of which fall below the optimum
curve. Interestingly, some superconductors have a critical temperature
reasonably close to their optimal value.

The last few paragraphs demonstrate the usefulness of functional derivatives
in understanding the systematics of $T_c$. A variant of these results can
easily be obtained, which may shed even more light on $T_c$ systematics.
As we have already mentioned, the functional derivative discussed involves
the moving around of spectral weight, {\em subject to the condition that
the area remain constant}. However, as Eq. (\ref{lam_mcm}) suggests, it is
not the area which likely remains constant while phonons soften, but rather
the area times a frequency. Hence, one can define a different spectral
function,
\be
\alpha^2G(\nu) \equiv \nu \alpha^2F(\nu),
\label{a2fnew}
\ee
and take functional derivatives with respect to this new function.
The result is easily obtained from that in Eq. (\ref{tcfun}), simply
by dividing by $\Omega$. Then ${\delta T_c \over \delta \alpha^2G(\Omega) }$ 
will peak at zero frequency, and it would seem that it is always
advantageous to decrease the phonon frequency. Continuing this process
will result in a spectrum for which the calculation which gives
Eq. (\ref{tcfun}) is no longer valid, and one would have to self-consistently
calculate the functional derivative, numerically. To our knowledge this
has not been done for $T_c$ or any other superconducting property.

\subsection{Isotope Effect}

As already remarked in the Historical Developments subsection, the
discovery of an isotope effect on $T_c$ played an important role in the
subsequent development of the theory. In the BCS equation the isotope effect
is clear from the prefactor; phonon frequencies for elemental superconductors
are inversely proportional to the square root of the ionic mass, and hence
the isotope coefficient $\beta$ is
\be
\beta \equiv - {d \, \ln T_c \over d \, \ln M} = {1 \over 2}.
\label{bcs_iso}
\ee
The last equality follows from Eq. (\ref{bcs_tc_weak}), using the fact that
$\lambda$, as defined there, is independent of ion mass. In the standard
Eliashberg theory, $\lambda$ as defined by Eq. (\ref{lambda}) remains
independent of ion mass, and, with $\mu^\ast \equiv 0$, we once again
obtain $\beta = 1/2$. Complications can arise, for example, from a 
finite electronic bandwidth \cite{marsiglio92}, or from a non-constant
density of states near the Fermi level 
\cite{tsuei90,schachinger90,markiewicz91}.

There are two other clear sources of deviation from $\beta = 1/2$. One is 
that in non-elemental superconductors, an isotopic substitution for one of
the elements will result in varying changes in $T_c$, depending on how
the element being substituted contributes to the important phonon modes.
One then has to define partial isotope coefficients, defined by \cite{rainer79}
\be
\beta_i \equiv - {d \, \ln T_c \over d \, \ln M_i},
\label{partial_iso}
\ee
where $M_i$ refers to the mass of the $i$th element. The total isotope
coefficient, $\beta_{\rm tot}
\equiv \sum_i \beta_i$, will sum to $1/2$ (in the absence of $\mu^\ast$).
The total isotope coefficient can also be broken down by frequency, with
\be
\beta(\omega) \equiv \alpha^2F(\omega) {d \over d \omega} \biggl(
{\omega \over 2 T_c} {\delta T_c \over \delta \alpha^2F(\omega)} \biggr),
\label{partial_iso_freq}
\ee
and then 
\be
\beta_{\rm tot} = \int_0^\infty d\omega \beta(\omega).
\label{total_iso}
\ee
Eq. (\ref{total_iso}) is useful when phonon modes coming from one of the
elements are well separated from those coming from the others, as exists, for
example, in the high temperature perovskites, since oxygen is much lighter
than the other elements, and hence is chiefly responsible for the high
frequency modes. Example calculations can be found in 
Refs. \cite{ashauer87,carbotte90}.

The second source of deviation from $\beta = 1/2$ is because 
$\mu^\ast(\omega_c)$ is non-zero. To understand why this causes less of a
reduction in $T_c$ (when a heavier mass is substituted) recall that
$\mu^\ast(\omega_c)$ is reduced from some larger value $\mu(\omega_B)$
through the pseudopotential effect. That is, it is through retardation
that a weaker electron-phonon interaction can overcome the stronger
direct Coulomb repulsion. In analytical treatments this is often modelled
by endowing a mass dependency to the Coulomb pseudopotential through the
cutoff \cite{mcmillan68}. For example, inspection of Eq. (\ref{pseudo})
shows a mass dependency {\em if } the cutoff frequency $\omega_c$ is made
to correspond to a phonon frequency. Then one can derive, from the
McMillan equation for $T_c$, Eq. (\ref{mcmill}), the following expression
for the isotope coefficient (assuming one element):
\be
\beta = {1 \over 2} \biggl(
1 - {1.04(1 + \lambda)(1 + 0.62\lambda) \over [\lambda - \mu^\ast (1
+ 0.62 \lambda)]^2} \mu^{\ast 2}
\biggr).
\label{mcmill_beta}
\ee
This result properly reduces to $1/2$ when $\mu^\ast = 0$, and shows that
the isotope coefficient is generally reduced when $\mu^\ast$ is finite.
In fact it is clear from Eq. (\ref{mcmill_beta}) that the isotope
coefficient is reduced for {\em both positive and negative} $\mu^\ast$.
The isotope coefficient is reduced for positive $\mu^\ast$ because, when
you lower the ionic mass, the increase in phonon frequency will raise
$T_c$, but not as much as would be the case if the Coulomb repulsion
were not present. This is because the discrepancy in frequency scales
has been reduced slightly, and the retardation-induced attractive
interaction is not as large as before the isotopic substitution. On the
other hand a negative $\mu^\ast$ represents some unknown attractive
non-phonon mechanism \cite{ashauer87}, which contributes to $T_c$. A
weakening of the phonon-induced attraction (through an isotopic substitution)
reduces $T_c$ only partially, resulting again in a reduced isotope
coefficient.

A more accurate determination of the isotope coefficient can be obtained
simply numerically, following the prescription of Rainer and Culetto
\cite{rainer79}. In this case a cutoff is imposed on the Eliashberg
equations, which is independent of (but much greater than) the maximum
phonon frequency. An isotopic substitution results in only a shift in
the phonon spectrum, and a subsequent calculation of $T_c$ will yield
the isotope coefficient. This is physically more transparent than the
analytical approach described above, as an isotope substitution does
not alter (at this level of theory) the direct Coulomb repulsion.
 
There is a substantial literature on the isotope effect; much of the older
results are summarized in Ref. \cite{meservey69}. By 1969 many low $T_c$
superconductors had been found, several of which had very low isotope
coefficients. These had, for the most part, been explained through detailed
calculations \cite{morel62,garland63}, due to the physics outlined above.
It is worth noting that this explanation of the sometimes low isotope
coefficient observed was not universally accepted \cite{matthias60}.

The discovery of the high temperature cuprate materials prompted considerable
activity concerning the isotope coefficient, as is reviewed in Ref.
\cite{franck94}. The isotope coefficient displays some unusal doping
dependence in the La$_{2-x}$\{Sr,Ba\}$_x$CuO$_{4 - \delta}$ systems, but
is essentially zero in the optimally doped 90 K YBa$_2$Cu$_3$O$_{7 - y}$
system. The question is, can a realistic (and conventional) 
electron-phonon interaction give rise to a 90 K superconductor 
with a near zero isotope coefficient ? A qualitative answer can be
obtained \cite{akis90} through the use of the McMillan equation 
(\ref{mcmill},\ref{mcmill_beta}). For a given electron-phonon
coupling, $\lambda$, and phonon frequency $\omega_E$, one can determine
the required value of $\mu^\ast(\omega_c)$ to fix $T_c$ 
from Eq. (\ref{mcmill}) (assuming $\omega_c$ refers
to a cutoff associated with the phonon spectrum). 
These parameters can then be used in Eq. (\ref{mcmill_beta}) to determine
the isotope coefficient, $\beta$. Some such results are plotted in Fig. 20.
To obtain the desired results for optimally doped YBa$_2$Cu$_3$O$_{7 - y}$
($T_c \approx 90$ K and
$\beta \approx 0$) would require high frequency phonons $\omega_E \approx
100$ meV with very strong electron-phonon coupling ($\lambda \approx 5$).
That such a coupling strength is unrealistic, particularly for such
very high frequency phonons, was discussed much earlier by Cohen and
Anderson \cite{cohen72}. 

The qualitative validity of Fig. 20 has been verified by several numerical
solutions to the Eliashberg equations \cite{akis90,barbee91,marsiglio92}.
In particular, in Ref. \cite{marsiglio92} a natural bandwidth cutoff was 
employed, with similar results.
In summary the conventional Eliashberg theory can yield a near-zero isotope
coefficient, {\it provided} $T_c$ is low. One must go beyond the conventional
framework to obtain a zero isotope coefficient with $T_c \approx 90$ K. 

\subsection{The Energy Gap}

The existence of a single particle energy gap, although not fundamental
to superconductivity \cite{abrikosov60}, nonetheless has played an important
role in our understanding of superconductivity. How an energy gap arises in the
I-V characteristic of a conventional superconductor has already been 
discussed in Section 3.3.2; there we focussed on extracting
detailed information about the mechanism. Here we turn our attention to the
gap, a much more prominent feature in the experimental result, and learn
what a particular value may imply about the superconductor.

The first step is to examine what occurs in BCS theory. The order parameter
is then given by a constant, as written in Eq. (\ref{bcs_zero}). Suitable
generalization to the model interaction given by Eq. (\ref{squarewell}) yields
\be
\Delta=2 \omega_c \exp{\biggl(-{1 + \lambda \over \lambda - \mu^\ast} \biggr)},
\label{delta_weak}
\ee       
in the weak coupling limit. The solution at finite temperature is somewhat
more complicated; it can be obtained numerically, and shows the typical
mean field behaviour near $T_c$ \cite{bardeen57}:
\be
\Delta(T) \propto (T_c - T)^{1/2}. 
\label{delta_near_tc}
\ee
Near $T=0$ the order parameter is exponentially flat \cite{abrikosov88}:
\be
\Delta(T) \approx \Delta(0) - [2 \pi \Delta(0) T]^{1/2} e^{-\Delta(0)/T}.
\label{delta_near_zero}
\ee
The order parameter is a real (i.e. not complex) number for all temperatures
\cite{remark9}. Thus, Eq. (\ref{didv}) shows that the $dI/dV$ curve (which
provides an image of the density of states) will show an energy gap at
$\Delta(T)$ at each temperature. An illustration of the temperature
dependence of the order parameter is given in Fig. (21a), along with
the density of states at several temperatures (Fig. (21b)).

Within Eliashberg theory, the calculation of the corresponding property
is much more complicated. First of all, a careful distinction between the
gap or pairing function (which is now a function of frequency at any given
temperature) and the energy gap is required. The energy gap is defined through
Eq. (\ref{gapedge}). The gap function is, in general, a complicated and complex
function of frequency, that results from a solution of the Eliashberg 
equations. These, in turn, can be solved either on the imaginary axis
(Eqs. (\ref{gb1}) and (\ref{gb3}) for the `standard' theory) or the
real axis (Eqs. (\ref{anal_zz}) and (\ref{anal_phi2})). Example solutions
for a real electron-phonon spectrum (Pb) are shown in Fig. 22 
and Fig. 23.
The solutions on the imaginary axis turn out to be real;
on the real axis they are complex.
The corresponding densities of states are shown in Fig. 24.

The low frequency behaviour of the various functions plotted is not clear
on the figures shown. A careful analysis \cite{karakozov75} leads to
\bea
{ {\rm Re} \Delta(\omega) \, = \, c \atop {\rm Im} \Delta(\omega) \,
= \, 0} \qquad T = 0
\nonumber \\
{ {\rm Re} Z(\omega) \, = \, d \atop {\rm Im} Z(\omega) \, = \, 0} \qquad
T = 0
\label{zerot_freq}
\eea
at zero temperature, where $c$ and $d$ are constants, 
whereas at any non-zero temperature, we obtain
\bea
{ {\rm Re} \Delta(\omega) \, \propto \, \omega^2 \atop {\rm Im} \Delta(\omega) 
\, \propto \, \omega} \qquad
T > 0
\nonumber \\
{ {\rm Re} Z(\omega) \, = \, d(T) \atop {\rm Im} Z(\omega) \, \propto \, 
1/\omega} \qquad T > 0.
\label{abovezerot_freq} 
\eea 
The latter result in particular implies that, strictly speaking, at finite
temperature there is always "gapless" superconductivity. However, as can
be seen from Fig. 24, in reality the ``finite temperature'' density of 
states at zero energy is generally quite small (except for very close
to $T_c$). The extent to which this is true depends on the electron-phonon
coupling strength; as this increases the zero frequency density of states
can be a significant fraction of the normal state value at temperatures
near $T_c$ \cite{marsiglio91}.

In the remaining subsections we wish to examine the dependence of the
energy gap on coupling strength. Since the electron-phonon interaction is
characterized by a spectral function, $\alpha^2F(\Omega)$, we first must
decide how to quantify the coupling strength of a particular superconductor.
Historically the mass enhancement parameter, given by Eq. (\ref{lambda}),
has played this role. However, depending on the material, the direct Coulomb
repulsion, characterized by $\mu^\ast(\omega_c)$, where $\omega_c$ is some
suitable cutoff frequency, can offset the effect of $\lambda$. Another
possible parameter is the ratio of the critical temperature to an average
phonon frequency, a quantity first advocated by Geilikman and Kresin
\cite{geilikman65,geilikman75,masharov74,kresin74}. This approach was
further quantified by Mitrovi\'c {\it et al.} \cite{mitrovic84}. In this
reference (see also Ref. \cite{marsiglio86}), the Allen-Dynes parameter
$T_c/\omega_{\ln}$  emerged naturally in the derivation of strong
coupling corrections, as an indicator of coupling strength. A large
number of superconducting properties were obtained in this way
(see Refs. \cite{marsiglio88a,carbotte90} for derivations and more details),
and semi-empirical fits were obtained based on accurate numerical solutions.
We discuss these further in the next section. 

\subsection{The Energy Gap: Dependence on
Coupling Strength $T_c/\omega_{\ell n}$}

As we have already emphasized, $T_c$ cannot be reliably calculated at
present. The first, perhaps simplest, test for the accuracy of Eliashberg
theory is then its ability to properly obtain the gap ratio, $2\Delta_\circ
/k_BT_c$, where, by $\Delta_\circ$, we mean the zero temperature gap edge.
In Ref. \cite{mitrovic84} (see also Ref. \cite{carbotte90}),
numerically calculated results were compared to experimental tunneling
results for $\Delta_\circ$, obtained for a variety of conventional 
elemental and alloy superconductors.
The deviations of the gap ratio from the BCS universal result, $2\Delta_\circ
/k_BT_c = 3.53$, are up to 50 \%; yet the level of error is about 5 \% ,
with one notable exception (Nb$_3$Sn). The theoretical results are obtained
from a solution of the imaginary axis equations (Eqs. (\ref{gb1}) and 
(\ref{gb3}), with the standard approximations of infinite bandwidth and
particle-hole symmetry), followed by an analytical continuation to the real
axis. To obtain the gap edge, a Pad\'e approximant suffices to get very
accurate results \cite{vidberg77}, as the more systematic continuation
\cite{marsiglio88} verifies. In any event it is desirable to have
an analytic form for these corrections. The result of Mitrovi\'c {\it et al.}
\cite{mitrovic84} is
\be
{2\Delta_\circ \over k_B T_c} = 3.53 \biggl[ 1 + 12.5 \bigl({T_c
\over \omega_{\ln} }\bigr)^2 \ln{({\omega_{\ln} \over 2 T_c})} \biggr].
\label{str_gap}
\ee
In obtaining this result the spirit of the McMillan equation was followed, and 
the coefficients $12.5$ and $2$ were chosen from fits to the numerical data
for a large number of superconductors. These results are plotted in Fig. 25.
From this Figure it is clear that Eq. (\ref{str_gap}) describes the overall
trend very well. As the electron-phonon interaction increases (i.e. becomes
more retarded), the gap ratio increases to values exceeding 5.0.

Figure 25 illustrates that a simple analytic form describes the trend of
the gap ratio as a function of ${T_c/ \omega_{\ln} }$ rather well for
a variety of conventional superconductors. In each case electron-phonon
spectral functions were used, as obtained from tunneling data, or, in some
cases, model calculations. On occasion, one sometimes uses a phonon
spectrum obtained from inelastic neutron scattering, scaled to give the
measured critical temperature. This latter process assumes that the
electron-phonon coupling is constant as a function of frequency (seen to be
reasonable in the case of Pb), and often assumes a value of
the Coulomb pseudopotential, $\mu^\ast(\omega_c = 6\omega_{\rm max})
\approx 0.1$ ($\omega_{\rm max}$ is the maximum phonon frequency). Specific
references to the sources of these spectra can be found in Refs.
\cite{marsiglio88a,carbotte90}. 

An important question, particularly when faced with a new superconductor
whose phonon characteristics may or may not be `typical', is to what
extent the trend modelled by the semi-empirical analytic form, 
Eq. (\ref{str_gap}) can be violated, for a given coupling study. This
question was considered in Ref. \cite{coombes86}. They took existing
electron-phonon spectra, $\alpha_\circ^2F(\nu)$, and scaled them to new
spectra, $\alpha_\circ^2F(\nu)^\ast = B\alpha_\circ^2F(b\nu)$, where
$B$ and $b$ are constants, chosen to span a continuum of values of
${T_c/ \omega_{\ln} }$. Thus, given some spectral shape, say that of
Pb, one can determine a curve of $2\Delta_\circ/k_BT_c$ vs.
${T_c/ \omega_{\ln} }$. In this way they were able to ascertain, for
a given value of ${T_c/ \omega_{\ln} }$, the shape dependence of the
gap ratio. They of course found more significant deviations from the
analytical form, Eq. (\ref{str_gap}); nonetheless, the deviations remained
small on the scale of Fig. 25. Larger deviations were obtained  
with the use of (somewhat artificial) delta-function model spectra
\cite{coombes86,combescot95,varelogiannis95}. Similarly, if the 
electron-phonon coupling strength is taken to be extremely high, 
large deviations occur from one spectral shape to another \cite{marsiglio87a}. 

The net conclusion is that, with physical spectra and physically relevant
coupling strength (${T_c/ \omega_{\ln} } {< \atop \sim} 0.2$), the
strong coupling corrections are quasi-universal, and are well described
by Eq. (\ref{str_gap}). We explore in the next subsection how this can
be used to optimize the gap and gap ratio.

\subsection{Optimal Phonon Spectra and Asymptotic Limits}

A functional derivative analysis similar to that described for $T_c$ yields,
for $\Delta_\circ$, an optimum phonon frequency for a given spectral area.
One finds that for a delta function spectral function,
the zero temperature gap edge obeys a scaling relation
just like $T_c$ given by Eq. (\ref{scaling}):
\be
\Delta_\circ/A = g({\bar{\omega}}_E,\mu^\ast),
\label{scaling_gap}
\ee
where all quantities are as defined following Eq. (\ref{scaling}).
As found there, for a given base spectrum, an optimum frequency
$\bar{\omega}_E^\ast$ exists whose value is generally {\em lower}
than the characteristic frequency of the base spectrum --- this
is particularly clear when the base spectrum itself is a delta
function. With $T_c$ one found that shifting the spectral
weight to that optimum frequency resulted in an enhancement of
$T_c$. Furthermore, an iteration of this procedure resulted in
convergence to the situation where, for a given spectral area, the
maximum $T_c$ had been achieved, with a frequency given by
$\bar{\omega}_E^\ast \approx 1.3$ (for $\mu^\ast(\omega_c) = 0.1$).
The functional derivative of $T_c$ with respect to $\alpha^2F(\nu)$
using this base spectrum is non-positive definite \cite{marsiglio87}
with a maximum at $\bar{\omega}_E^\ast$,
showing that $T_c$ could no longer be increased.

The situation with the gap edge is similar, but differs in the following
crucial point. Upon iteration one finds that the optimum frequency
continues to decrease, as the Einstein frequency of the base spectrum
decreases. Thus, the implication is that the gap edge, and therefore
the gapedge ratio, $2 \Delta_\circ / k_BT_c$, will be maximized in the
limit as $\bar{\omega}_E^\ast \rightarrow 0$. Alternatively, since
these calculations are for fixed spectral area, $A$, this will occur
as $\lambda \rightarrow \infty$.

What is the maximum value of $2 \Delta_\circ / k_BT_c$ allowed within
`standard' Eliashberg theory~? Carbotte {\it et al.} \cite{carbotte86}
answered this question through a scaling theorem, and backed up with
numerical work. They found that the gap ratio increased monotonically
as $\lambda$ increased, finding (numerically) a value close to 10
(recall BCS gives 3.53) for values of $\lambda \approx 30$. In doing so
they proved that $\Delta_\circ \propto \sqrt{\lambda} \omega_E$ as
$\lambda \rightarrow \infty$, just like $T_c$ does (Eq. (\ref{tc_asy})).
Claims were made to the contrary, but these were definitively put to
rest in Ref. \cite{marsiglio91}. By solving a set of Eliashberg equations
written specifically for $\lambda \rightarrow \infty$, they found a maximum 
value of the gap ratio equal to 12.7. A variety of other properties were
explored in the asymptotic limit, $\lambda \rightarrow \infty$, as can 
be found in the previous references and in 
Ref. \cite{bulaevskii87,kresin87,marsiglio89a,combescot95a,combescot96}.

\section{Thermodynamics and Critical Magnetic Fields}

These topics have been amply covered in previous reviews \cite{carbotte90}.
Nonetheless, we include here for completeness a brief summary of the
impact of the electron-phonon interaction on these properties in the
superconducting state.

\subsection{The Specific Heat}

To calculate the specific heat one requires the free energy. For an
interacting electron system, a practical formulation of this problem
was first proposed by Luttinger and Ward \cite{luttinger60}, and further
pursued by Eliashberg \cite{eliashberg62}. A simpler calculation
requires the free energy {\it difference} between the superconducting
and normal state, for which an expression due to Bardeen and Stephen
\cite{bardeen64} is
\be
{\Delta F \over N(0)} = -\pi T \sum_m \biggl(
\sqrt{\omega_m^2 + \Delta^2(i\omega_m)}  - | \omega_m | 
\biggr) 
\biggl(Z^S(i\omega_m) - Z^N(i\omega_m) {| \omega_m |
\over \sqrt{\omega_m^2 + \Delta^2(i\omega_m)}}
\biggr),
\label{bsfree}
\ee
where, for clarity, we include the two Eliashberg equations from
Eqs.(\ref{gb1} - \ref{gb4}):
\bea
Z_m & = & 1 + \pi T \sum_{m^\prime} \lambda(i\omega_m - i\omega_{m^\prime})
{( \omega_{m^\prime} / \omega_m ) Z_{m^\prime} \over
\sqrt{ \omega_{m^\prime}^2 Z^2_{m^\prime} + \phi^2_{m^\prime} }}
\label{gc1}
\\
\phi_m & = & \pi T \sum_{m^\prime}
\biggl( \lambda(i\omega_m - i\omega_{m^\prime}) - N(0) V_{\rm coul} \biggr)
{\phi_{m^\prime} \over
\sqrt{ \omega_{m^\prime}^2 Z^2_{m^\prime} + \phi^2_{m^\prime} }}.
\label{gc2}
\eea
These equations ignore band structure effects entirely (except through the
electron density of states at the Fermi level, denoted here by $N(0)$), and
we again have adopted the shorthand $Z(i\omega_m) = Z_m$ etc., and used the
gap function $\Delta(i\omega_m) \equiv \phi(i\omega_m)/Z(i\omega_m)$.
For the free energy expression we have used superscripts `S' or `N' to
denote the superconducting or normal state, respectively. In the normal state
$Z^N(i\omega_m)$ reduces to the expressions obtained in subsection
(2.3), which is easily seen if one uses the relation
\be
\Sigma(z) = z \bigl( 1 - Z(z) \bigr),
\label{sigz}
\ee
where $z$ is a frequency anywhere in the upper half plane, and $\Sigma(z)$
is the electron self-energy.

Equation (\ref{bsfree}) can easily be evaluated, once the imaginary axis
Eliashberg equations (\ref{gc1}-\ref{gc2}) are solved. From this the specific
heat difference,
\be
\Delta C (T) = -T {d^2\Delta F \over dT^2},
\label{spec}
\ee
and the thermodynamical critical field,
\be
H_c(T) = \sqrt{-8\pi \Delta F},
\label{hc}
\ee
can be computed. The former displays a jump at $T_c$, characteristic of 
a mean field theory, which is the level of approximation of
Eliashberg theory. At low temperatures the specific heat in the
superconducting state should be exponentially suppressed. This is generally
observed \cite{daams81}, and deviations that do occur at very low temperatures
can readily be explained by anisotropy in the gap parameter \cite{weber77}.

Because properties like the electron density of states at the Fermi
level are difficult to measure or calculate reliably, one would like to focus
on observables that are independent of these properties. For the specific
heat difference, one way of accomplishing this is to normalize the specific
heat to the normal state result, which presumably contains the same
electron density of states. The result is then independent of $N(0)$,
and can be compared directly to the measured results. A textbook example 
was provided in the case of Al \cite{phillips_n59}; the data
is reproduced in Fig. 26, along with the BCS prediction.
The normal state specific heat for a weakly interacting electron gas
is given by
\be
C_N (T) = \gamma T,
\label{specnorm}
\ee
where $\gamma$ is the Sommerfeld constant given by
\be
\gamma = {2 \over 3} \pi^2 k_B^2 N(0) (1 + \lambda).
\label{gamm0}
\ee
Here, $\lambda$ is the electron-phonon enhancement parameter, already
referred to on many occasions. The electron-phonon interaction alters
the low temperature specific heat through the mass enhancement parameter,
$1 + \lambda$. In fact, a more careful treatment 
\cite{prange64,grimvall69,grimvall81} yields a temperature-dependent
$\gamma(T)$ for the specific heat coefficient (which, at {\it very}
low temperature, reduces to the Sommerfeld $\gamma$). Besides providing 
quantitative corrections to the electronic specific heat in the normal
state, this correction also provides a properly physical contribution from
the low frequency phonon modes, as found in Ref. \cite{marsiglio86a}.

For a variety of conventional superconductors, like Al, the normal
state low temperature specific heat is easily measured by suppressing
the superconducting state with a magnetic field. Then the ratio
$\Delta C(T) /\gamma T_c$ can be determined. At $T_c$, the BCS result 
for this ratio
is universal, like the gap ratio: it is 1.43. Strong coupling corrections
can be derived \cite{marsiglio86} as before, as a function of the 
strong coupling parameter, $T_c/\omega_{\ln}$. The result is
\be
{\Delta C (T_c) \over \gamma T_c} = 1.43 \biggl[ 1 + 53 \bigl({T_c
\over \omega_{\ln} }\bigr)^2 \ln{({\omega_{\ln} \over 3 T_c})} \biggr].
\label{str_spec}
\ee                                
Again, the coefficients $53$ and $3$ were determined semi-empirically by
fits to numerical data. A plot of this result, along with some of the
numerical data, is shown in Fig. 27.
We already remarked about Al --- its calculated value is indicated by
the point nearest the ordinate, and agrees very well with experiment. 
The result for Pb is also shown; the experimental
value is 2.65, almost a factor of 2 greater than the BCS result. The
theoretical result, based on a numerical solution of 
Eqs. (\ref{bsfree}-\ref{gc2}), is in good agreement.

The result for stronger coupling has also been calculated \cite{marsiglio87a}.
In particular, the asymptotic limit can be computed, following standard
procedures. The result is \cite{marsiglio89a} 
\be
{\Delta C (T_c) \over \gamma T_c} = {19.9 \over \lambda},
\label{spec_asy}
\ee
showing that the relative magnitude of the jump decreases for large
$\lambda$, and therefore, as is already becoming apparent in Fig. 27,
the specific heat will have a maximum as a function of coupling strength.

Similar results can be derived for other thermodynamic properties as
well. These have been summarized in Ref. \cite{carbotte90} and will
be omitted here.

\subsection{Critical Magnetic Fields}

In a type-I superconductor, a critical magnetic field ($H_c$) exists, given
by Eq. (\ref{hc}). In a type-II superconductor, a lower critical field,
$H_{c1}$, and an upper critical field, $H_{c2}$, exist; the former
signals the departure from the Meissner state to one in which one vortex
penetrates the system, while the latter occurs at the normal/superconducting
transition. The thermodynamic critical field continues to exist as a
thermodynamic property, but not one that can be measured by application of
a magnetic field. 

A theory of $H_{c1}$ has been worked out within the BCS approximation
in Ref. \cite{usadel70,wattstobin74} (in the dirty limit). This work was
extended to the level of Eliashberg theory in Ref. \cite{rammer87}.
It is traditional to calculate the reduced field, $h_{c1}(T/T_c) \equiv
{H_{c1}(T) \over T_c H^\prime_{c1}(T_c)}$ as a function of $T/T_c$. Such
a curve has a slope of $-1$ near $T_c$, and saturates to some value
at $T=0$. Rammer \cite{rammer87} found that the low temperature value
{\it decreased} with coupling strength (there characterized by a
particular spectrum).

A detailed theory has also been provided for the upper critical field,
$H_{c2}$. In 1957 Abrikosov essentially created the subject of type II
superconductivity \cite{abrikosov57}. Both experimental and theoretical
work in this exciting area continued to flourish throughout the 1960's.
Applications of superconductivity in the mixed state require type II
superconductivity in order to sustain high magnetic fields. Abrikosov's
solution used the phenomenology of Ginzburg-Landau theory \cite{ginzburg50}.
Further theoretical developments utilized the microscopic theory of
Gor'kov \cite{gorkov58}. The first of these was by Gor'kov \cite{gorkov59}
for clean superconductors,
followed by five papers by Werthamer and collaborators 
to include impurity
effects \cite{helfand64,helfand66}, spin and spin-orbit effects
\cite{werthamer66}, Fermi surface anisotropy effects \cite{hohenberg67},
and retardation effects \cite{werthamer67}. All of these papers used
an instantaneous attractive potential (i.e. as in BCS theory), except
for the last. Further developments to include retardation effects
were carried out in Refs. \cite{eilenberger67,usadel70,bergmann73,rainer74}
and others. Finally, in Ref. \cite{schossmann86} the Eliashberg theory
of $H_{c2}$, including Pauli paramganetic limiting and arbitrary impurity
scattering, was formulated and solved. 

Without retardation effects or Pauli limiting, the zero temperature
upper critical field, when expressed in terms of the slope near $T_c$,
takes on universal values, dependent only on the elastic impurity scattering
rate, given by $1/\tau$. For example, the quantity $h_{c2}(0) \equiv
H_{c2}(0)/(T_c|H^\prime_{c2}(T_c)|)$ is given by 0.693 in the dirty limit
($1/\tau >> \Delta$) and 0.727 in the clean limit ($1/\tau = 0$). For 
intermediate scattering rates the result falls somewhere in between. It
is worth mentioning that the absolute value of the upper critical field
increases with increased impurity scattering. We often use the ratio
because the slope near $T_c$ is measured, and then the zero
temperature value is obtained by using the universal number quoted
above. The value at zero temperature is of special interest because
the Ginzburg-Landau coherence length can then be extracted through
\be
H_{c2} = {\Phi_0 \over 2 \pi \xi_{\rm GL}^2}.
\label{cohgl}
\ee
Here $\Phi_0$ is the fluxoid quantum, and we have used the subscript
`GL' to denote the Ginzburg-Landau coherence length, which, at zero
temperature, is often close to the BCS coherence length, and gives us
an indication of the Cooper pair size \cite{marsiglio90a}. 
Hence, deviations from 0.693 (or 0.727) due to retardation effects
are of interest for this reason.

For completeness, we quote the equations which govern $H_{c2}$, taking
into account electron-phonon interactions in the Eliashberg sense,
and Pauli limiting. The gap equation is linear in the order parameter
\cite{schossmann86}
\be
\tilde{\Delta}(i\omega_n) = \pi T \sum_m \biggl[ \lambda(i\omega_n - i\omega_m)
- \mu^\ast \biggr] {\tilde{\Delta}(i\omega_m) \over 
\chi^{-1}(\tilde{\omega}(i\omega_m)) - 1/2\tau},
\label{hc2_1}
\ee
with
\be
\tilde{\omega}(i\omega_n) = \omega_n + \pi T \sum_m
\lambda(i\omega_n - i\omega_m) {\rm sgn} \omega_m + {1 \over 2\tau}{\rm sgn}
\omega_n.
\label{hc2_2}
\ee
The factor $\chi(\tilde{\omega}(i\omega_n))$ is given by
\be
\chi(\tilde{\omega}(i\omega_n)) = {2 \over \sqrt{\alpha}}
\int_0^\infty dq \, e^{-q^2} {\rm tan}^{-1} \bigl( {\sqrt{\alpha} q \over
|\tilde{\omega}(i\omega_n)| + i \mu_B H_{c2} {\rm sgn}\tilde{\omega}(i\omega_n)
} \bigr).
\label{hc2_3}
\ee
Here $\alpha(T) \equiv {1 \over 2} |e|H_{c2}(T)v_F^2$, with $e$ the
charge of the electron and $v_F$ the electron Fermi velocity. $\mu_B$ is the
Bohr magneton. Eq. (\ref{hc2_1}) can be written as an eigenvalue equation,
just like $T_c$. It is linear because the solution is valid only on the
phase boundary between the normal and the superconducting states.

We have carried out extensive numerical investigations of $h_{c2}(0)$ as
a function of coupling strength. In the conventional regime, the dependence
on coupling strength is very weak \cite{marsiglio90b}; in the dirty limit
$h_{c2}(0)$ decreases initially (as a function of $T_c/\omega_{\ln}$)
to about 0.65, and then increases to beyond 0.70. In the clean limit
there is first a barely discernible decrease, followed by an increase to
values of approximately 0.80. These are all theoretical results, and,
in many cases have not been carefully investigated with experiment. On
the other hand the expected changes are of order 10\% or less, and
may well be masked by other effects. A thorough investigation was provided
for Nb by Schachinger {\it et al.} \cite{schachinger88}. The agreement with
the available data was excellent (although they did invoke, in addition
to the theoretical framework described here, anisotropy effects). For
further information the interested reader is directed to the 
aforementioned references.

Before leaving this section we should also mention that optimum spectrum
analysis \cite{akis88} and asymptotic limits (3rd reference in 
\cite{marsiglio89a}) have also been investigated for $H_{c2}$; the result
is very dependent on elastic impurity content, except in the asymptotic
limit. In that case, the results approach a universal value, i.e.
$h_{c2}(0) \rightarrow 0.57$ as $\lambda$ increases.

\section{Response Functions}

In the previous sections we have seen effects due to the inclusion
(through the Eliashberg formalism) of the detailed electron-phonon coupling.
The result is in many cases a large quantitative correction to the 
corresponding BCS result. In this way one can infer, from experiment, the
necessity of taking into account the dynamics of the electron-phonon
interaction. Nonetheless, as we saw in
Section 3 (particularly in the Tunneling and Optical Conductivity subsections)
dynamical interactions manifest themselves more clearly in dynamical
properties. For this reason we now focus on various response functions.

\subsection{Formalities}

A theory of linear response can be approached from two very different
frameworks, the Kubo formalism, and the Boltzmann equation. The two
frameworks often lead to the same result; their connection is discussed
at length in Ref. \cite{mahan81}.
Early treatments \cite{bardeen57,mattis58} of the various response functions 
in a superconductor neglected the electron-phonon interaction,
except insofar as it provided the mechanism for the superconductivity
in the first place. The main interest was the investigation of a new
state which apparently had a single electron energy gap, which
would manifest itself either directly in spectroscopic methods (optical
and tunneling) or more indirectly as a function of temperature (NMR relaxation
rate, acoustic attenuation, etc.).
Sometime later two seminal papers appeared \cite{holstein64,prange64}, 
both of which discussed the impact of the electron-phonon interaction
on transport in the electron gas. These dealt specifically with the normal
state. Work at a similar level but in the superconducting state appeared
a little later \cite{nam67}; this latter work was generalized to apply
for arbitrary elastic impurity scattering only much later \cite{lee89}.
These authors used quasiclassical techniques; below we will sketch
an alternative derivation based on the Kubo \cite{kubo57} formula.
 
We should preface this work with some remarks about vertex corrections.
They are generally ignored in calculations of response functions, so
that a particle-hole `bubble', consisting of one single electron
Green function and one single hole Green function, requires evaluation
\cite{remark10}.
Older work \cite{shaw68,swihart71} investigated the need for vertex
corrections and found that they contributed very little; later
work in the normal state \cite{scher70,allen71} suggested that
their contribution could be summarized by substituting a
`transport' electron-phonon spectral function, $\alpha_{\rm tr}^2F(\nu)$,
for the usual spectral function, $\alpha^2F(\nu)$, in the transport
equations. This alteration discriminated in favour of back scattering
as being particularly effective in depleting the current, as one would
expect. Over a large frequency range, however, these spectral functions
are not expected to differ substantially; nonetheless, quantitative
investigations are currently lacking \cite{mahan81}, particularly in
the superconducting state.

The contribution to the conductivity consists of two components:
the paramagnetic and diamagnetic responses. The diamagnetic
response is straightforward \cite{hirsch92}; the paramagnetic
response is determined by the evaluation of a current-current response
function. A standard decoupling of this function (ignoring, as noted
above, vertex corrections) yields
\begin{equation}
\sigma(\nu) = {i \over \nu + i\delta} \Bigl( \Pi(\nu + i\delta) +
{ne^2 \over m} \Bigr),
\label{cond}
\end{equation}
\noindent where $\Pi(\nu + i\delta)$ is the paramagnetic response function
whose frequency dependence (on the imaginary frequency axis) is given by
\be
\Pi (i\nu_n) = {1 \over N\beta} \sum_{{\bf k},m} {\rm Tr} (ev_x)^2 
\hat{G}({\bf k},i\omega_m) \hat{G}({\bf k},i\omega_m+i\nu_n),
\label{para}
\ee
where $\hat{G}({\bf k},i\omega_m)$ is actually a matrix in the Nambu formalism 
\cite{nambu60}. It is given, in terms of functions with which we are
already familiar, by
\be
\hat{G}({\bf k},i\omega_m) = - {i\omega_m Z(i\omega_m) + (\epsilon_{\bf k} - 
\mu) \hat{\tau}_3 +\phi(i\omega_m) \hat{\tau}_1 \over 
(\epsilon_{\bf k} - \mu)^2 - (i\omega_m Z(i\omega_m))^2 +\phi^2(i\omega_m) },
\label{gnambu}
\ee
where the Pauli spin matrices are given by
\be
\hat{\tau}_0 \equiv \left(\matrix{1&0\cr
                            0&1\cr}\right), \,\,\,\, 
\hat{\tau}_1 \equiv \left(\matrix{0&1\cr
                            1&0\cr}\right), \,\,\,\,
\hat{\tau}_2 \equiv \left(\matrix{0&-i\cr
                            i&0\cr}\right), \,\,\,\,
\hat{\tau}_3 \equiv \left(\matrix{1&0\cr
                            0&-1\cr}\right).
\label{pauli}
\ee
In Eq. (\ref{para}) the trace is over the Pauli spin space. The presence
of the factor $(ev_x)^2$ shows explicitly that the dressed vertex has
been replaced with a bare vertex; $v_x$ is the component of the electron
velocity in the x-direction. The momentum sum is over the entire Brillouin
zone; the factors preceding the summations include the total number of atoms
in the crystal, $N$, and the inverse temperature, $\beta \equiv 1/k_BT$.
The diamagnetic piece in Eq. (\ref{cond}) contains the electron density
$n$ and the electron mass, $m$. Finally, the single electron energy is denoted
by $\epsilon_{\bf k}$, and, as before, we use a notation where we 
explicitly subtract off the chemical potential, $\mu$. 

The paramagnetic kernel denoted by $\Pi (i\nu_n)$ in Eq. (\ref{para}) is
a special case of a more general `bubble' diagram. A similar calculation,
for example, is required for the NMR relaxation rate 
\cite{fibich65,allen91,akis91a}, 
or the phonon self-energy \cite{zeyher88,marsiglio92a,zeyher91,marsiglio93},
except that the vertices are not proportional to $\hat{\tau}_0$ (as was
the case in Eq. (\ref{para})), but to some other Pauli matrix. This has 
the effect that the so-called `coherence factors' will differ, depending
on the particular response function; some will result in a cancellation
with singularities arising from the single electron density of states, 
whereas others will result in a potentially singular response, at low
frequencies, as we shall see below. An early review outlining these
differences in the context of Eliashberg theory is given in Ref.
\cite{scalapino69}.

Returning to Eq. (\ref{para}), the standard procedure is as follows;
one would like to evaluate the Matsubara sum --- only then can one perform
the proper analytic continuation to real frequencies required for the
optical conductivity. This is straightforward, through the spectral
representation, which is the Nambu generalization of Eq. (\ref{spect_ele}).
The cost is that two new frequency integrals are required, one of which
can be done immediately by making use of the Kramers-Kronig-like relation
(see Eq. (\ref{speca}))
\be
\hat{G}({\bf k},z) = \int_{-\infty}^{\infty} \,d\omega 
{\hat{A}({\bf k},\omega) \over z - \omega},
\label{spect_ele2a}
\ee
with $z$ anywhere in the upper half plane. Finally, we would like to
perform the Brillouin zone integration analytically; to do so, we
note that the only dependence on wavevector ${\bf k}$ in Eq. (\ref{para})
occurs through $\epsilon_{\bf k}$ (this is not so for more complicated
response functions, such as is required for neutron scattering, for example,
where the momentum dependent kernel, $\Pi({\bf q},\nu+i\delta)$, is
required). This feature of the optical response allows us to make the
usual replacement, given already by Eq. (\ref{dos}):
\be
{1 \over N} \sum_{\bf k} \rightarrow \int \, d \epsilon N(\epsilon),
\nonumber \\
\ee
where $N(\epsilon)$ is the single electron density of states. 
As in that case $N(\epsilon)$ can be taken as constant ($=N(\epsilon_F)$)
and, along with the electron velocity, $v_x$, taken out of the integration
as an overall constant, $2N(\epsilon_F)e^2v_x^2 \equiv {\omega_P^2 \over
4 \pi} \equiv ne^2/m$, where $\omega_P$ is the electron plasma frequency.
However, one would normally like to extend the integration over 
single electron energy from $-\infty$ to $+\infty$, as is often done
within Eliashberg theory. Here, however, one has to be slightly more
careful, and first subtract the normal state contribution to
the kernel. This makes the integral sufficiently convergent that
extension to an infinite bandwidth (effectively) is possible. Then the
integral can be readily performed by contour integration. The integration
over the normal state contribution alone must be done separately; an
integration cutoff $\pm D$ must be used, the effect of which is an additional
(imaginary) contribution. The final result is \cite{remark11}
\begin{eqnarray}
\sigma(\nu) & = & {ine^2 \over m\nu}
\Biggl\{ \int_0^\infty d\omega \tanh({\beta\omega \over 2})  
\Bigl(
h_1(\omega,\omega + \nu) -  h_2(\omega,\omega + \nu)
\Bigr)
\nonumber \\
& & \phantom{{ne^2 \over m}\Biggl\{ -1} + \int_{-\nu}^D
d\omega \tanh({\beta (\omega + \nu) \over 2})
\Bigl(
h_1^\ast (\omega,\omega + \nu) +  h_2(\omega,\omega + \nu)
\Bigr)
\Biggr\}
\label{kernel}
\end{eqnarray}
\noindent with
\begin{eqnarray}
h_1(\omega_1,\omega_2) & = & {1 - N(\omega_1) N(\omega_2) -
P(\omega_1) P(\omega_2) \over 2(\epsilon(\omega_1) + \epsilon(\omega_2)) }
\nonumber \\
h_2(\omega_1,\omega_2) & = & {1 + N^\ast(\omega_1) N(\omega_2) +
P^\ast(\omega_1) P(\omega_2) \over 2(\epsilon(\omega_2) -
\epsilon^\ast(\omega_1)) }
\nonumber \\
N(\omega) & = & { \wtilde(\omega +i\delta) \over \epsilon(\omega + i\delta) }  
\nonumber \\
P(\omega) & = & { \phi(\omega +i\delta) \over \epsilon(\omega + i\delta) }
\nonumber \\
\epsilon(\omega) & = &
\sqrt{\wtilde^2(\omega +i\delta) -\phi^2(\omega +i\delta )}
\label{definitions}
\end{eqnarray}
where $D$ is the large cutoff mentioned above, to be taken to infinity 
for large electronic bandwidth, and $\wtilde(\omega +i\delta) \equiv
\omega Z(\omega +i\delta)$. 

Various limits can be extracted from these expressions; for example
the normal state results of Section 3.3.3 can be readily obtained, as well
as the simple Drude result, obtained by assuming only elastic scattering
characterized by a frequency independent rate, $1/\tau$. When inelastic
scattering is included (here through electron-phonon scattering), low 
frequency Drude-like fits can be obtained  through simple expansions
\cite{marsiglio95}. We will turn to these later.

Equation (\ref{kernel}) represents the `standard' theory of the optical
conductivity with Eliashberg theory. As already mentioned, this 
characterization includes the caveats discussed above about vertex
corrections and $\alpha^2F(\nu) \rightarrow \alpha_{\rm tr}^2F(\nu)$ 
replacements. It is valid for both inelastic scattering and
elastic scattering processes (within the Born approximation). The impact 
of elastic scattering on the Eliashberg equations have not yet been discussed,
so we turn to these now. Equations (\ref{gc1},\ref{gc2}), on the imaginary
axis, along with Eqs. (\ref{anal_zz},\ref{anal_phi2}), on the real
frequency axis, are written for the clean limit (`clean limit' is here
defined to mean that the elastic scattering rate is zero, $1/\tau = 0$).
When elastic scattering is included, new terms appear on the right hand
side of these equations. (As an aside, one way of using the existing equations
to include elastic scattering is to include a component of $\alpha^2F(\nu)$
(called $\alpha_{\rm imp}^2F(\nu)$ for simplicity) which models the elastic
scattering part. At any non-zero temperature it will be given by a zero 
frequency contribution
\be
\alpha_{\rm imp}^2F(\nu) = {\nu \over 2 \pi \tau T} \delta(\nu).
\label{phon-imp}
\ee
Substitution of this expression into Eqs. (\ref{gc1},\ref{gc2}), for example,
will yield simple expressions on the right hand side proportional to $1/\tau$.)
In principle one would think that Eqs. 
(\ref{gc1},\ref{gc2},\ref{anal_zz},\ref{anal_phi2}) require iteration to 
a solution for every new value of impurity scattering. In actual fact,
however, they need be solved only in the clean limit. Then, the pairing
$\phi(\omega+i\delta)$, and renormalization $\wtilde(\omega +i\delta) 
\equiv \omega Z(\omega +i\delta)$ functions can be modified by the simple
contribution
\def\csqrt{\sqrt{\wtilde^2(\omega+i\delta) - \phi^2(\omega+i\delta)}}
\begin{equation}
\phi(\omega+i\delta) \rightarrow \phi(\omega+i\delta) + {i \over 2\tau} 
{\phi(\omega+i\delta)
\over \csqrt}
\label{phidirt}
\end{equation}
\begin{equation}
\wtilde(\omega+i\delta) \rightarrow \wtilde(\omega+i\delta) + {i \over 2\tau} 
{\wtilde(\omega+i\delta)
\over \csqrt}.
\label{wtildedirt}
\end{equation}
Equations (\ref{kernel},\ref{definitions}) remain the same with 
impurity scattering. This is a consequence of the so-called Anderson's 
`theorem' \cite{anderson59}.
The modifications are all implicitly contained in the pairing and
renormalization functions. Note that the gap parameter, $\Delta(\omega+i\delta)
\equiv \phi(\omega+i\delta)\bigg/ Z(\omega+i\delta)$, remains the same, 
independent of the impurity scattering rate.         

\subsection{BCS results}

The purpose of this chapter is to examine effects specifically due to
the electron-phonon interaction. Nonetheless, it is best to first see what
occurs in the BCS limit, and then examine the differences. The means
for achieving the BCS limit from Eliashberg theory
was examined in Section 4.1; in general we mean by the `BCS limit'
that limit which corresponds to taking $\alpha^2F(\nu)$ to be non-zero
only for some very high frequency component (so that the ``strong coupling''
indicator $T_c/\omega_{\ln} \rightarrow 0$). As a result, the renormalization
function, $Z(\omega+i\delta) \rightarrow 1$, and the gap function
$\Delta(\omega+i\delta) \rightarrow \Delta$, a constant, as a function of
frequency. This allows one to explicitly break up the integrals in
Eq. (\ref{kernel}) into portions involving the BCS gap parameter, $\Delta$,
and the electromagnetic frequency, $\nu$. A very efficient FORTRAN program
has been provided in Ref. \cite{zimmermann91} in this case.

\subsubsection{Far-Infrared: Dirty Limit}

A historically important case is the dirty limit. This is defined by
$1/\tau >> \Delta$, and was first treated by Mattis and Bardeen
\cite{mattis58}. An analytical expression can be obtained at zero
temperature \cite{mattis58}:
\begin{eqnarray}
{\sigma_1 \over \sigma_n} & = & \bigl(1 + {2\Delta \over \nu} \bigr)
E(k) - {4 \Delta \over \nu}K(k) \,\,\,\,\,\,\,\, \nu > 2\Delta
\label{realbcs}
\\
{\sigma_2 \over \sigma_n} & = & {1 \over 2}\bigl(1 + {2\Delta \over \nu} \bigr)
E(k^\prime) - {1 \over 2}\bigl(1 - {2\Delta \over \nu} \bigr)K(k^\prime),
\label{imagbcs} 
\end{eqnarray}
where $\sigma_n \equiv {ne^2 \tau \over m}$ is the normal state
conductivity (pure real) and the real part of the conductivity is
identically zero for frequencies, $\nu < 2\Delta$.
In these expressions
\be
k = | {2 \Delta - \nu \over 2\Delta + \nu} | \,\,\,\,\, {\rm and} \,\,\,\,\,\,
k^\prime = \sqrt{1 - k^2}, 
\label{nothing}
\ee
and $E(k)$ and $K(k)$ are the complete elliptic integrals of the first
and second kind. For other cases (finite temperature and/or lower impurity
scattering rate) one must integrate numerically \cite{zimmermann91}.
Figure 28 shows (a) the real part and (b) the imaginary part of the
conductivity in the zero temperature BCS superconducting state, for various
impurity scattering rates.
We have used some definite values for the impurity scattering rates and the
coupling strength. The latter has been chosen to yield an absorption
edge, $2\Delta = 10.4$ meV, which, because of the insensitivity of
superconductivity to elastic impurity scattering \cite{anderson59}, holds
for all scattering rates. A well-defined absorption onset is evident
in Fig. (28a); otherwise the curves simply deviate from what would have
been Drude-like curves in the normal state. In Fig. (28b) the frequency
times the imaginary part of the conductivity is shown for the same
scattering rates. Such a combination is shown because the zero frequency
limit gives a direct measure of the London penetration depth:
\be
1/\lambda^2(T) = \lim_{\nu \rightarrow 0} {4\pi \over c^2} \nu \sigma_2(\nu).
\label{penet}
\ee
As is evident from the figure, the penetration depth increases as the impurity
scattering rate increases.

Another feature stands out in Fig. (28b); there is a notable `dip' in
$\nu \sigma_2(\nu)$ at $2\Delta$, particularly in the clean limit. Otherwise
the curves all approach the Drude limit at high frequency, which, for this
property, is unity (conductivities are in units of 
$ne^2/m \equiv \omega_P^2/4\pi$). 

\subsubsection{Penetration Depth}

Before we examine the effects of the electron phonon interaction on
the real and imaginary parts of the conductivity, we first summarize
the `BCS' results for the penetration depth as a function of impurity
scattering, which can be extracted analytically \cite{berlinsky93,marsiglio96}
from the zero frequency limit of the conductivity. The result is, with
$\alpha \equiv {1 \over 2 \Delta \tau}$,
\bea
{1 \over \lambda^2(T=0)} & = & {1 \over \lambda_{cl}^2(T=0)} \Biggl\{
{\pi \over 2 \alpha} - {1 \over \alpha \sqrt{1 - \alpha^2}}\sin^{-1}
(\sqrt{1 - \alpha^2}) \Biggr\} \phantom{bbbbbbbbbb} \alpha < 1
\nonumber \\
& & {1 \over \lambda_{cl}^2(T=0)} \Biggl\{
{\pi \over 2 \alpha} - {1 \over 2 \alpha \sqrt{\alpha^2 - 1}}\ln
\Bigl( {\alpha + \sqrt{\alpha^2 - 1} \over a - \sqrt{\alpha^2 - 1} }
\Bigr) \Biggr\} \phantom{bbbbbb} \alpha > 1.
\label{londirt1}
\eea
Here, the zero temperature London penetration depth in the
clean limit is given by
\be
\lambda_{cl}^2(T=0) = {mc^2 \over 4\pi ne^2}.
\label{lonclean}
\ee 
In the weak scattering limit Eq. (\ref{londirt1})
reduces to the more familiar form,
\be
{1 \over \lambda^2(0)} \approx {1 \over \lambda_{cl}^2(0)} {1 \over
1 + {\pi \over 4} \alpha}.
\label{londirt2}
\ee
This expression can be written in terms of the
zero temperature coherence length, $\xi_0$, and the mean free path,
$\ell$, using $\Delta = {v_F \over \pi \xi_0}$ and $v_F = \ell/\tau$,
where $v_F$ is the Fermi velocity:
\be
{1 \over \lambda^2(0)} \approx {1 \over \lambda_{cl}^2(0)} {1 \over
1 + {\pi^2 \over 8} {\xi_0 \over \ell}}.
\label{londirt3}
\ee

\subsubsection{Microwave Regime: Coherence Factors}

The microwave regime (1 - 60 GHz) corresponds to very low energies
(1 GHz = 0.0041 meV). This energy scale is much lower than that of
the superconducting energy gap. Measurements of the microwave response of
a superconductor have been used in recent years to determine
the penetration depth and optical conductivity in the high $T_c$ cuprates 
\cite{hardy93,bonn96},
but, historically, either the real or the imaginary component of
the surface impedance was measured, making a determination of the
complex conductivity impossible. It is of interest to examine
the conductivity in this case, because BCS theory makes a highly
non-trivial prediction that the real part of the conductivity shows
a so-called coherence peak just below $T_c$. This coherence peak was
almost simultaneously predicted \cite{bardeen57} and observed
\cite{hebel57,redfield59} in measurements of the 
NMR relaxation rate \cite{remark12}.
We will briefly discuss the source of these coherence factors, and return
to a description of the microwave conductivity, since a detailed
discussion of the NMR relaxation rate \cite{slichter78} is outside the
scope of this review, and the final expression relevant to superconductors
is a special case of the microwave conductivity.

Within BCS theory the transition probabilities between an initial and final
state that enter the expression for various linear response functions are
of the form \cite{bardeen57,schrieffer64}
\be
F_{\bf kk^\prime } = (u_{\bf k}u_{\bf k^\prime } \mp  v_{\bf k}
v_{\bf k^\prime })^{2}
\label{coh1}
\ee
\noindent with $u_{\bf k},v_{\bf k}$ the amplitudes that relate quasiparticle
operators to electron operators
\par
\medskip
\be
u_{\bf k} = \left({1 \over 2} \left (1 + {\epsilon_{\bf k}-\mu \over 
E_{\bf k}}\right ) \right
)^{1/2}
\label{amp1}
\ee
\par
\be
v_{\bf k} = \left({1 \over 2} \left(
1 - {\epsilon _{\bf k}-\mu \over E_{\bf k}}\right )\right)^{1/2}
\label{amp2}
\ee
and $E_{\bf k}$ is the usual quasiparticle energy:
\be
E_{\bf k} = \sqrt{(\epsilon_{\bf k}  -\mu )^{2} + \Delta_{\bf k}^{2}}
\label{quasi}
\ee
where $\epsilon_{\bf k}$ is the electron band energy, $\mu $ is
the chemical
potential and $\Delta_{\bf k}$  is  the  gap
function.  In Eq. (\ref{coh1}) the upper (lower) sign
corresponds to case I (case II) observables. These signs have important
consequences for the response, particularly just below $T_c$. A case in
point is the electromagnetic absorption; the temperature-dependent
result (derived from Eqs. (\ref{kernel},\ref{definitions})) in the
dirty limit ($1/\tau >> \Delta$) is \cite{mattis58}
\def\Delt{\Delta_0}
\bea
{\sigma_{1}\over \sigma_{n}} & = & {2\over \nu } \int^{\infty }_
{\Delt }
dE{E(E+\nu ) + \Delt ^{2}\over (E^{2}-\Delt ^{2})^{1/2}((E+\nu
)^{2}-\Delt ^{2})^{1/2}} [f(E)-f(E+\nu )]
\nonumber \\
& + & \theta ( \nu -2\Delt ) {1\over \nu } \int^{-\Delt }
_{\Delt -\nu}dE{E(E+\nu) + \Delt ^{2}\over
(E^{2}-\Delt ^{2})^{1/2}((E+\nu )^{2}-\Delt ^{2})} [1-2f(E+\nu)],
\label{condtempbcs}
\eea
where $\Delt \equiv \Delta(T)$ is the temperature-dependent gap function.
The second plus sign in $E(E+\nu ) + \Delt$ which appears in this
expression is due to the fact that the electromagnetic absorption
is a case II observable. In a case I observable this would be a minus
sign; it is then readily seen that whereas Eq. (\ref{condtempbcs})
contains a divergence as $\nu \rightarrow 0$, the corresponding
case I observable would not, as the numerator (coming from the
coherence factor given in Eq. (\ref{coh1})) would then cancel the
density of states factors, which are explicit in Eq. (\ref{condtempbcs}),
and which contain square-root divergences. In both cases the `freezing
out' of excitations as the temperature is reduced leads to a low temperature
suppression of the response function --- this is simply a consequence
of the gap. On the other hand, near $T_c$ an enhancement is expected for
type II observables, while, for type I observables, the response is
immediately suppressed as the temperature is lowered below the
superconducting transition temperature. In the limit that the frequency
is zero, one obtains from Eq. (\ref{condtempbcs}),
\be
{\sigma_{1}\over \sigma_{n}}  =  2 \int^{\infty }_{\Delt }
dE{E^2 + \Delt^{2} \over E^{2}-\Delt^{2}} \bigl( -{\partial f \over \partial E}
\bigr) \equiv (1/T_1)_s/(1/T_1)_n,
\label{cond_zero}
\ee
which is formally divergent (at all temperatures). The divergence is in
fact eliminated in practice by anisotropy in the gap or retardation
effects. As noted by the second equality, this is the expression for the
superconducting to normal ratio of the NMR relaxation rate.

For a type I observable (like the ultrasonic attenuation) the numerator
in Eq. (\ref{cond_zero}) has a minus sign, so that numerator and denominator
cancel, and the remaining integral is trivial. One obtains
\be
\alpha_s/\alpha_n = 2 f(\Delta(T)),
\label{ultra}
\ee
where $\alpha_{s(n)}$ is the ultrasonic attenuation in the superconducting
(normal) state, and $f$ is the Fermi function. This is a monotonically
decreasing function as the temperature decreases from $T_c$ to zero.

\subsubsection{Far-Infrared Regime --- Arbitrary Impurity Scattering}

The expressions for the optical conductivity provided in the last 
three subsections apply only in the dirty limit. As already mentioned
earlier, a comprehensive expression
(for all values of elastic impurity scattering), along with a very
efficient FORTRAN program, was provided in Ref. \cite{zimmermann91}.
For completeness, we illustrate here the temperature dependence
for two extreme cases, close to the clean limit ($1/\tau = 1$ meV), and
the dirty limit ($1/\tau \rightarrow \infty$), in Fig. 29 and Fig. 30,
respectively.
As noted earlier, the optical gap ($= 2\Delta(T)$) is clearly evident
in both the real and imaginary part of the conductivity. The evolution
from the normal state to the superconducting state is clearly evident
as well; note, in particular, that in the real part of the conductivity,
the missing area is taken up as a delta function at the origin (not
shown).
 
\subsection{Eliashberg Results}

Within Eliashberg theory, changes occur for two related reasons. First,
even in the normal state the self-energy acquires a frequency dependence
(no wavevector dependence, because of the simplifying assumptions made
at the start); secondly, the gap function in the superconducting state
acquires a frequency dependence {\em and} acquires an imaginary part.
This latter fact tends to smear many of the `sharp' results shown in
the last section, a feature which is already evident in comparing
the single electron densities of states in Fig. (24) to those in
Fig. (21b), for example.
For this reason, it is important to re-examine the impact of retardation
on a variety of observables. 

\subsubsection{NMR Relaxation Rate}

In the first few years following the discovery of the high temperature
superconductors \cite{bednorz86}, several anomalous features were
measured in the superconducting state. One of these was the absence
of the coherence peak (the so-called `Hebel-Slichter' peak) in the NMR
spin relaxation rate, $1/T_1$, just below $T_c$ \cite{pennington90}. 
Motivated by the possibility that this `anomaly' could be explained by
damping effects due to retardation, Allen and Rainer \cite{allen91}
and Akis and Carbotte \cite{akis91a} calculated the ratio of the
relaxation rate in the superconducting state to that in the normal state
with several hypothetical electron-phonon spectra (obtained by scaling
known spectra from conventional superconductors). Both groups found
that sufficiently strong coupling (as measured by $\lambda$ or
$T_c/\omega_{\ln}$) smears out the coherence peak entirely. An example
is shown in Fig. 31 (taken from Ref. \cite{allen91}), 
which shows the theoretical and
experimental \cite{williamson73} results for a conventional superconductor
(Indium) along with data from YBCO \cite{hammel89}, and theoretical
results obtained using scaled spectra. While the present consensus
is that the lack of a coherence peak is {\em not} solely due to
damping effects, the lesson learned from these calculations is clear:
retardation effects damp out the coherence peak in the NMR relaxation rate.
It is worth noting here that even within a BCS framework (i.e. no
retardation), the coherence peak can be suppressed in the dilute electron
density limit \cite{marsiglio91b}.

\subsubsection{Microwave Conductivity}

A natural extension of this argument applies to the microwave conductivity.
In this case, even within BCS theory, a divergence does not occur since
the experiment is conducted at some definite non-zero microwave frequency
(see Eq. (\ref{condtempbcs})). Before discussing retardation effects,
however, it is important to realize the amount of impurity scattering
(as characterized by $1/\tau$) also influences the height and presence of the
coherence peak \cite{marsiglio91a,akis91}. In Fig. (32a)
we show, within the BCS framework, the conductivity ratio for a small
but finite frequency as a function of reduced temperature, for a variety
of elastic scattering rates, ranging from the dirty limit to the clean limit.
Quite clearly the coherence peak is reduced and then eliminated as a function
of $1/\tau$.

To see how retardation effects also serve to reduce and eliminate the
coherence peak (just as in NMR) we focus on the dirty limit ($1/\tau
\rightarrow \infty$) where the peak is largest without retardation.
In Fig. (32b) we show results obtained from a Pb spectrum (Fig. 11), scaled
by varying degrees to increase $\lambda$ from 0.77 to 3.1. For the largest
coupling considered the coherence peak has essentially vanished. This is
the same effect seen in the NMR relaxation rate. In Fig. (32c)
we illustrate the impact of changing the microwave frequency. Clearly, in the
limit of very weak coupling (BCS) one expects the strongest variation,
since, as $\nu \rightarrow 0$, the BCS result will diverge logarithmicly.
However, as the coupling strength increases, the damping due to retardation
reduces the peak far more effectively than an increase in microwave
frequency would, so that the conductivity ratio (at some temperature near
where a maximum would occur in the BCS limit) is essentially constant
as a function of frequency. This is clearly illustrated by the two lowest
curves in the Figure, representing the strongest coupling situations.

A measurement of the coherence peak in the microwave wasn't actually
performed until the early 1990's, in Pb \cite{holczer91} and in
YBCO \cite{holczer91a} (although the peaks observed in these
latter measurements are now thought {\em not} to be the BCS coherence
peak \cite{nuss91,bonn92,bonn96}.

Several other groups have since examined the microwave response
in conventional superconductors. In Ref. \cite{marsiglio94}
Nb was examined in detail. The experiment was performed at 17 GHz,
and a prominent coherence peak was observed, as shown in Fig. 33.
Also shown are theoretical curves obtained from Eliashberg
calculations; they all fall significantly {\em below} the experimental
results. We have also included the BCS result (dotted curve) computed
for this frequency; it is not very different from one of the curves
obtained using the full Eliashberg formalism. The BCS result represents
probably the highest achievable coherence peak; other alterations of the
standard theory (anisotropy, finite bands, non-dirty limit, etc.) would tend
to {\em decrease} the theoretical result further. Hence, at present the
coherence peak observed in Nb remains anomalous because {\em it is too big.}
Other measurements in Nb and Pb \cite{klein94} showed agreement with
Eliashberg theory, but they were carried out at a much higher frequency 
(60 GHz). Another measurement of the electrodynamic response
(using simultaneous measurement of the amplitude and phase of the
transmission in Nb thin films) \cite{pronin98} supported our results.
A more recent measurement of the coherence peak in Nb$_3$Sn \cite{hein99}
also finds a large discrepancy with Eliashberg theory --- the experimental
results show a peak which is far too large compared to theory.

\subsubsection{Far-Infrared Regime}

While more recent investigations of the far-infrared (and slightly lower
Terahertz) regime in superconductors utilize transmission techniques which 
simultaneously measure amplitude and phase information \cite{nuss91,pronin98}, 
the more conventional Fourier-transform spectroscopy \cite{wooten72}
requires Kramers-Kronig relations, as outlined in Section (3.3.3). 
For this reason the entire spectrum needs to be measured, often with an
assortment of spectrometers \cite{timusk89}. How do the real and
imaginary parts of the conductivity change as a function of the coupling
strength $\lambda$~? In Fig. 34 we show real ((a) and (b)) and imaginary
((c) and (d)) parts of the conductivity with $1/\tau = $ 2 meV and 25
meV, respectively. In all four figures it is clear that an increased
coupling strength decreases the real and imaginary parts of the conductivity,
at least in the low frequency regime. In fact, at low temperatures, in
the normal state, one can derive a Drude-like expression \cite{marsiglio95}
\be
\sigma_{\rm Drude}(\nu) \approx {ne^2 \over m^\ast} {1/\tau^\ast \over \nu^2 +
[1/\tau^\ast]^2 }
\label{drudelike}
\ee
where $ m^\ast/m = 1 + \lambda $ and $\tau/\tau^\ast = 1/(1 + \lambda)$.
This expression clearly indicates that, while the zero frequency
conductivity remains unaffected, the rest of the conductivity is
diminished by the electron-phonon interaction \cite{prange64}. In fact
integration of Eq. (\ref{drudelike}) yields the result
\be
\int_0^\infty d\nu \,\, \sigma_{\rm Drude}(\nu) = {\pi \over 2}
{ne^2 \over m^\ast} {1 \over 1 + \lambda}.
\label{partsum}
\ee
This is lower than the Kubo sum rule \cite{kubo57} by the factor of
$1/(1 + \lambda)$, which says that the rest of the area is taken up in
the phonon-assisted absorption, which occurs at higher frequency (in
the phonon range). Also note that one effect of an increased electron-phonon
interaction strength is to {\em decrease} the impurity scattering rate:
$1/\tau \rightarrow 1/\tau{1 \over 1 + \lambda}$. This occurs because 
the inelastic scattering reduces the spectral weight of the quasiparticle
undergoing the elastic scattering. Further discussion of the Drude-like
behaviour at low frequency but for non-zero temperature can be found in
Ref. \cite{marsiglio95,dolgov91}.

Returning to Fig. 34, we note that except for small corrections to the
gap edge as $\lambda$ increases ($2\Delta$ tends to increase as well),
the occurrence of an abrupt onset of absorption in the real part
((a) and (b)) exists for all coupling strengths. While a cusp remains
in the imaginary part ((c) and (d)), its size is clearly diminished as
the coupling strength increases. Note that the penetration depth
(given by the square-root of the inverse of the intercept in the
imaginary part --- see Eq. (\ref{penet})) tends to increase 
as the coupling strength increases. 
Also note that, while not apparent on the frequency scale shown in
(c) and (d), the frequency times the imaginary part of the conductivity
approaches unity (in units of $ne^2/m$) as the frequency approaches
large values. This fact was utilized in the case of Ba$_{0.6}$K$_{0.4}$BiO$_3$,
which we briefly discuss next.

Fig.~35 shows the imaginary part of the conductivity obtained from reflectance
measurements on Ba$_{0.6}$K$_{0.4}$BiO$_3$ 
\cite{puchkov94,puchkov95,marsiglio96}. A prominent dip occurs near 
12 meV, which has been roughly fit by two models as indicated. The occurrence
of this dip fully supports the existence of a superconducting state with
s-wave symmetry, with a gap value that is high compared to that expected
from BCS theory ($2\Delta/k_BT_c \approx 5$ compared with $3.5$). 
This value is somewhat higher than that obtained previously with infrared
\cite{schlesinger89} or tunneling \cite{zasadzinski90,sharifi91} measurements.
Nonetheless, a thorough analysis of the temperature dependence of the
Drude fits at low frequency \cite{marsiglio95} and the frequency dependence
illustrated in Fig. 35 \cite{marsiglio96} shows that the electron-phonon
interaction must be weak in this material, too weak to support $30$ K
superconductivity. Two model calculations are shown with the data in 
Fig.~35. The data is clearly consistent with an electron-phonon coupling
strength $\lambda \approx 0.2$ (which requires an additional mechanism to
produce $T_c = 30$ K), and entirely inconsistent with $\lambda \approx 1$. 

As is clear from the preceding paragraph, either the real or the imaginary
part of the conductivity contains all the relevant information about the
absorption processes in the system. This is due to the fact that they obey
Kramers-Kronig relations, which ultimately can be traced to requirements of
causality and analyticity \cite{wooten72}. In an effort to make these
absorption processes more explicit, one can also favour other functions;
a particular example is the effective dynamical mass, $m^\ast(\nu)$, and
the effective scattering rate, $1/\tau(\nu)$, introduced through
\cite{dolgov91}
\be
\sigma(\nu) = {\omega_P^2 \over 4 \pi} {1 \over 1/\tau(\nu) - i \nu
m^\ast(\nu)/m},
\label{extended_drude}
\ee
where $\omega_P$ and $m$ are the bare electron plasma frequency and mass,
respectively. Then, one can define an effective scattering function,
$1/\tau(\nu)$, which can be extracted (say, from experiment) through
\be
1/\tau(\nu) \equiv {\omega_P^2 \over 4 \pi} {\rm Re} { 1 \over \sigma(\nu)},
\label{taueff}
\ee
bearing in mind that $\sigma(\nu)$ itself has been obtained through 
Kramers-Kronig relations from, say, reflectance data.
This is precisely the function required to invert normal state conductivity
data to extract $\alpha^2F(\nu)$ (see Eq. (\ref{explicit})). A plot of
$1/\tau(\nu)$ vs. frequency is nevertheless revealing. It tends to illustrate
at roughly what energies absorption process `turn on' 
\cite{dolgov91,puchkov96}. For example, we show in Fig. 36 the function
$1/\tau(\nu)$ derived from conductivity results of model calculations for
\BKBO and \YBCO \,\, \cite{marsiglio97}. The former uses a model 
phonon spectrum extracted from neutron scattering measurements \cite{loong91} 
while the latter uses a model spin fluctuation spectrum \cite{jiang96}.
The fact that the \YBCO \,\, result continues to rise at 300 meV reflects the
frequency scale of the spin fluctuation spectrum. In contrast, the \BKBO
\, \, result has almost saturated by 100 meV, since the phonon spectrum extends
only to 80 meV. More detailed comparisons with self-energy-derived
scattering rates have been provided in Refs. \cite{dolgov91,marsiglio97}.

The results shown in Fig. 36 were obtained in the normal state. In the
superconducting state the presence of a gap will modify the low
frequency behaviour of the scattering rate, $1/\tau(\nu)$.
Results within BCS theory (elastic scattering rate only --- no inelastic
scattering) are shown in Fig. 37. At low frequencies the overall scale
of the effective scattering rate is set by the elastic scattering rate
(2 and 25 meV, respectively). Note that in the gap region (below 32 meV)
the effective scattering rate is zero (at zero temperature), while
slightly above the gap the effective scattering rate below $T_c$ 
is actually enhanced with respect to the normal state value. In Fig. 38
we show the effective scattering rate vs. frequency using the model
\BKBO \, \, spectrum in (a) the clean limit and (b) with significant impurity
scattering. The results are qualitatively similar to those in Fig. 37.

\subsection{Phonon Response}

Much of this review has focused on various properties whose determination
allows one to infer the degree of electron-phonon coupling that exists in
the material under study. The majority of properties that fall in this 
category refer to a modification of the electronic structure or response
due to a coupling with phonons. To a much lesser extent the phonons
themselves are modified because of the electron-phonon coupling, and in
this section we briefly address a few examples in this category. 

The impact of the superconducting state on the phonons was first
investigated using ultrasound experiments \cite{morse57}. Sound waves are
attenuated due to their absorption in the solid. The absorption requires 
interaction with electrons with energies very close to the Fermi energy
(the phonon energy is typically very low for sound waves --- in the
$100$ MHz $= 0.0004$ meV range). These electron states are gapped in
the superconducting state, so the attenuation is expected to be
suppressed to zero as $T \rightarrow 0$. The BCS result, given by
Eq. (\ref{ultra}), is valid for
an order parameter with s-wave symmetry.
A similar law can be derived for other symmetry types \cite{lupien01},
which results in some sort of power law decay rather than exponential
at low temperatures.

Of main interest here is how Eq. (\ref{ultra}) is modified when
retardation effects are accounted for. An early calculation \cite{ambegaokar66}
found that retardation effects did not alter the result Eq. (\ref{ultra}).
Therefore, little can be learned about the electron-phonon interaction
through ultrasonic experiments; instead, one should examine higher energy
phonons.

The classic experiment of this type was performed using neutron scattering
on Nb and Nb$_3$Sn \cite{axe73}. The idea is simply that the electron charge
susceptibility modifies the phonon spectrum. Within the normal state this
modification is hardly noticeable in metals over a temperature range of 300
K or so. However, when the material goes superconducting, the electron
density of states is profoundly modified at energy scales of order the
gap; this in turn will affect phonons whose energy is on the same scale.
In particular, a low energy phonon (energy less than $2\Delta$) that had
a finite lifetime because it could decay into an electron-hole pair will
be unable to do so in the superconducting state because no states exist
at energies below the gap, $\Delta$. Therefore its lifetime will lengthen
considerably in the superconducting state, resulting in a narrower
lineshape below $T_c$. Fig. 39 shows the experimental result from Nb$_3$Sn
\cite{axe73} where the lineshape has clearly become narrower in the
superconducting state. Similarly, if the phonon energy is slightly above
$2\Delta$, then, under the right conditions, the linewidth will increase,
since the electron density of states increases in this energy regime in
the superconducting state.

A detailed theory of these effects was first given in Ref. \cite{bobetic64},
within BCS theory. The theory consists of a calculation of a response function
corresponding to a Case I observable. Similar calculations were performed
much later by Zeyher and Zwicknagl \cite{zeyher88} to understand the 
frequency shifts and linewidth changes (due to superconductivity) in 
the ${\bf q} = 0$ Raman spectra for various optical modes in \YBCO . 
They found, using the BCS approximation,
\be
{Re\ \Delta \Pi ({\bf q}=0,\nu + i\delta) \over N(0)} = \cases{
-{2 \over \bar{\nu} \sqrt{1-\bar{\nu}^2} } tan^{-1} \Bigl( {\bar{\nu}
\over \sqrt{1-\bar{\nu}^2} } \Bigr)  & for $\bar{\nu}< 1$ \cr
{1\over \bar{\nu} \sqrt{\bar{\nu}^2 -1} } ln \Bigl( 2\bar{\nu}^2 -1 +
2\bar{\nu} \sqrt{\bar{\nu}^2-1} \Bigr) & for $\bar{\nu}> 1$. \cr}
\label{phononbcs}
\ee
The imaginary part is given for all temperatures by:
\be
{Im\ \Delta \Pi ({\bf q}=0,\nu + i\delta) \over N(0)} = 
-\pi \theta ( \bar{\nu} -1)
{ tanh\  \beta \nu/4 \over \bar{\nu}\sqrt{\bar{\nu}^2 - 1} }, 
\label{phononbcs_im}
\ee
where $\bar{\nu} = \nu/(2\Delta(T))$. 
Here, $\Delta \Pi ({\bf q},\nu + i\delta)$ is the 
change in the phonon self energy
between the superconducting state and the normal state. A positive 
(negative) real part means that phonons harden (soften) in the 
superconducting state, while a positive (negative) imaginary part means
that the phonon linewidths narrow (broaden). Thus, phonons below the
gap edge ($2\Delta$) soften while those above harden. Also, above the
gap edge they broaden while below their linewidth does not change. The
broadening above $2\Delta$ can be understood as being due to the enhanced
scattering with electrons, since the electron density of states now
has a square-root singularity in the energy range of $\Delta$, and the phonon
self energy is essentially a convolution of two single electron Green
functions (see Eq. (\ref{para})). 

Eqs.~(\ref{phononbcs},\ref{phononbcs_im}) have been derived assuming single
particle Green functions without impurity scattering. The ${\bf q} = 0$
limit is somewhat anomalous in this case, in that the phonon width is
already zero in the normal state. Hence, no change can occur in the
linewidth in the superconducting state, for frequencies below $2\Delta$.
A calculation with impurities \cite{marsiglio92a} provides a non-zero
linewidth in the normal state. Because of the gap in the single electron
density of states in the superconducting state, this linewidth is reduced to
zero when the system enters the superconducting state, so the change in the
imaginary part of the phonon self energy is positive. These results are
summarized in Fig. 40. 
Note that the softening below the gap edge is significantly reduced with
impurity scattering present, and the phonons above $2\Delta$ also soften
when a significant degree of impurity scattering is present. As Fig. 40b
shows, phonons whose energy lies below $2\Delta$ acquire a narrower
linewidth in the superconducting state, as noted above. 

The effects of retardation on the phonon self energy are not very significant.
The changes that do occur follow the changes already discussed due to
including elastic scattering; high energy phonons soften rather than harden,
and the broadening that accompanies this softening is reduced compared
to the clean BCS case. More detailed changes are documented in
Refs. \cite{zeyher88,marsiglio92a}. 

Because these phonon changes can be observed through neutron scattering
experiments, it is of interest to examine the phonon self energy at
non-zero momentum, ${\bf q}$ \cite{zeyher91,marsiglio93}.
In this case the phonon has a non-zero linewidth in the normal state, and
so line narrowing is observed in superconducting state at low frequencies,
due to the development of a single electron gap. The detailed frequency
dependence is a function of the band structure; in particular, with
two dimensional nesting phonon changes due to superconductivity are
enhanced \cite{marsiglio93}.

\section{Summary}

We have examined a variety of ways in which the retarded electron phonon
interaction influences the properties of a conventional superconductor.
The first and simplest effect is through a renormalization of Fermi
Liquid parameters, like the effective mass. While this effect appears
in a number of normal state properties (for example, the low temperature
electronic specific heat capacity, where the Sommerfeld $\gamma$ is
enhanced by $1 + \lambda$ --- see Eq. (\ref{gamm0})), it also appears
in many superconducting properties.  The most obvious (but least
measurable) example is in the $T_c$ equation, Eq. (\ref{tc_weak}),
where $1 + \lambda$ appears in the exponent. Another (perhaps more
detectable) occurrence is in the slope of the upper critical magnetic
field. In each of these cases, the renormalization occurs in the normal
state --- its occurence in the superconducting state is because the
property in question depends on the normal state effective mass, or
Fermi velocity, etc. One should also bare in mind that the factor
$1 + \lambda$, comes from a weak coupling approach. In 
a strong coupling approach, an electron phonon
renormalization is still present, but may be much more significant than
suggested by the weak coupling approach, and polaron-like physics
may dominate \cite{alexandrov95}.

The most important manifestation of the electron phonon interaction is
the superconducting state itself. In fact, according to our present
understanding of Cooper pairing, the electron phonon-induced attraction
between two electrons would not overcome their direct Coulomb repulsion,
except for the fact that the former is retarded whereas the latter is not.
This gives rise to the pseudopotential effect; in some sense the
pseudopotential effect is the true mechanism of superconductivity, rather
than the electron phonon interaction per se. This is perhaps emphasized
in the cuprate materials, where presumably the electrons could not utilize
the difference in energy (and hence time) scales between the attractive
mechanism (whatever it is) and the direct Coulomb repulsion to overcome
the latter. Instead the pairing has apparently adopted a different
symmetry (d-wave) to avoid the direct Coulomb repulsion.

Nonetheless a minimal accounting for these retardation effects accounts
fairly well for the superconducting ground state. This was accomplished by
BCS theory. A more accurate theory with retardation effects (Eliashberg
theory) quite clearly accounts for {\em quantitative} discrepancies
with experiment. Here, Pb and Hg are held up as paradigms for retardation
effects, the simplest occurring in a measurement of the gap ratio, for
example. The BCS theory predicts a universal number for this ratio,
$2\Delta /k_BT_c = 3.53$. With Eliashberg theory a value for Pb is found
close to 4.5, in excellent agreement with experiment. We have characterized
the discrepancy with BCS theory through a retardation parameter, 
$T_c/\omega_{\ln}$. Various properties have been quantitatively 
accounted for through simple analytical expressions with this parameter,
as given in Sections 4 and 5 (see Ref. \cite{carbotte90} and references therein
for many more).

Finally, various dynamical properties exhibit `signatures' of the 
electron-phonon pairing. These tend to manifest themselves as `wiggles'
in the data, the most famous of which occurs in the tunneling data, and
allows an inversion to extract the electron phonon spectral function,
$\alpha^2F(\nu)$. 
As we saw briefly in Section 3, and then again in Section
6, these `wiggles' occur in various two-electron response functions, most
prominent of which is the optical conductivity. An accurate measurement of
these response functions allows one to infer a significant electron-phonon
coupling.

We have focussed on very conventional superconductors, and have,
for example, avoided any analysis of the high temperature superconductors.
Signs of electron phonon interactions have occurred in these new materials
as well, but the relation to the superconductivity in them is yet unclear.
Moreover, such effects will no doubt be covered in other chapters.
Nonetheless, we wish to add a few remarks about other classes of 
superconducting materials that have been discovered over the last twenty
years.

Cubic Perovskites, beginning with strontium titanate (SrTiO$_3$)
\cite{cohen64,schooley64}, have already been discussed in Section 2.
As mentioned there, these compounds (including 
BaPb$_{0.75}$Bi$_{0.25}$O$_3$ \, ($T_c \approx 12$ K) \cite{sleight75}
and Ba$_{1-x}$K$_x$BiO$_3$ \,\, ($T_c \approx 30$ K) \cite{mattheiss88})
are generally regarded as in a distinct class from the high $T_c$
cuprates. This has left them, somewhat by default, as electron-phonon
driven superconductors. On the other hand, there is strong optical
evidence \cite{puchkov94,marsiglio96} that the electron phonon
interaction is very weak in these materials. Hence, as far as we are concerned,
the mechanism of superconductivity in these perovskites 
is not understood at all.
Tunneling studies \cite{zasadzinski90,sharifi91} are divided on this
issue.

One- and two-dimensional organic superconductors were discovered in
1979 \cite{jerome80}. The subject had developed sufficiently so that,
by 1990, a book devoted to the topic was written \cite{ishiguro90}. 
Organic superconductivity represents another interesting idea that was
first presented by theorists \cite{little64,ginzburg64}, on the basis
of a phonon-mediated interaction, but that now is considered by
most practitioners {\em not to be due to electron phonon interactions}.
Many of the organics abound in physical phenomena, with several containing,
on the same phase diagram, charge density wave (CDW) and spin density
wave (SDW) instabilities, juxtaposed with superconductivity 
\cite{bourbonnais00}. The nature of the superconducting state has not really
been sharply defined by experiments, to the extent that both singlet and
triplet pairing may be present \cite{shimahara00}, and the presence of a gap
has not been unequivocally established. While it is probably fair to say that
the electron phonon interaction has not been ruled out as the mechanism
for superconductivity, spin fluctuation-mediated pairing seems to be
favoured \cite{bourbonnais00}.

Heavy Fermion systems were discovered to be superconducting also in 1979
\cite{steglich79}. While $T_c$ has remained low, these compounds have remained
of interest because (i) the root cause of the heavy electron mass is not
completely understood, and (ii) the superconducting ground state coexists 
in a number of cases with antiferromagnetic order. It has now
been established through thermal conductivity measurements that the order
parameter contains nodes \cite{lussier96}, and the circumstantial
evidence points towards an unconventional magnetically mediated mechanism
for superconductivity \cite{mathur98}. There is very little indication
that superconductivity in this class of compounds has anything to do with
the electron-phonon interaction.

Superconductivity in alkali-doped buckminster fullerene (A$_3$C$_{60}$,
with A = K, Rb, Cs) was briefly mentioned earlier in this chapter. On the
basis of optical measurements \cite{degiorgi94}, a sizable electron phonon
coupling was inferred, and, in fact $\alpha^2F(\nu)$ was extracted by
an inversion procedure outlined in Section (3.3.3) \cite{marsiglio98}.
Evidence for electron phonon-mediated superconductivity was also presented
in earlier reviews \cite{gunnarson97}. On the other hand, doubts remain
concerning the validity of a weak coupling framework
\cite{chakravarty00}. One would like to understand the `bigger picture',
i.e. the progression from insulator with pure C$_{60}$ through the
superconducting phase with A$_3$C$_{60}$, and back to insulator with
A$_6$C$_{60}$. In fact, band structure calculations \cite{satpathy92}
suggest (simplisticly) that A$_2$C$_{60}$ should be superconducting with
a higher $T_c$ than A$_3$C$_{60}$, when, in fact, that compound does not
readily form.

The electron phonon theory can be subjected to even more tests, now that
workers have managed to fabricate a field effect transistor which allows
electron \cite{schon00} and hole \cite{schon00b} doping of C$_{60}$.
$T_c$ is much higher for hole doping ($T_c = 52$ K), and spans a wide
range of dopant concentration. In fact this peculiar asymmetry between
electron and hole doping finds a natural explanation through the hole
mechanism of superconductivity \cite{hirsch89}. An explanation in terms of
a dopant-dependent electron phonon coupling strength appears somewhat
unnatural.

The borocarbides (RNi$_2$B$_2$C, where R denotes a rare earth element)
were found to be superconducting in 1993 \cite{nagarajan94,cava94}.
In addition to having a sizeable transition temperature ($T_c \approx 20$ K),
some of these compounds exhibit coexistent superconductivity and
antiferromagnetic order, and indeed, share some similarities with the
heavy fermion compounds \cite{canfield98}. Nonetheless, tunneling has
determined that a well-defined gap exists at low temperatures, and this
and other measurements have established these compounds to have very
BCS-like properties \cite{suderow01}. A detailed comparison of various
superconducting properties with results based on Eliashberg theory
(including some small anisotropy) \cite{manalo01} yields excellent
agreement. A model spectrum was used for the electron phonon interaction,
and, at present, it remains unclear to what extent this agreement points
unequivocally to the electron phonon mechanism for superconductivity in these
compounds.

Very recently, superconductivity with $T_c = 39$ K has been discovered 
in the very simple binary compound, MgB$_2$ \cite{nagamatsu01}. 
Preliminary results indicate a gap in the single electron density of
states \cite{rubiobollinger01,karapetrov01,sharoni01}, and an isotope
effect has been observed \cite{budko01}. Calculations of the electron
phonon coupling strength, not quite consistent with $T_c = 39$ K, have
been reported \cite{kortus01}, as has a competing non-electron phonon
mechanism, based on the hole mechanism \cite{hirsch01}. More experimental
results will be required before a real assessment of the electron phonon
mechanism can be provided.

Finally, Sulfur has been found to exhibit a
high superconducting transition temperature ($T_c = 17$ K) \cite{struzhkin97}.
Very little work has been carried out regarding the
mechanism; a notable exception is Ref. \cite{rudin00}, where ab initio
calculations are performed to estimate the electron phonon coupling
strength for Sulfur. They find that under pressure, in a different
structural phase, the electron phonon coupling is enhanced, consistent with
the increase in $T_c$.

As is evident by the foregoing examples, a steady search for new 
superconductors is being rewarded with discoveries of materials
with high critical temperatures, now in the same category as those of the high
temperature cuprates. The A15 compound record of $T_c \approx 23$ K would
have been broken many times by now, even if the layered cuprates had not
been discovered. Most intriguing is the fact that many of these compounds
may be driven to the superconducting state through the electron phonon
mechanism. As far as future developments in this area is concerned, an
obvious question to be addressed is the soundness of the original
Cohen-Anderson estimate \cite{cohen72} for the maximum electron phonon
mediated critical temperature. It may simply be a matter of quantitative
assessment, or perhaps some more exotic effect (within the electron phonon
picture) has been overlooked. An intermediate or strong coupling approach 
\cite{alexandrov95} may yet provide new insights. Finally, one can't help but
notice the recent resurgence of investigations in the high temperature
cuprates themselves, that indicate strong electron phonon effects
\cite{lanzara01}. To paraphrase \cite{anderson87}, 'The fat lady probably
hasn't yet sung'.

\section{Appendix: Microscopic Developments}

In this Appendix, we will first outline a derivation of Eliashberg theory,
based on a weak coupling approach. By this we mean that we start with
momentum eigenstates. While other derivations may be given in other chapters,
we include one here to keep this chapter somewhat self-contained.
Migdal theory follows by simply dropping the anomalous amplitudes in what
follows.
We will then outline various other attempts to
understand electron phonon interactions, particularly in the strong
coupling regime.

\subsection{Migdal-Eliashberg Theory}

We begin with the definition of the one electron Green function, defined
in momentum space, as a function of imaginary time \cite{mahan81},
\be
G(\bk,\tau-\taup) \equiv - < T_\tau c_{\bk\sigma}(\tau)
c^\dagger_{\bk \sigma}(\taup) >,
\label{green_t}
\ee
where ${\bf k}$ is the momentum and $\sigma$ is the spin. The angular
brackets denote, as usual, a thermodynamic average. With this definition
such a Green function can be Fourier expanded in imaginary frequency:
\bea
G(\bk,\tau) = {1\over \beta} \sum_{-\infty}^\infty e^{-i\omega_m \tau}
G(\bk,i\omega_m)
\nonumber \\
G(\bk,i\omega_m) = \int_0^\beta d\tau G(\bk,\tau) e^{i\omega_m \tau}.
\label{fourier_green}
\eea
\noindent The frequencies $i\omega_m$ are known as the Matsubara
frequencies, and are given by $i\omega_m = i\pi T(2m-1)$,
$m= 0,\pm1,\pm2,...$, where $T$ is the temperature.
Because the $c$'s are Fermion operators, the Matsubara frequencies
are {\em odd} multiples of $i\pi T$. As is evident from these equations,
the imaginary time $\tau$ takes on values from 0 to $\beta$ ($\equiv
{1 \over k_B T}$).

Similar definitions hold for the phonon Green function:
\be
D(\bq,\tau-\taup) \equiv -<T_{\tau}A_\bq(\tau)A_{-\bq}(\taup)>,
\label{phonon_green}
\ee
\noindent where $A_\bq(\tau) \equiv a_\bq(\tau) \phantom{a} + \phantom{a}
a_{-\bq}^\dagger(\tau)$. The Fourier transform is similar to that given
in Eq. (\ref{fourier_green}) except that the Matsubara frequencies are
$i\nu_n \equiv i\pi T2n$, $n = 0,\pm1,\pm2,...$ i.e. they occur at {\em
even} multiples of $i\pi T$.

To derive the Eliashberg equations, we use the equation-of-motion
method, taken from Ref. \cite{rickayzen65}. The starting point is 
the (imaginary) time derivative of eq. (\ref{green_t})
\be
{\partial \over \partial \tau}G(\bk,\tau) = -\delta(\tau)
\phantom{a} - \phantom{a} <T_\tau\bigl[H- \mu N,c_{\bk\sigma}(\tau)\bigr]
c^\dagger_{\bk\sigma}(0)>,
\label{deriv1}
\ee
\noindent where, without loss of generality, we have put $\taup = 0$.
For definiteness, we use the Hamiltonian (\ref{ham_BKF_mom}), and, in addition,
assume, for the Coulomb interaction, the simple Hubbard model,
$H_{Coul} = U\sum_i n_{i\uparrow}n_{i\downarrow}$. 
The sum result is
\be
H  =  \sum_{\bk \sigma}\epsilon_\bk c^\dagger_{\bk \sigma} c_{\bk \sigma}
 + \sum_\bq \hbar \omega_\bq a^\dagger_\bq a_\bq
 + {1 \over \sqrt{N}} \sum_{\bk \bk^\prime \atop \sigma}
g(\bk,\bk^\prime) \bigl(a_{\bk - \bk^\prime} + a^\dagger_{-(\bk - \bk^\prime)}
\bigr) c^\dagger_{\bk^\prime \sigma} c_{\bk \sigma} \, \, +
\, \, {U \over N} \sum_{\bf k,k^\prime,q} c^\dagger_{\bk \uparrow}
c^\dagger_{{\bf -k + q} \downarrow} c_{{\bf -k^\prime + q} \downarrow} 
c_{{\bf k^\prime} \uparrow} ,
\label{ham_BKF_mom_hubb}
\ee               
where the various symbols have already been defined in the text.
Working out the
commutator on Eq. (\ref{deriv1}) is then straightforward. We obtain
\bea
\Biggl({\partial \over \partial \tau} + \epsilon_{\bk} \Biggr)
G_\uparrow (\bk,\tau) & = & -\delta(\tau) - {1 \over \sqrt{N}}
\sum_{\bk^\prime} g_{\bk \bk^\prime} <T_\tau A_{\bk - \bk^\prime}
(\tau) c_{\bk^\prime\uparrow}(\tau) c^\dagger_{\bk\uparrow}(0)>
\nonumber \\
& & + {U \over N}\sum_{\bf p p^\prime} <T_\tau c_{{\bf p^\prime - k + p}
\downarrow}^\dagger (\tau) c_{{\bf p^\prime}\downarrow}(\tau)
c_{{\bf p}\uparrow}(\tau) c^\dagger_{\bk\uparrow}(0)>,
\label{deriv2}
\eea
\noindent where for definiteness we are considering the Green function
with $\sigma = \uparrow$.
On the right-hand side of Eq. (\ref{deriv2}) various higher order propagators
appear; to determine them an equation of motion would have to be written,
which would, in turn, generate even higher order propagators, eventually 
leading to a set of equations with hierarchical structure.
This infinite series is normally truncated at some point by the process of
decoupling, which is simply an approximation procedure. For example,
in Eq. (\ref{deriv2}) the Coulomb term is normally not expanded further;
instead a decoupling procedure is employed. Thus, under normal
circumstances, the last term would become
\bea
<T_\tau c_{{\bf p^\prime - k + p}
\downarrow}^\dagger (\tau) c_{{\bf p^\prime}\downarrow}(\tau)
c_{{\bf p}\uparrow}(\tau) c^\dagger_{\bk\uparrow}(0)> & \rightarrow &
<T_\tau c_{{\bf p^\prime - k + p}
\downarrow}^\dagger (\tau) c_{{\bf p^\prime}\downarrow}(\tau)>
<T_\tau c_{{\bf p}\uparrow}(\tau) c^\dagger_{\bk\uparrow}(0)>,
\nonumber \\
& \rightarrow &-\delta_{\bk{\bf p}}G_\downarrow({\bf p^\prime},0)
G_\uparrow(\bk,\tau).
\label{uterm}
\eea
The case of the electron-phonon term is a little more subtle, however.
In this case we define a Green function,
\be
G_2(\bk,\bk^\prime,\tau,\tau_1) \equiv <T_\tau A_{\bk-\bk^\prime}(\tau)
c_{\bk^\prime\uparrow}(\tau_1)
c^\dagger_{\bk \uparrow}(0) >,
\label{green2}
\ee
\noindent and write out an equation of motion for it.
We get
\be
{\partial \over \partial \tau}G_2(\bk,\bk^\prime,\tau,\tau_1) =
-\omega_{\bk-\bk^\prime} <T_\tau P_{\bk-\bk^\prime}(\tau)
c_{\bk^\prime\uparrow}(\tau_1)
c^\dagger_{\bk \uparrow}(0) >,
\label{green2_deriv1}
\ee
\noindent where $P_\bq(\tau) = a_\bq(\tau) - a_{-\bq}(\tau)$. Taking
another derivative yields
\be
\Biggl[ {\partial^2 \over \partial \tau^2} - \omega_{\bk-\bk^\prime} \Biggr]
G_2(\bk,\bk^\prime,\tau,\tau_1)
= \sum_{\bk^{\prime \prime} \sigma}2\omega_{\bk-\bk^\prime}
g_{\bk-\bk^\prime}<T_\tau c^\dagger_{\bk^{\prime \prime} - \bk + \bk^\prime
\sigma}(\tau) c_{\bk^{\prime \prime}\sigma}(\tau) c_{\bk^\prime \uparrow}
(\tau_1)c^\dagger_{\bk \uparrow}(0)>.
\label{green2_deriv2}
\ee
\noindent One might be tempted to decouple Eq. (\ref{green2_deriv2})
and thus close the hierarchy that begins with Eq. (\ref{deriv2}).
However, retardation effects are properly included only when the phonon
propagator is taken into account. While the electron-phonon interaction
affects the phonons as well as the electrons, the influence on the phonons
occurs most at higher temperatures. For many materials the phonons
have reached their ground state configurations by about room
temperature. As a result, for low temperatures the phonons remain
virtually unaffected by the electron-phonon interaction, and it suffices
to disregard the electron-phonon interaction as far as the phonons
are concerned provided they have been properly renormalized due               
to effects which took place at higher temperature. To put this another
way, inelastic neutron scattering measurements of the phonon
dispersion curves show a dependence on
temperature only at temperatures well above room temperature
\cite{brockhouse62,stedman67}.

As already mentioned in the text, the phonons are normally taken
from experiment, and hence the ``calculation'' of the phonon 
propagator is greatly simplified. One simply assumes that the 
phonons are non-interacting. The equation of motion for the 
phonon propagator is then
\be
\Biggl( {\partial^2 \over \partial \tau^2} - \omega_\bq^2 \Biggr)
D(\bq,\tau - \taup) = 2\omega_\bq \delta(\tau - \taup).
\label{phonon_prop}
\ee
\noindent Utilizing this expression in eq. (\ref{green2_deriv2}) then
yields
\be
G_2(\bk,\bk^\prime,\tau,\tau) =
{1 \over N} \sum_{\bk^{\prime \prime}\sigma}\int_0^\beta d\taup
g_{\bk \bk^\prime}D(\bk - \bk^\prime, \tau - \taup)
<T_\tau c^\dagger_{\bk^{\prime \prime} - \bk + \bk^\prime
\sigma}(\taup) c_{\bk^{\prime \prime}\sigma}(\taup) c_{\bk^\prime \uparrow}
(\tau)c^\dagger_{\bk \uparrow}(0)>,
\label{green_deriv3}
\ee
\noindent where now $\tau_1$ has been set equal to $\tau$ as is required
in Eq. (\ref{deriv2}). This can now be substituted into Eq. (\ref{deriv2}),
and the whole result can be Fourier transformed (from imaginary time
to imaginary frequency). Before stating the result  of this exercise,
however, we note that the superconducting state is specially characterized
by the existence of anomalous amplitudes, attributed to
Gorkov \cite{gorkov58} and often referred to as Gorkov amplitudes. Thus,
in the Wick decomposition \cite{mahan81} of the various two-particle Green
functions, the anomalous amplitudes also must be taken into account, in
addition to the normal amplitudes given, for example, in Eq. (\ref{uterm}).

The anomalous amplitudes take the form
\be
F(\bk,\tau) \equiv -<T_\tau c_{\bk \uparrow}(\tau)c_{-\bk \downarrow}(0)>
\label{f_amplitude}
\ee
\noindent and
\be
\bar{F}(\bk,\tau) \equiv -<T_\tau c^\dagger_{-\bk \downarrow}(\tau)      
c^\dagger_{\bk \uparrow}(0)>.
\label{fbar_amplitude}
\ee
Now it is necessary to go through the same procedure with $F$ and $\bar{F}$
as with $G$. The methodology is the same, so we skip the necessary steps.
\par

We then define two self-energies, the usual one (generalized to the
superconducting state), denoted by $\Sigma(\kiwm)$, and an anomalous
self-energy, often called the pairing function, $\phi(\kiwm)$, and
we arrive at Eqs.(\ref{g1}-\ref{g5}).

\subsection{The Polaron Problem}

A rather different and less developed approach to the electron phonon
problem focuses on the effect of the phonons on a {\em single} electron.
A review is provided in Ref. \cite{alexandrov95}, and we merely
highlight some of the important points here. 

There are many kinds of polarons, i.e. {\em small} vs. {\em large},
{\em weakly coupled} vs. {\em strongly coupled}, {\em Fr\"ohlich}
vs. {\em Holstein}, etc. As far as we can tell these classifications
are merely qualitative, so that, in most cases, distinctions can
be readily drawn for extreme parameters only. A case in point is the
distinction between an itinerant vs. self-trapped polaron. It seems
clear that no such transition exists, but nonetheless a crossover
occurs to a regime in which the polaron acquires a very large effective
mass. 

In thinking about the polaron problem, there is the usual
competition between kinetic energy (measured by the hopping integral,
$t$, or the bandwidth, $D = 2zt$, where $z$ is the coordination number for
a cubic lattice ($z = 2,4$ and $6$ in 1,2, and 3 dimensions, respectively))
and the potential energy (measured by $g$ --- see Hamiltonian 
(\ref{ham_Hol_mom})). In addition the phonon frequency represents
a third energy scale. In the case of the Holstein model, 
Eq. (\ref{ham_Hol_mom}), this scale is conveniently represented by
a single number, the Einstein oscillator frequency, $\omega_E$. A
dimensionless coupling constant, $\lambda \equiv 2g^2/(D \omega_E)$,
corresponds roughly to the enhancement parameter
introduced in section 2.3 (see Eq. (\ref{lambda_ein})). Note that in
terms of the parameters of the original Holstein Hamiltonian, 
Eq. (\ref{ham_Hol}), $\lambda \equiv \alpha^2/(KD)$. An increase in
$\lambda$ signifies an approach to the strong coupling limit. On
the other hand the adiabatic (anti-adiabatic) limit is represented 
by $\omega_E/t \rightarrow 0 (\infty)$. The values of both ratios
strongly influence the number of phonons present. An early
review that clearly delineates these different regimes is provided 
by Ref. \cite{feinberg90}.

There have been many approaches to solving the polaron problem (as governed by
a Hamiltonian like Eq. (\ref{ham_Hol})). Some of the early techniques are
amply covered in Ref. \cite{kuper63}; these are exemplified by weak
and strong coupling perturbation theory, and variational methods.
A review of the perturbation approaches is given in Appendix B and C
of Ref. \cite{marsiglio95a}. Weak coupling follows the Migdal approach,
while strong coupling utilizes the celebrated Lang-Firsov transformation.

This transformation immediately results in a narrow band, with effective
hopping parameter, $t_{\rm eff} = t\exp{(-g^2/\omega^2)}$, along with
exponential increases in effective mass and, in the adiabatic regime,
number of phonons in the ground state \cite{feinberg90,alexandrov92}.

With the advent of considerable computing capabilities over the last
two decades, exact methods have been used, that, in various cases, can
span the entire parameter regime. The first is Monte Carlo for a single
electron, pioneered in Ref. \cite{deraedt82}. Trugman {\it et al.} 
\cite{trugman88,bonca95,bonca00} utilized exact diagonalizations based
on a variational Hilbert space obtained from repeated applications of
the Hamiltonian on a trial state vector; their most recent results
are capable of achieving very high precision. In the meantime, Proetto
and Falicov \cite{proetto89} and 
Ranninger and Thibblin \cite{ranninger92} used a truncated Hilbert
space for a two-site problem, and performed a straightforward
numerical diagonalization. This was followed by work on larger
(one-dimensional) lattices (for one electron) in 
Refs. \cite{marsiglio93a,alexandrov94,marsiglio95a}. Most of this work
was performed for a specific model --- the Holstein model of electron phonon
coupling, already referred to in the text. Further work was carried
out also for the BLF (SSH) model, in Ref. \cite{capone97}. Yet
another technique utilizes the density-matrix renormalization group
(DMRG) \cite{white92} method \cite{jeckelmann98}, which has also been
extended to many electrons \cite{jeckelmann99}. Another variational technique
known as the Global-Local variational method \cite{romero99} also
provides very accurate results for the polaron problem. Finally, two
new Monte Carlo methods \cite{prokofev98,kornilovitch98} appear to
be particularly powerful in obtaining polaron properties.

In all
cases a clearer understanding is emerging; there is no self-trapping
transition, in any dimension, although there is a
farily abrupt (but still smooth) crossover from weak coupling-like
to strong coupling-like. This crossover has now been investigated
in 1,2, and 3 dimensions as well as with dynamical mean-field
theory, which is exact in infinite dimensions \cite{ciuchi97}.

An actual transition can be observed in higher dimensions in the
adiabatic limit ($\omega_E = 0$) \cite{kabanov93}. However, this limit
is regarded as somewhat pathological, and {\em not representative}
of the general case \cite{romero99}.

Finally, some work has been performed on the {\em bipolaron}
problem, i.e. whether two polarons bind or not. Much of this work
is summarized in Ref. \cite{alexandrov95}. Various discussions
of the supporting evidence and difficulties of these theories can
be found in Refs. \cite{losalamos90,chakraverty98,alexandrov99}.  
A related problem has been asked and partially answered in Refs.
\cite{marsiglio95a,freericks95,bonca00}, which is: to what extent will
two electrons interacting through phonon exchange {and Coulomb repulsion}
form a Cooper pair, particularly as the stength of the Coulomb repulsion
is increased well in excess of the effective strength of the attractive
phonon-induced interaction ? In other words, to what degree does the
pseudopotential effect play a role in pairing ?

In Ref. \cite{marsiglio95a} one of us found that pairing persists even
when the Coulomb interaction strength exceeds that of the electron phonon
attraction, and in Ref. \cite{bonca00} this statement was made more
precise (see also Ref. \cite{proville98}). In particular, binding persists
only up to a point; for sufficiently large Coulomb repulsion, the pair is
no longer bound. While more work is required, this finding implies that
the usual pseudopotential reduction, given by Eq. (\ref{pseudo}),
may be too strong. Eq. (\ref{pseudo}), for example, achieves a large
reduction in the limit $\mu(E_F) \rightarrow \infty$, whereas the
result of Ref. \cite{bonca00} says that for two electrons, at least, the
binding is lost in this limit. 

\subsection{Many Electrons on a Lattice}

The problem of many interacting electrons is, in many ways, significantly
more difficult than that of one or two electrons. The dimension of the
Hilbert space grows exponentially, so that exact diagonalizations
become prohibitive. A review of methods and results can be found in Ref.
\cite{dagotto94}. As far as the electron phonon problem is concerned,
there is some limited work which utilizes direct diagonalization,
usually in the context of the t-J model \cite{dobry95,wellein96}. Mainly, 
however, this problem has been approached through Monte Carlo methods,
and a variety of (somewhat uncontrolled) Green function techniques.
 
Monte Carlo methods have an illustrious history \cite{metropolis53}.
While they are not formally exact (because, for example, of a Trotter
\cite{trotter59} breakup), the error introduced by such a decomposition
can be {\em controlled}. Hence, in principle, and even in practice
through extrapolations, one can obtain results which are exact to within
some known error. 

Some of the first papers to utilize Monte Carlo methods in many body
fermion problems (in the condensed matter context) addressed the
electron phonon problem \cite{scalapino81,blankenbecler81,scalapino81b}.
This particular methodology integrated out the fermion degrees of
freedom analytically, leaving the boson degrees of freedom to which
Monte Carlo algorithms were applied. Various modifications immediately
arose, and were used to address the same electron phonon problem
\cite{hirsch81,hirsch82,fradkin83,hirsch83} as well as electron-electron
problems \cite{hirsch85b}. Much of this work is reviewed in Ref.
\cite{scalapino88}; a more comprehensive review of the many variants
of the Monte Carlo method (in condensed matter) is provided in Ref. 
\cite{vonderlinden92}.

Studies in two dimensions became more feasible in the late 1980's;
an immediate question that was addressed was the competition
between superconductivity and the charge density wave (CDW) instability
\cite{scalettar89,marsiglio89b,marsiglio90}. At half-filling
(where simulations are easiest) the CDW instability overwhelms the
tendency towards superconductivity, in part because the tight-binding
model with nearest neighbour hopping exhibits nesting at half-filling.
Veki\'c {\it et al.} \cite{vekic92} explored the 
impact of next nearest neighbour hopping
(to remove the nesting) but found it was difficult to discern whether
an incommensurate CDW instability or superconductivity dominates.

Another means of eliminating the CDW is through doping; again most
of the work is inconclusive. A third means is through the use of
a Hubbard $U$. A study \cite{berger95} of the
so-called Hubbard-Holstein model in two dimensions found that both the CDW
and the superconductivity susceptibilities are suppressed as $U$ grows.
To our knowledge, however, the pseudopotential effect (where the $U$
would essentially cancel the electron phonon interaction as far as the
CDW was concerned, but not as far as superconductivity was concerned) has
never been detected in many-electron Monte Carlo studies.

One of the reasons for exact studies of these lattice models 
(on small lattices) is for use as a benchmark to which diagrammatic
methods can be compared. Thus, for example, the conclusion in Ref.
\cite{marsiglio90} was that the Migdal formalism, without vertex
corrections, described the Monte Carlo results fairly accurately,
{\em provided phonon renormalization was taken into account}. A model
system is required to determine this, since, in real systems, the phonons
are often taken from experiment, and already contain renormalization 
effects. This conclusion was confirmed in Ref. \cite{berger95}, as well
as in Ref. \cite{niyaz93}. In this latter reference, the authors developed
the formalism even further to accommodate a CDW gap, and found
good agreement with Monte Carlo results. 

Nonetheless, it is probably safe to say that a reliable formalism
has not yet been developed to investigate low temperature properties
of electron phonon systems, particularly slightly away from half-filling.
A number of attempts have been made, particularly in the case of
electron-electron interactions \cite{bickers89,bickers91}, although
a comprehensive treatment has not yet been achieved (the many-body
approaches are also becoming almost as numerically intensive as the 
Monte Carlo methods, and so one of their advantages is diminishing).

Attempts have also been made to incorporate specific kinds of corrections
to the Migdal-Eliashberg formalism. One of these categories is the
inclusion of vertex corrections. Many feel that they may be necessary
because the adiabatic ratio $\omega_D/E_F$ is not small in
some cases (eg. high $T_c$ cuprates, and doped buckyballs). In the
cuprate materials two-dimensional effects may enhance vertex corrections 
as well.  Calculations showing an enhancement of $T_c$ due to vertex 
corrections have been reported in Ref. \cite{kostur93} (for a two-dimensional
gas). In Ref. \cite{freericks98}, a different tack is taken; $T_c$
is kept fixed, and calculations with vertex corrections included can
mimic those without through an adjusted $\mu^\ast$ (except for 
the isotope effect). In Ref. \cite{perali98} a two-dimensional tight-binding
model is used and once again, the conclusion is that vertex corrections
enhance the pairing interaction. To our knowledge, however, these effects
have never been observed in exact or controlled calculations.

\begin{figure}
\includegraphics[width=0.7\textwidth]{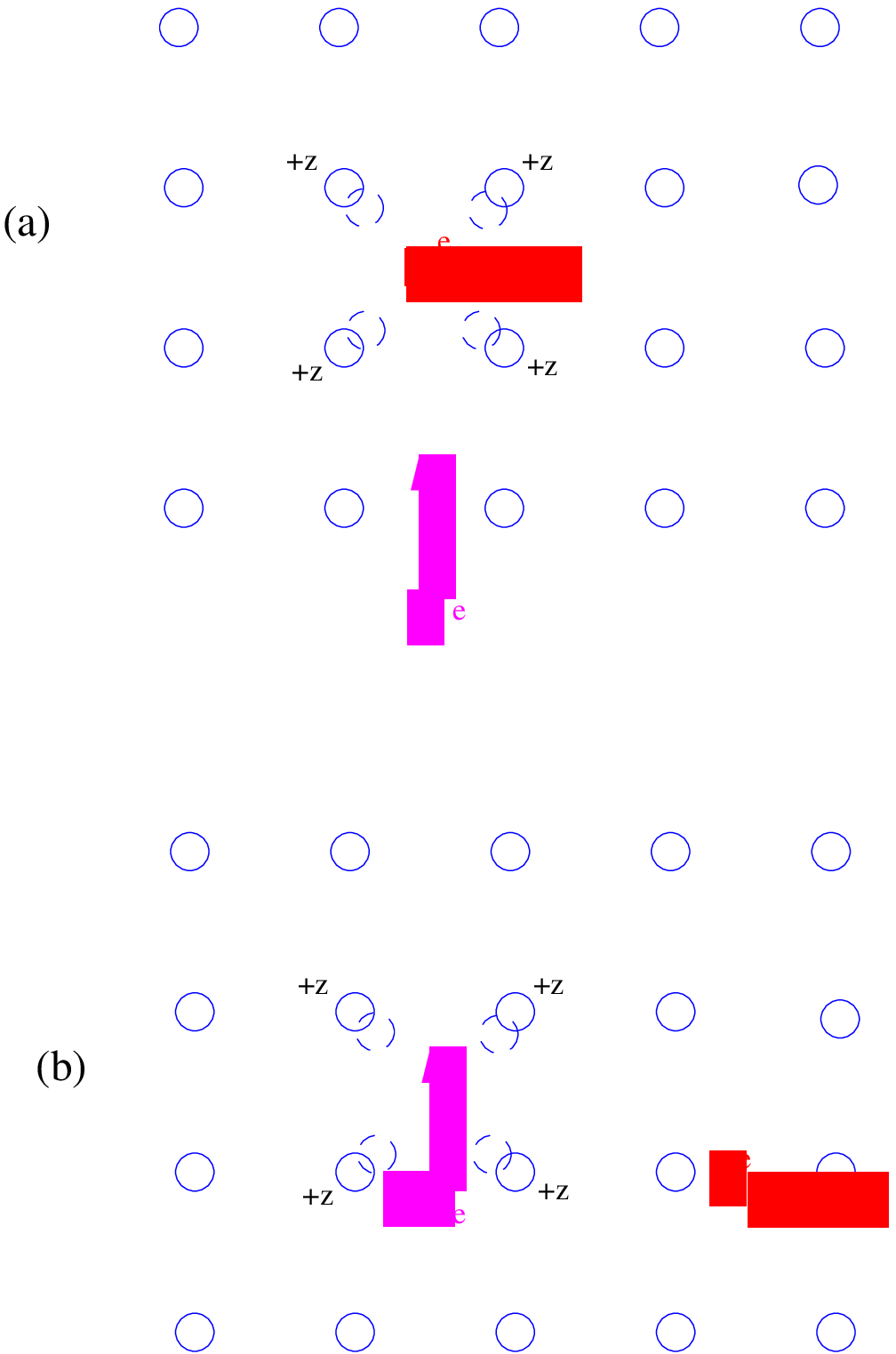}
\caption[]{In (a) one electron polarizes the lattice (indicated
by dashed circles displaced towards uppermost electron); in (b) that
electron has moved away. In the meantime a second electron (seen below in (a))
is attracted to the polarized region, {\em which has remained polarized
long after the first electron has left the region}. Figure is schematic
only, and does not, for example, properly convey the opposite momenta
such a pair should possess.}
\label{ff1}
\end{figure}  

\begin{figure}
\includegraphics[width=0.7\textwidth]{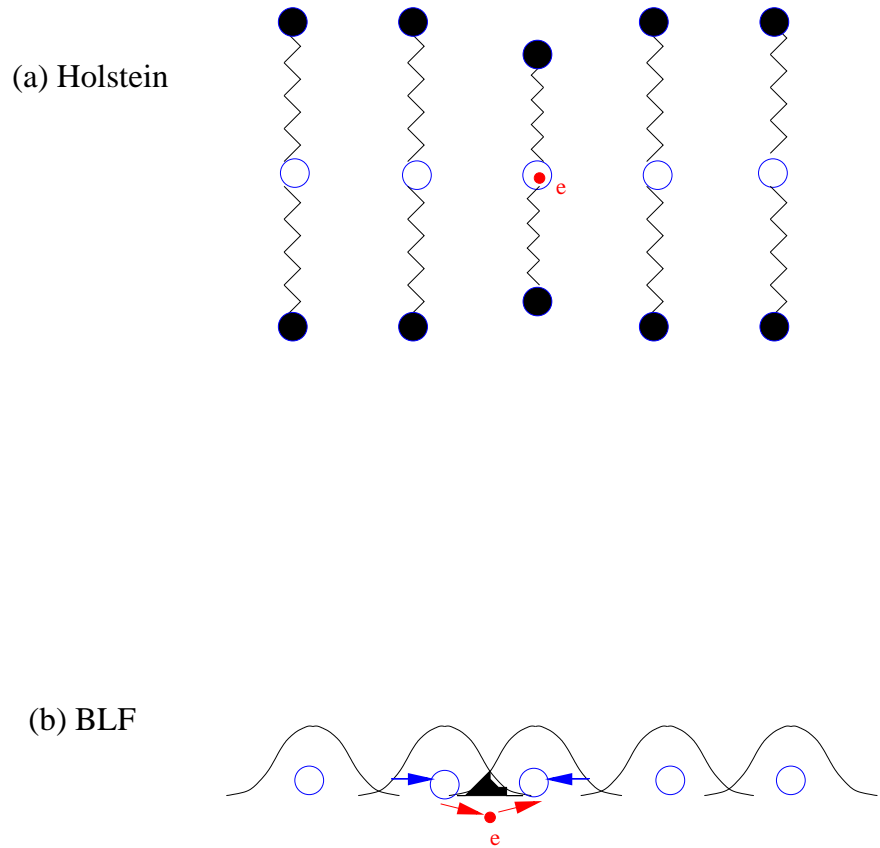}
\caption[]{Schematic of ionic displacements in (a) the Holstein model,
and (b) the BLF model. In (a) neighbouring chains are distorted in
the vicinity of the electron, and in (b) neighbouring ions, when displaced
while undergoing oscillations, lead to an increased (or decreased) overlap
region (shaded in black), which leads to an altered hopping amplitude for
the electron.}
\label{ff2}
\end{figure}  

\begin{figure}
\includegraphics[width=0.7\textwidth]{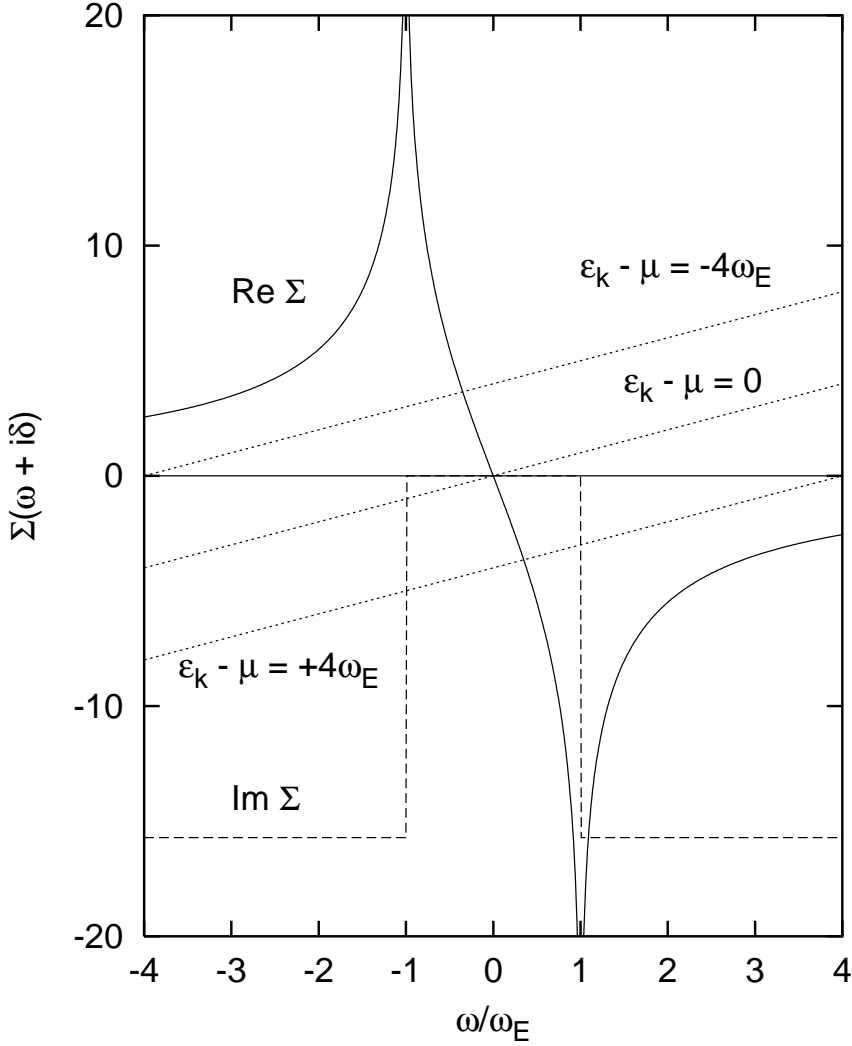}
\caption[]{Real and Imaginary parts of the electron self energy
in the normal state, for an Einstein spectrum ($\lambda = 1$).
The dotted lines are the inverse non-interacting electron Green
functions, $\omega - (\epsilon_{\bf k} - \mu)$, 
for $(\epsilon_{\bf k} - \mu)/\omega_E = -4, 0$, and $4$, from
top to bottom, respectively.}
\label{ff3}
\end{figure}

\begin{figure}
\includegraphics[width=0.85\textwidth]{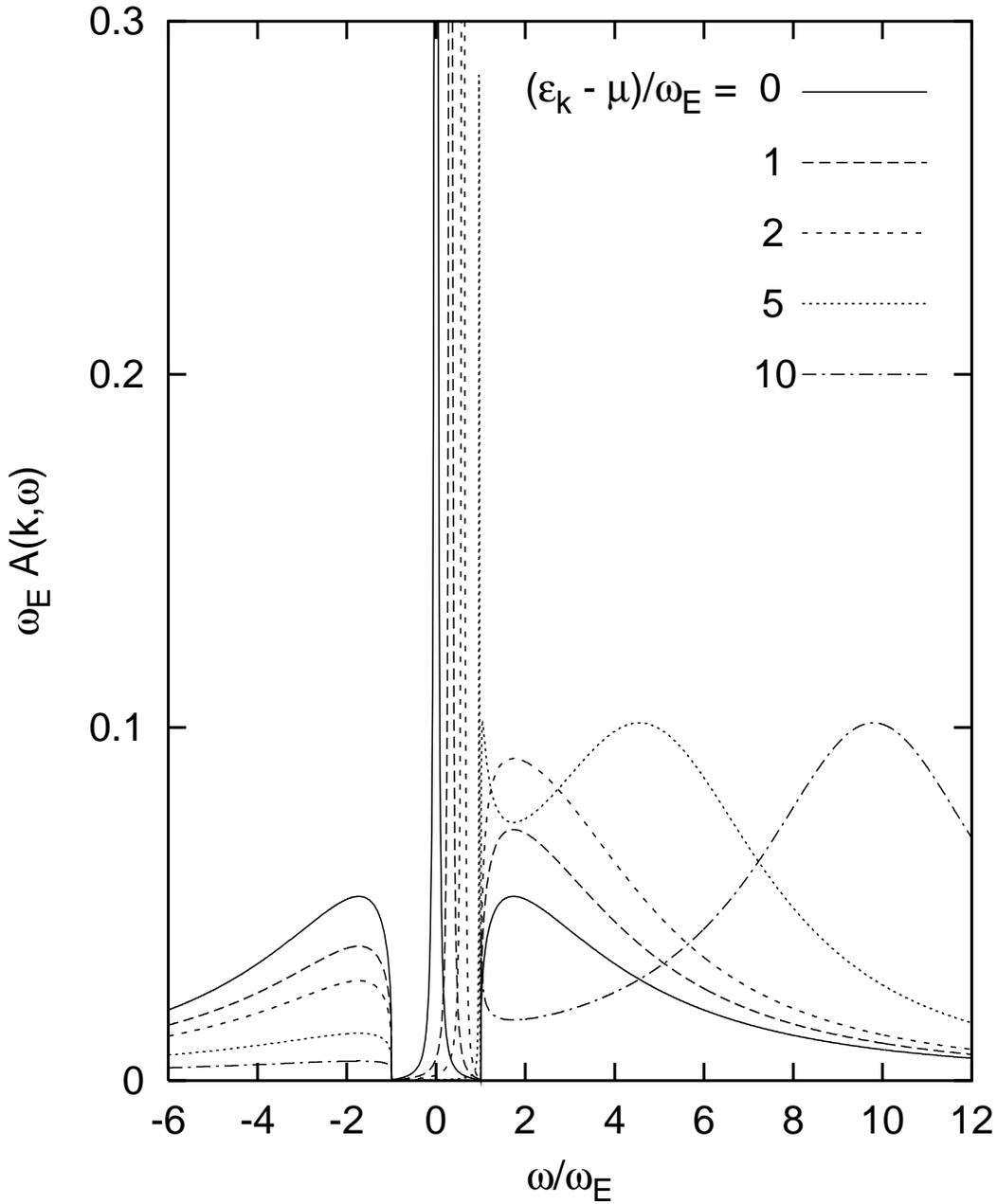}
\caption[]{The spectral function for an electron interacting with phonons (Einstein spectrum
with $\lambda = 1$) for various momenta as labelled. Note that for each
momentum there is a delta function contribution (artificially broadened in this figure)
whose weight diminishes as one moves away from the chemical potential and whose frequency approaches
the Einstein phonon frequency. The incoherent component grows with increasing $\epsilon_{\bf k} - \mu$,
and approaches a reasonably well-defined peak centered around $\epsilon_{\bf k} - \mu$ for large
values (eg. dot-dashed curve).}
\label{ff4}
\end{figure}  

\begin{figure}
\includegraphics[width=.7\textwidth]{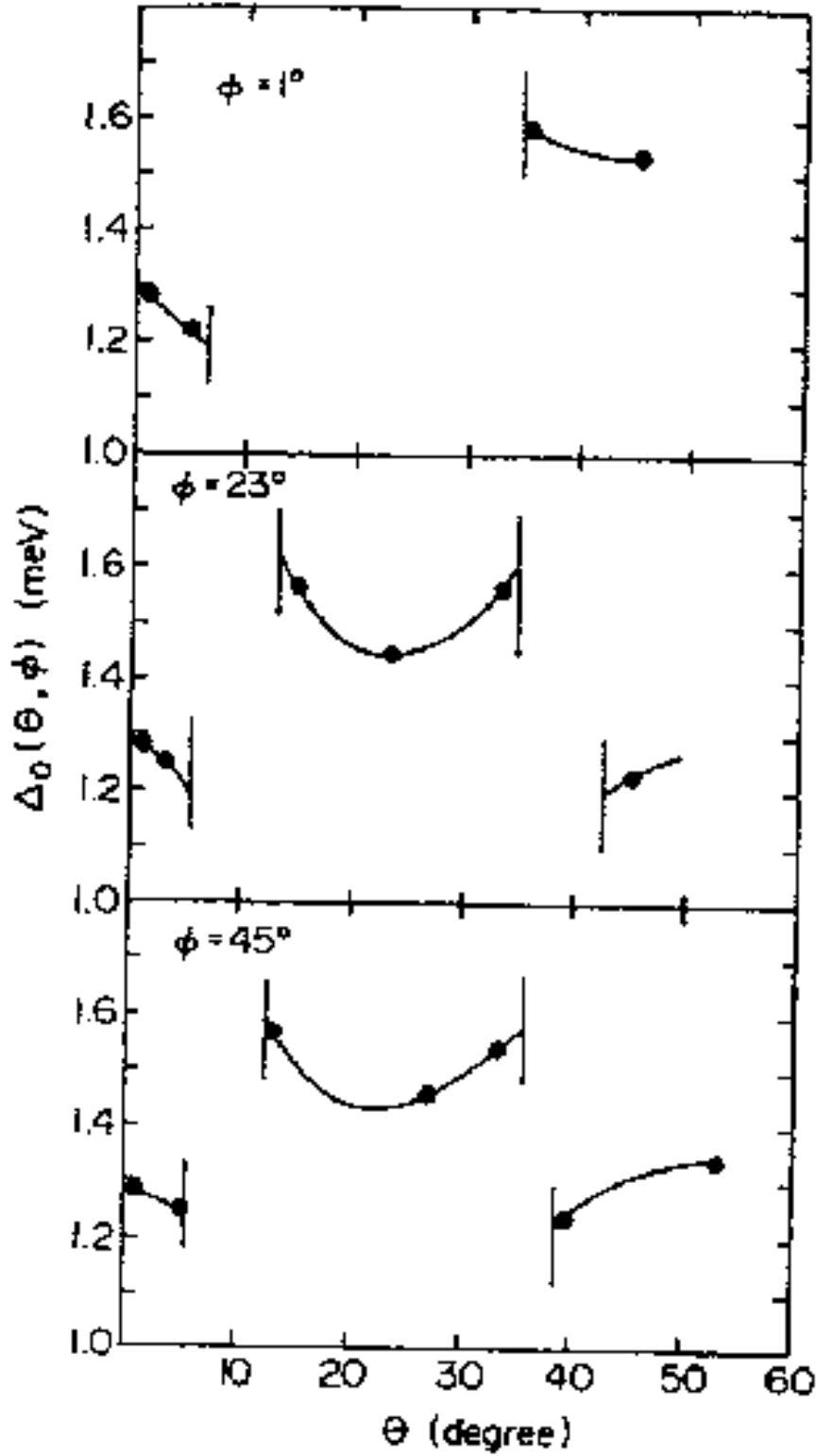}
\caption[]{Gap anisotropy for Pb as a function of angle, $\theta$, for three different values of
azimuthal angle, $\phi$. Regions where the Fermi surface of Pb does not exist are indicated
by vertical lines. Figure reproduced from Ref. \protect\cite{tomlinson76}.}
\label{ff5}
\end{figure}  

\begin{figure}
\includegraphics[width=.9\textwidth]{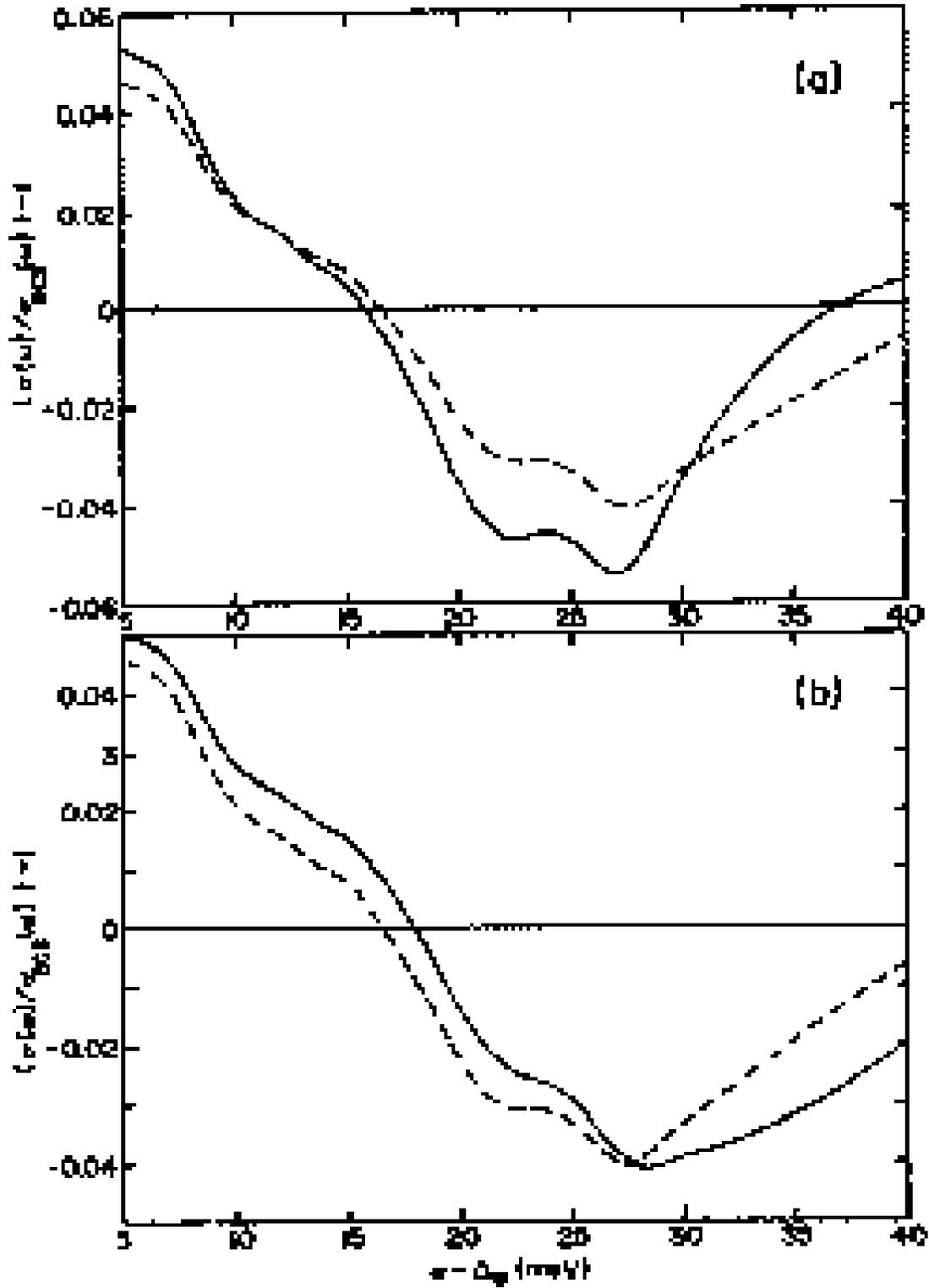}
\caption[]{Normalized tunneling conductance reduced to the corresponding BCS expression,
for (a) a peak, and (b) a valley in the electron density of states (solid curves). The dashed curves
were obtained with a constant density of states. Reproduced from Ref. \protect\cite{mitrovic83}.}
\label{ff6}
\end{figure}  

\begin{figure}
\includegraphics[width=0.8\textwidth]{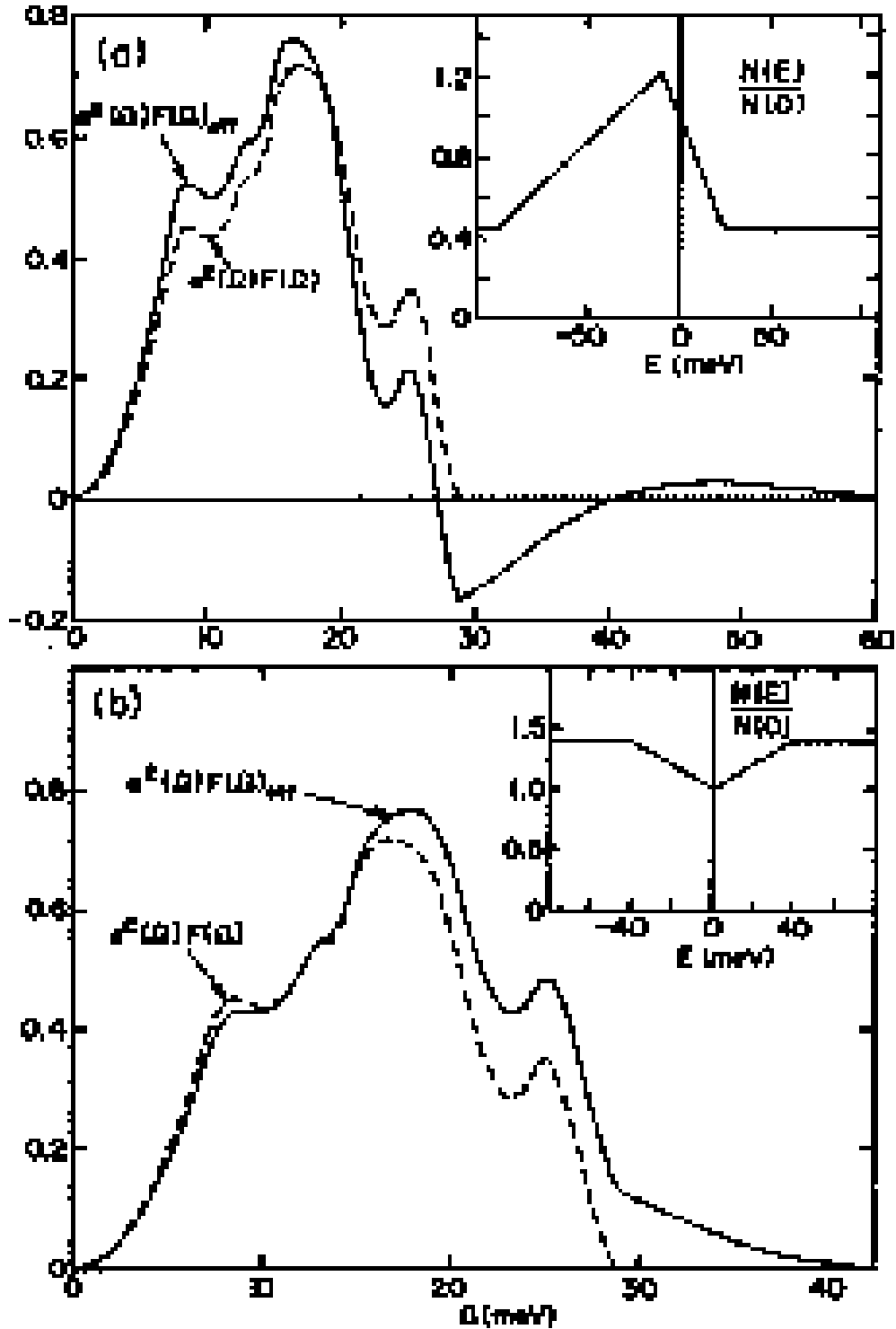}
\caption[]{Effective electron phonon spectral functions obtained by the inversion of the
calculated normalized tunneling conductances within the usual (i.e. constant electron density
of states) Eliashberg theory (solid curves). The input spectral functions are shown with the
dashed curves. The insets contain the corresponding electron densities of states used.
Reproduced from Ref. \protect\cite{mitrovic83}.}
\label{ff7}
\end{figure}  

\begin{figure}
\includegraphics[width=.6\textwidth]{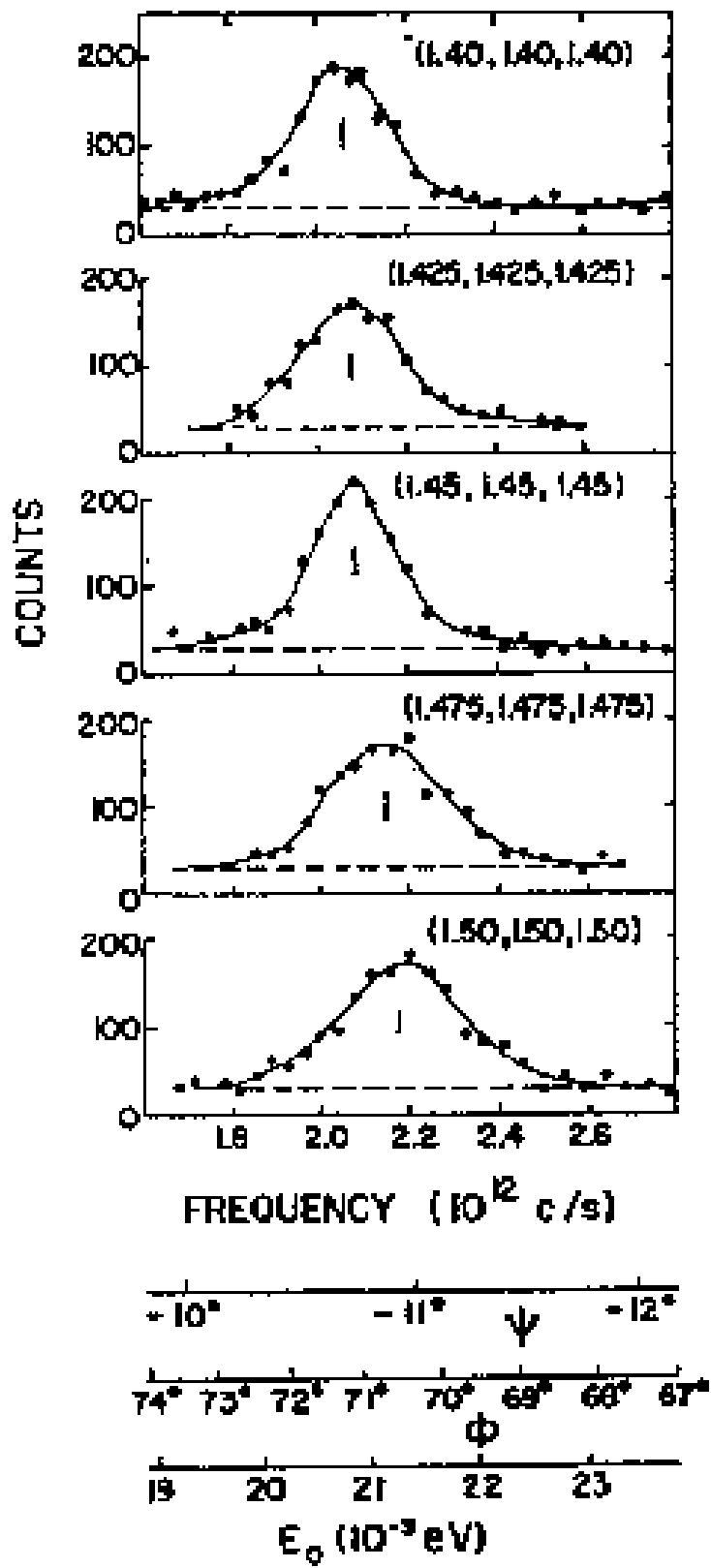}
\caption[]{A set of `constant ${\bf q}$' scans in Pb taken at various points along the
diagonal in the Brillouin zone. Reproduced from Ref. \protect\cite{brockhouse62}.}
\label{ff8}
\end{figure}   

\begin{figure}
\includegraphics[width=.9\textwidth]{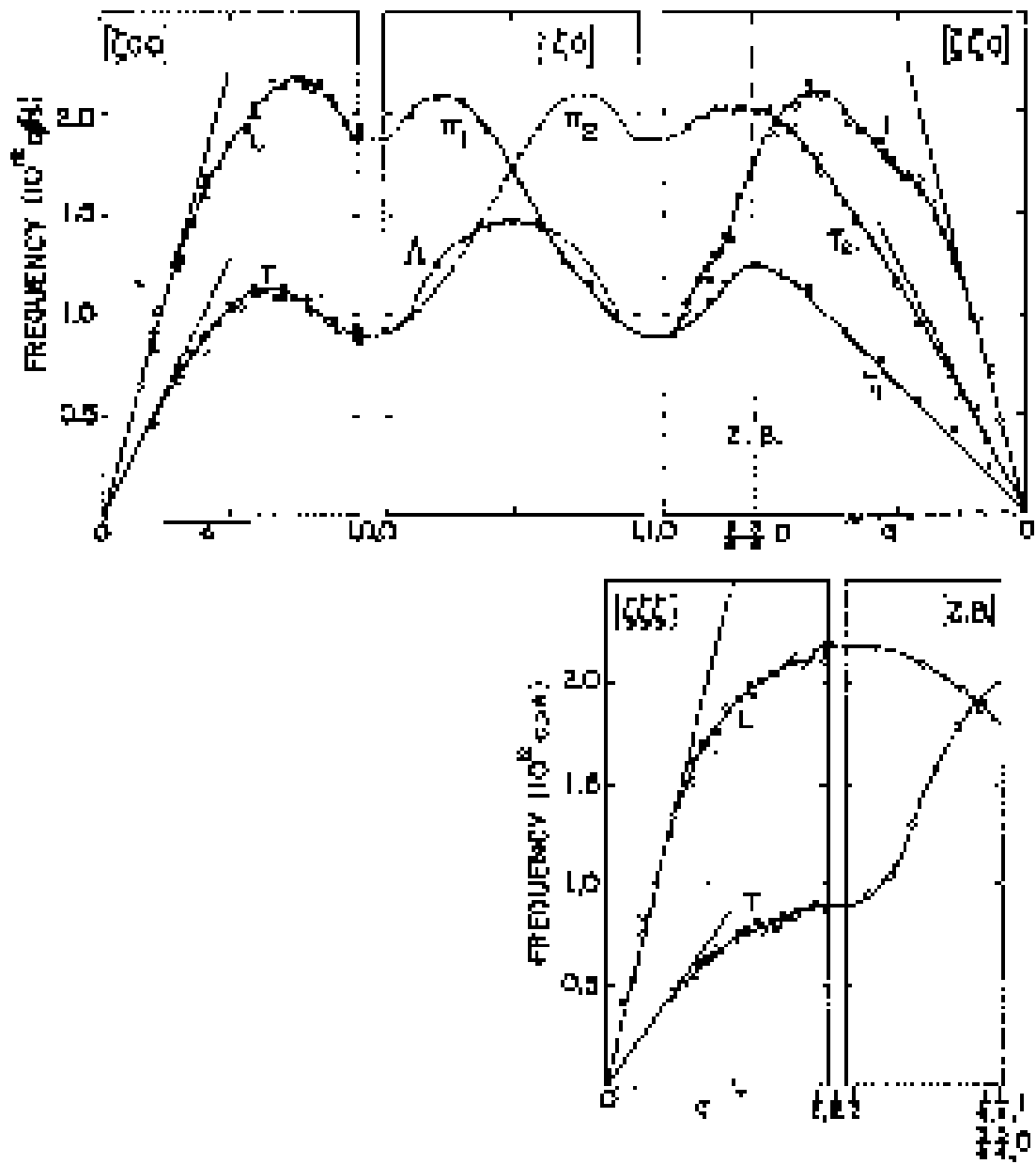}
\caption[]{The dispersion curves for Pb at 100 K, as a function of momentum along various
high symmetry directions. Reproduced from Ref. \protect\cite{brockhouse62}.}
\label{ff9}
\end{figure}   

\begin{figure}
\includegraphics[width=.68\textwidth]{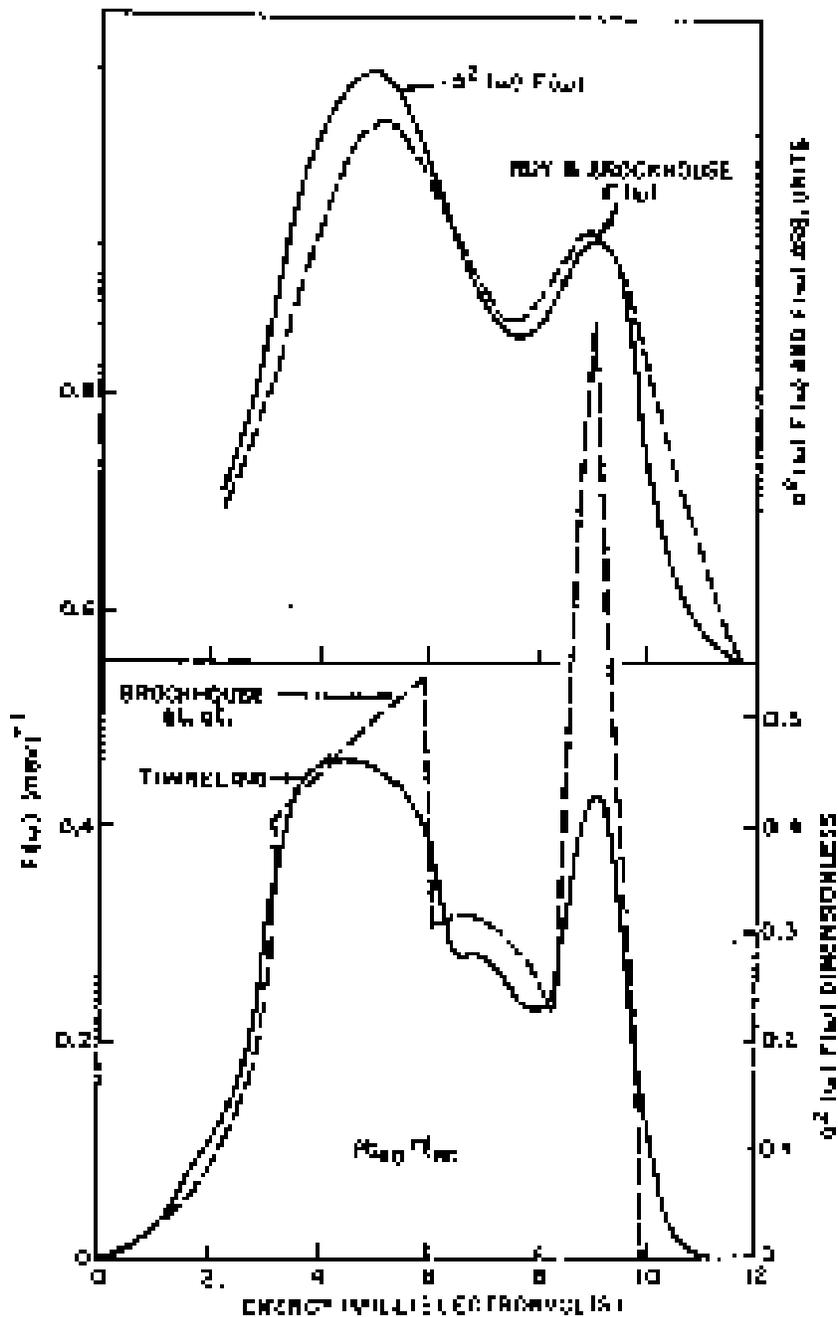}
\caption[]{The electron-phonon spectral function $\alpha ^2F(\omega )$ (solid curve) for
$Pb_{.40}T\ell _{.60}$ determined from tunneling experiments and convoluted by
instrument resolution of the neutron spectrometer compared with the neutron
results for the phonon frequency distribution $F(\omega )$ (dashed curve)
measured by incoherent inelastic neutron
scattering \protect\cite{roy70} (upper frame).  The lower frame shows the tunneling results (solid
curve) compared with the phonon frequency distribution (dashed curve)
determined from a Born von Karman analysis of the phonon dispersion curves in
$Pb_{.40}T\ell _{.60}$ \protect\cite{brockhouse68}.   }
\label{ff10}
\end{figure}
                
\begin{figure}
\includegraphics[width=.9\textwidth]{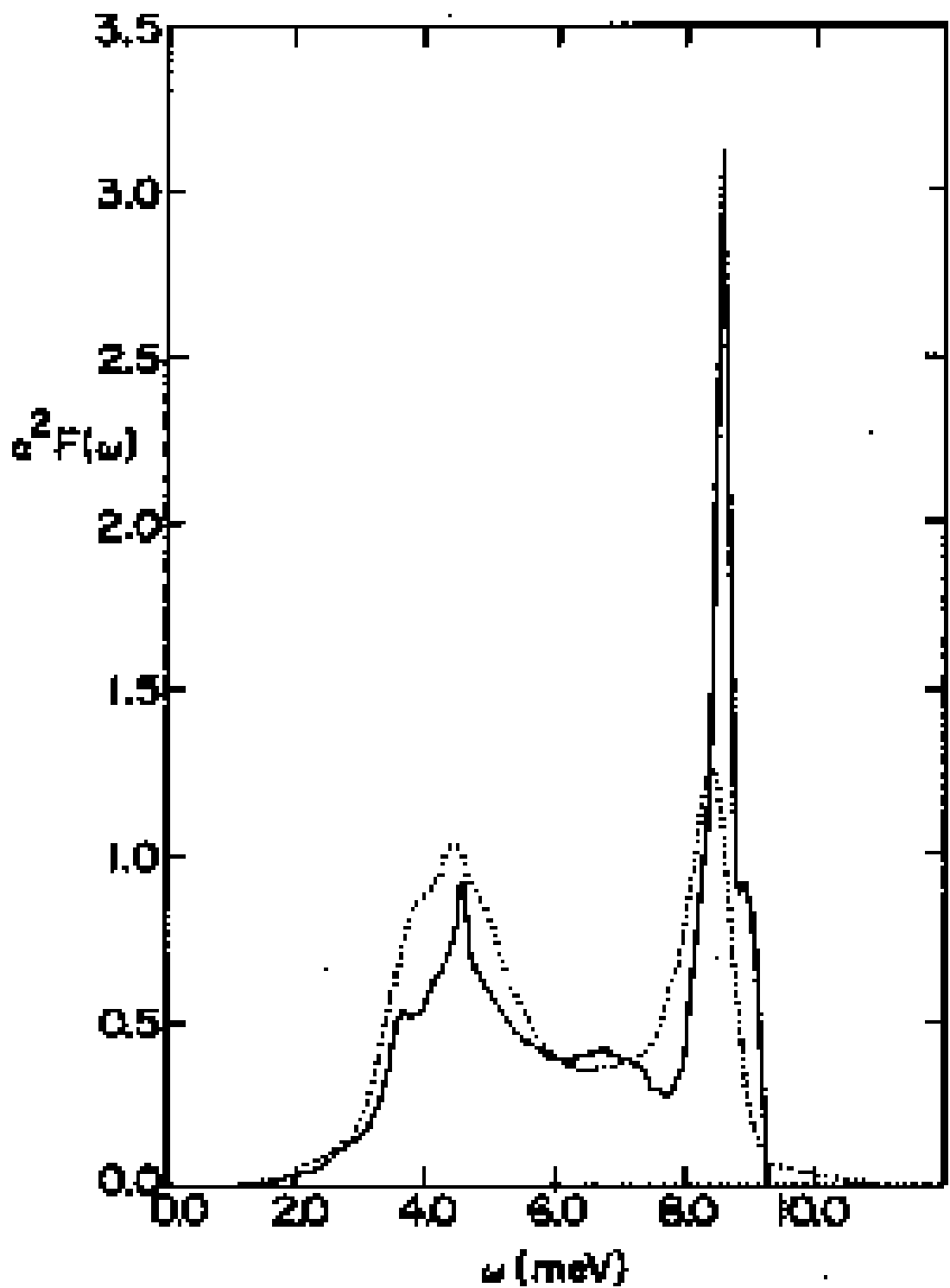}
\caption[]{The electron phonon spectral function $\alpha ^2F(\omega )$ measured in
tunneling experiments (dotted curve) compared with that which is calculated
from first principles (solid curve) \protect\cite{tomlinson77}. }
\label{ff11}
\end{figure}
                
\begin{figure}
\includegraphics[width=.9\textwidth]{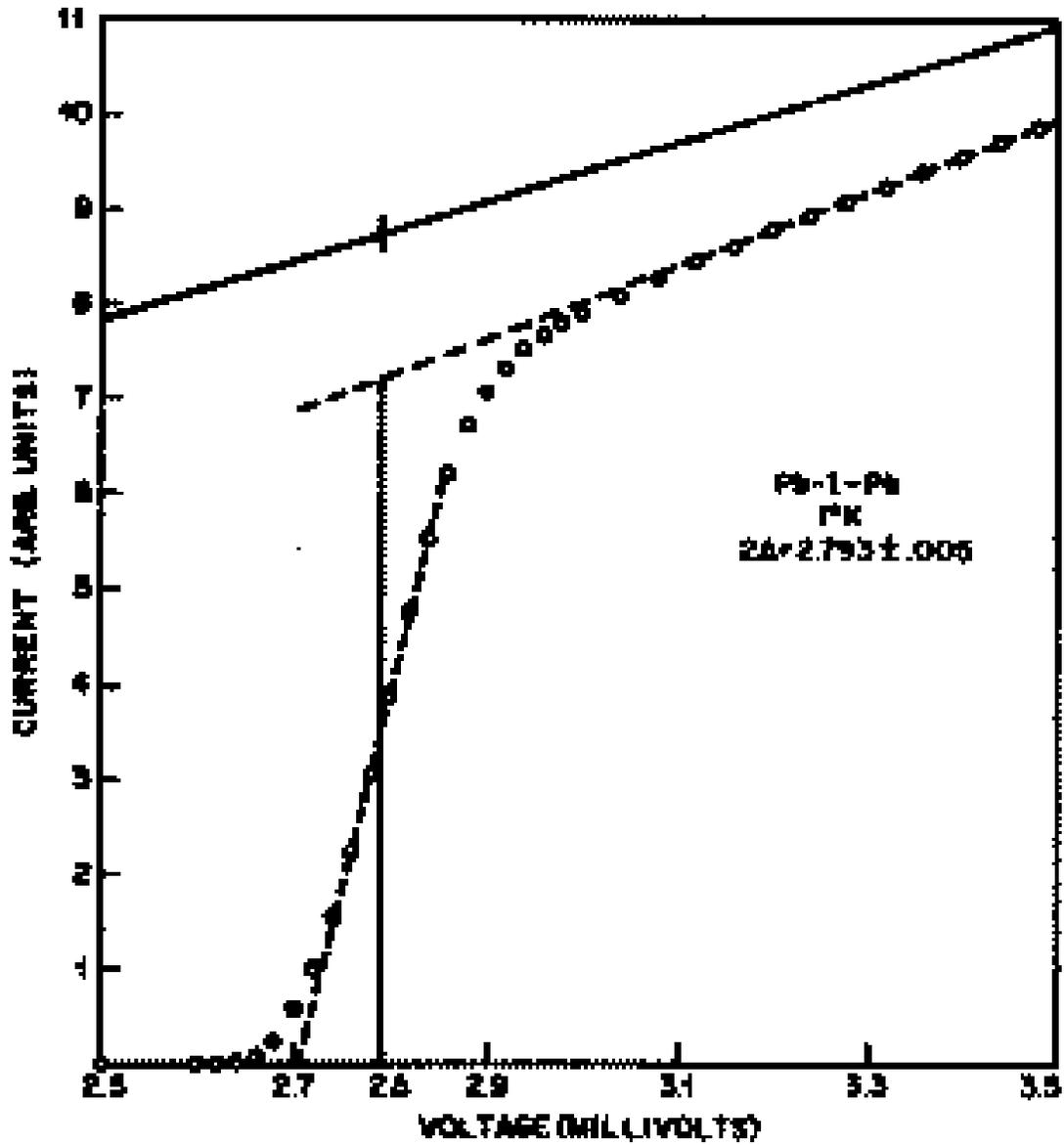}
\caption[]{I-V characteristic of a Pb-I-Pb junction showing the construction used
to find the energy gap. The solid line and open circles are the current in the normal
and superconducting states, respectively. Reproduced from Ref. \protect\cite{mcmillan69}.}
\label{ff12}
\end{figure}
                
\begin{figure}
\includegraphics[width=.9\textwidth]{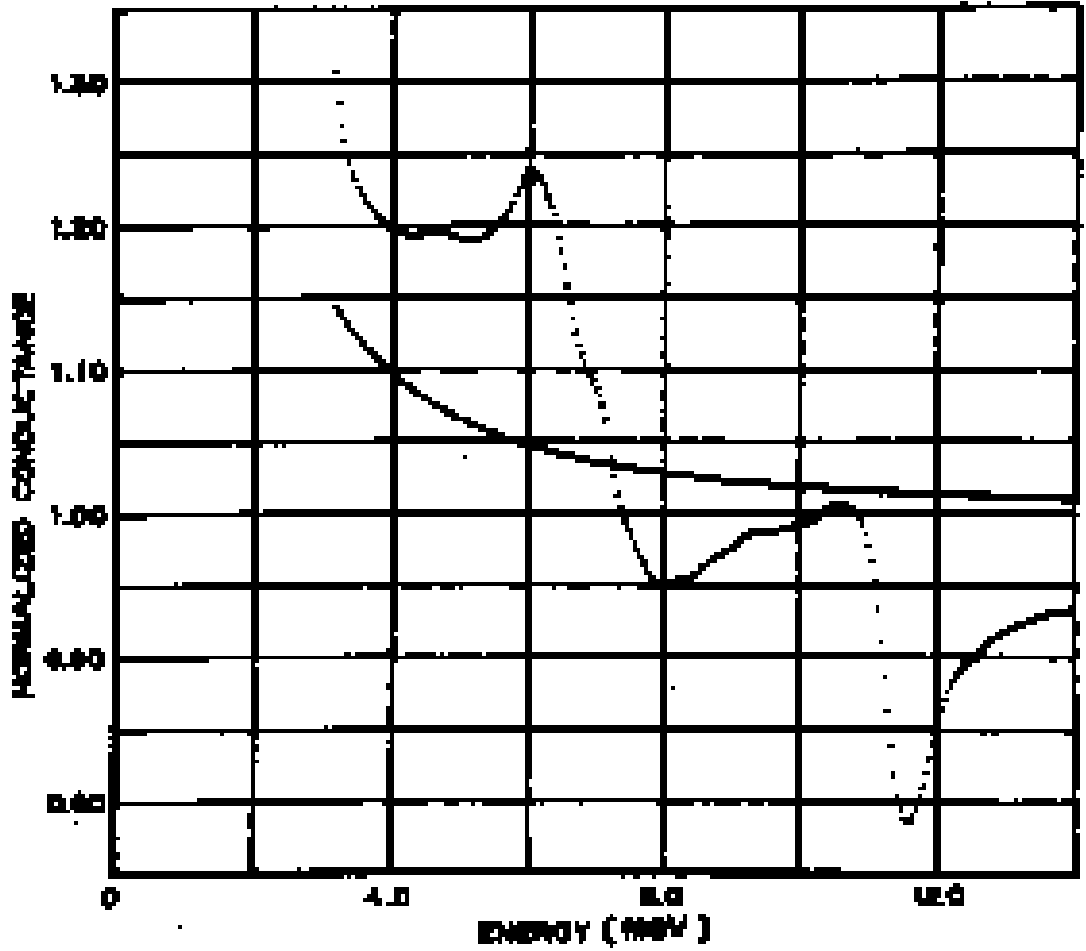}
\caption[]{Conductance $dI/dV$ of a Pb-I-Pb junction in the superconducting state normalized
by the conductance in the normal state vs. voltage. Also shown is the two-superconductor
conductance calculated from the BCS density of states which contains no phonon structure.
Reproduced from Ref. \protect\cite{mcmillan69}.}
\label{ff13}
\end{figure}
                
\begin{figure}
\includegraphics[width=.9\textwidth]{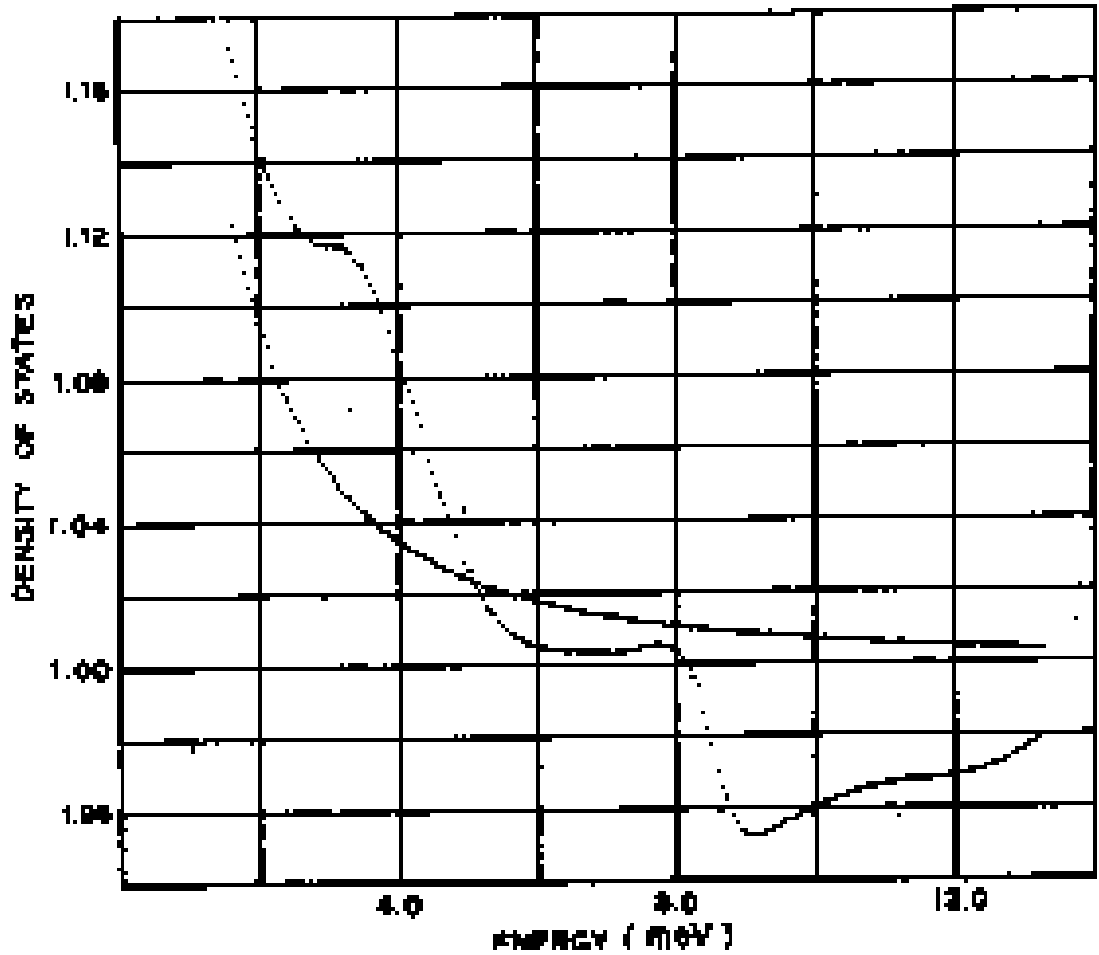}
\caption[]{Electronic density of states $N(E)$ vs. $E- \Delta_\circ$
for Pb, obtained from the data of Fig. 13. The smooth curve is the BCS
density of states.
Reproduced from Ref. \protect\cite{mcmillan69}.}
\label{ff14}
\end{figure}
                
\begin{figure}
\includegraphics[width=.9\textwidth]{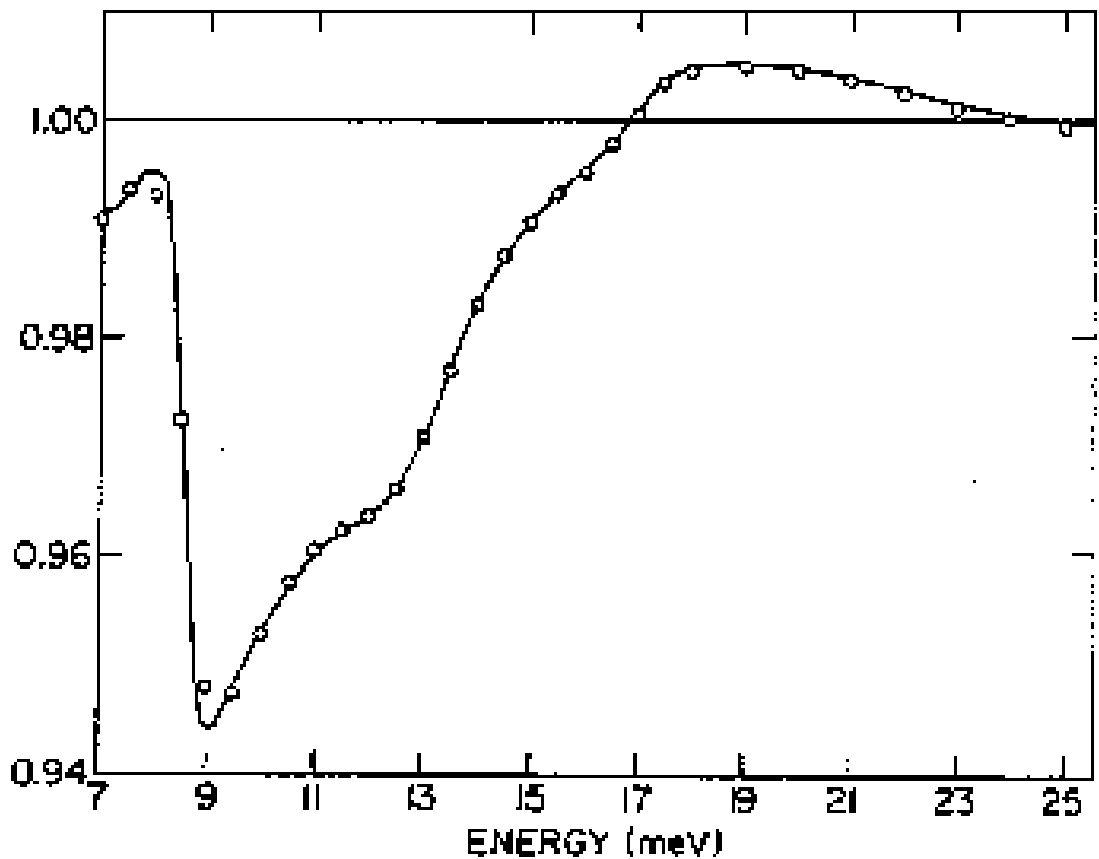}
\caption[]{The predicted (solid curve) normalized density of states in Pb as a function of
energy $\omega $ compared with measured values (open dots) as a function of
energy measured from the gap edge.  The measured density of states divided by
the BCS density of states above 11 meV was not used in the fitting procedure
that produced $\alpha ^2F(\omega )$ and a comparison of theory and experiment
in the multiple-phonon region is a valid test of the theory.
Reproduced from Ref. \protect\cite{mcmillan69}.}
\label{ff15}
\end{figure}
                
\begin{figure}
\includegraphics[width=.9\textwidth]{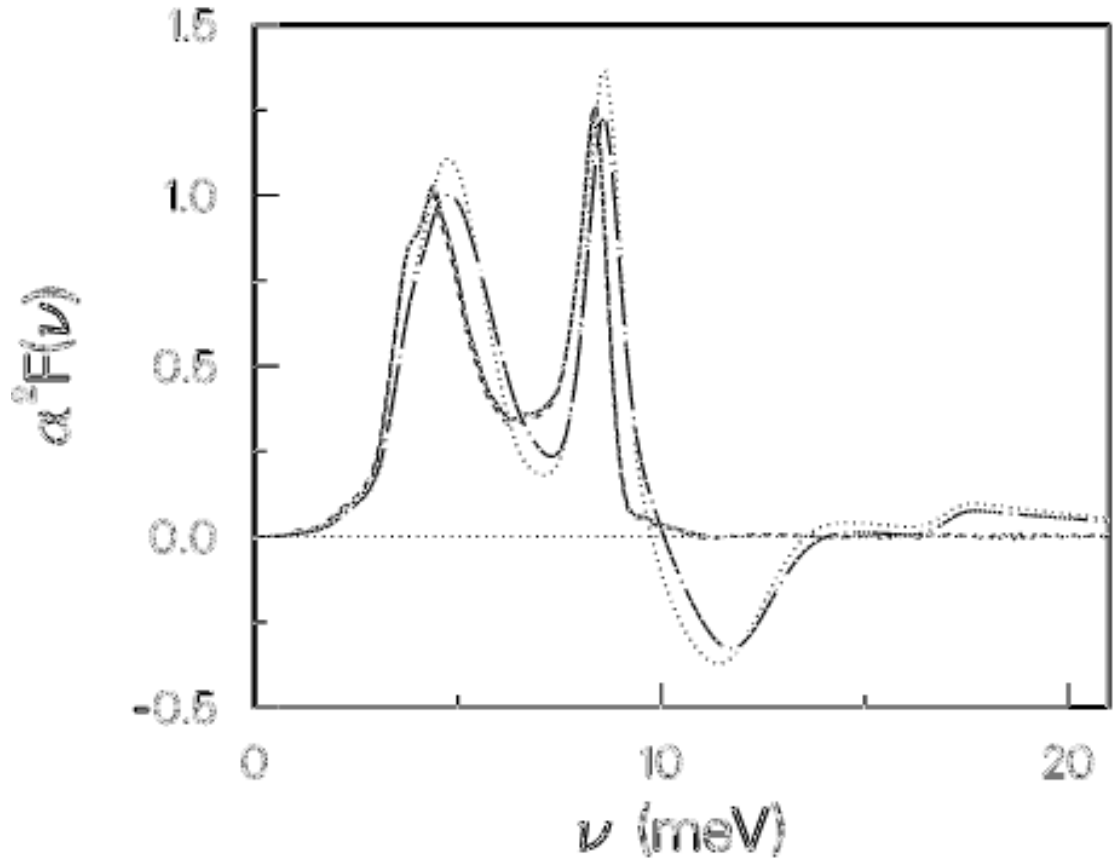}
\caption[]{$\alpha^2F(\nu)$ for Pb (solid curve) vs.
$\nu$, along with the estimates obtained from Eq. (\protect\ref{explicit})
with an impurity scattering rate, $1/\tau = 1$ meV (dotted) and
10 meV (dot-dashed).
These are both qualitatively quite accurate, before they
become negative at higher frequencies. Also plotted is the result
(dashed curve, indiscernible from the solid curve)
obtained from a full numerical inversion, as described in the text.
Taken from the second reference in Ref. \protect\cite{marsiglio98}.}
\label{ff16}
\end{figure}
                
\begin{figure}
\includegraphics[width=.9\textwidth]{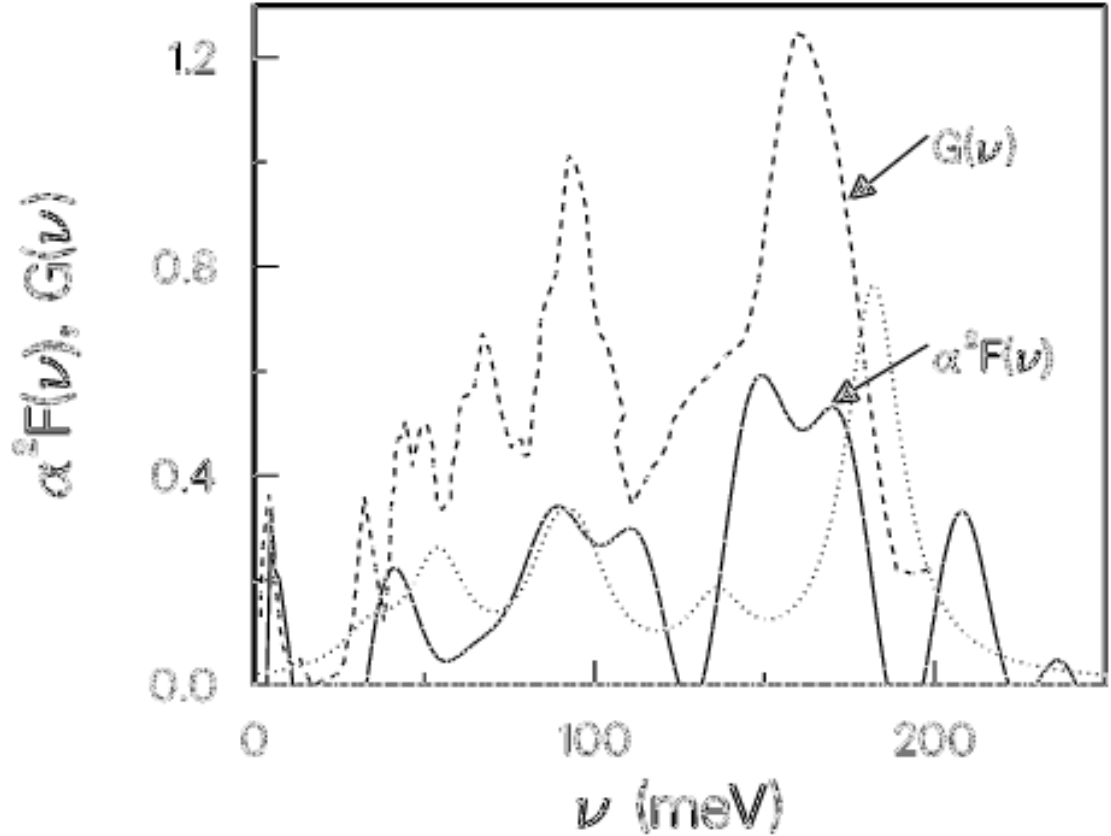}
\caption[]{The $\alpha^2F(\nu)$ for K$_3$C$_{60}$ (solid curve)
extracted from the reflectance data of Degiorgi {\it et al.}
\protect\cite{degiorgi94},
using Eq. (\protect\ref{explicit}). For purposes of analysis we have omitted
the negative parts. The neutron scattering
results from Ref. \protect\cite{pintschovius96} (dashed curve) are also shown.
Clearly the energy
scale in  $\alpha^2F(\nu)$ matches that of the phonons, and some of the
peaks even line up correctly. Finally, the dotted curve comes from an
analysis of photoemission data \protect\cite{alexandrov98}, where we have
arbitrarily broadened the phonon spectrum with Lorentzian lineshapes.
Taken from the second reference in Ref. \protect\cite{marsiglio98}.}
\label{ff17}
\end{figure}

\clearpage

\begin{figure}
\includegraphics[width=.99\textwidth]{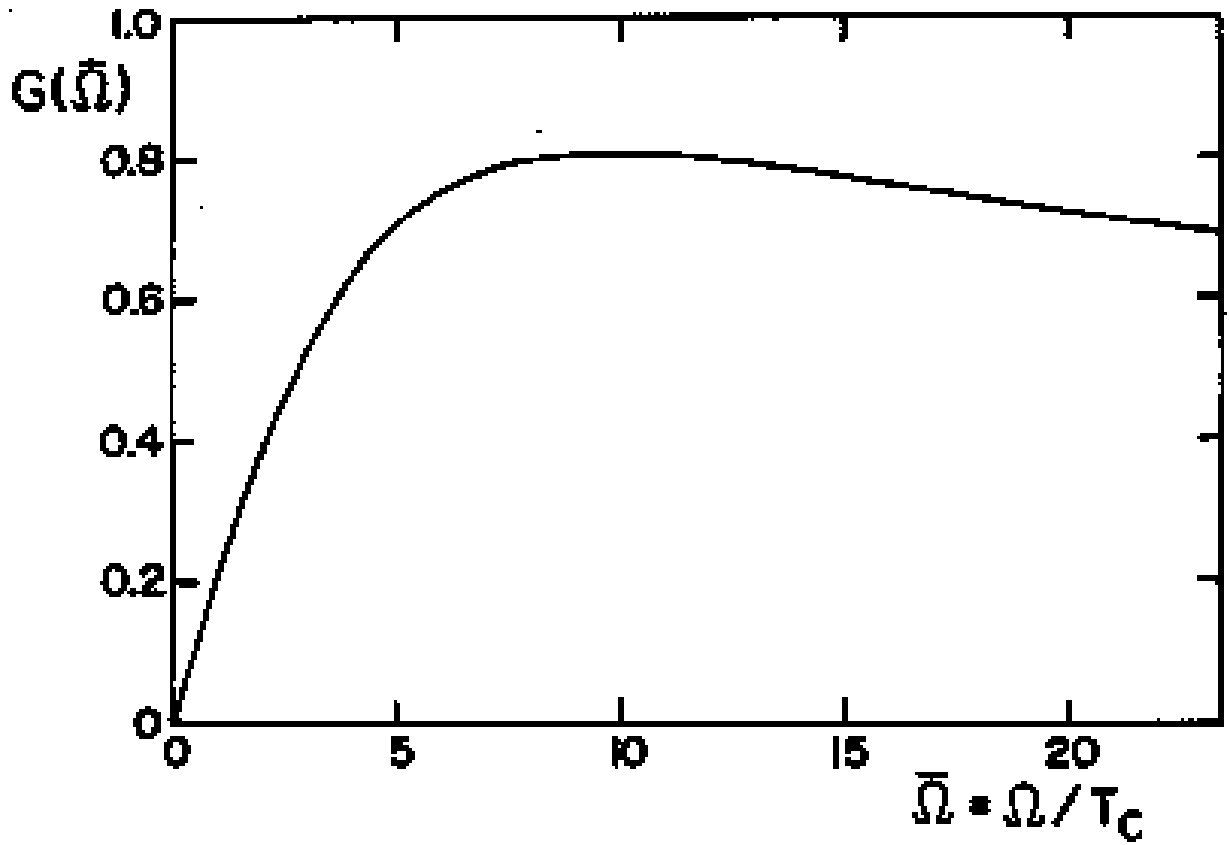}
\caption[]{The universal function $G(\bar \Omega )$ as a function of normalized phonon
energy $\bar \Omega =\Omega /T_c$ which enters the curve for the functional
derivative of $T_c$ with respect to $\alpha ^2F(\omega )$ in the
$\lambda ^{\Theta \Theta }$ model of Ref. \protect\cite{mitrovic81b}, from which this figure
was taken.}
\label{ff18}
\end{figure}

                
\begin{figure}
\includegraphics[width=.99\textwidth]{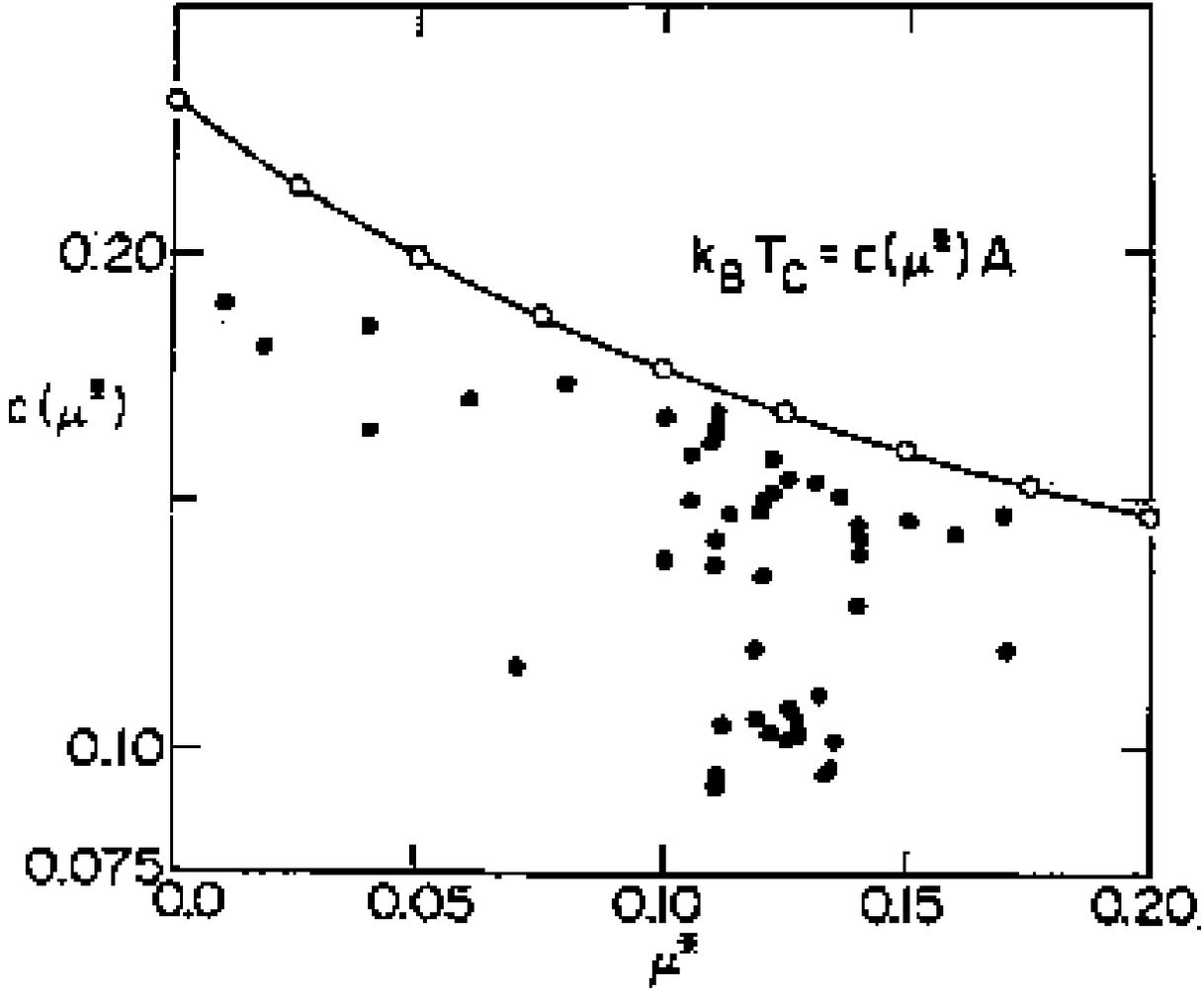}
\caption[]{The constant $c(\mu ^*)$ in the relation $k_BT_c=c(\mu ^*) A$ for the maximum
$T_c$ associated with a given A as a function of $\mu ^*$.  Placed on the same
figure (solid dots) are the results for $T_c/A$ obtained in the case of many
strong coupling superconductors for which $\alpha ^2F(\omega )$ is known from
tunneling spectroscopy.  The solid points all fall below the maximum curve as
they must. Adapted from Ref. \protect\cite{leavens75}. }
\label{ff19}
\end{figure}
                
\begin{figure}
\includegraphics[width=.9\textwidth]{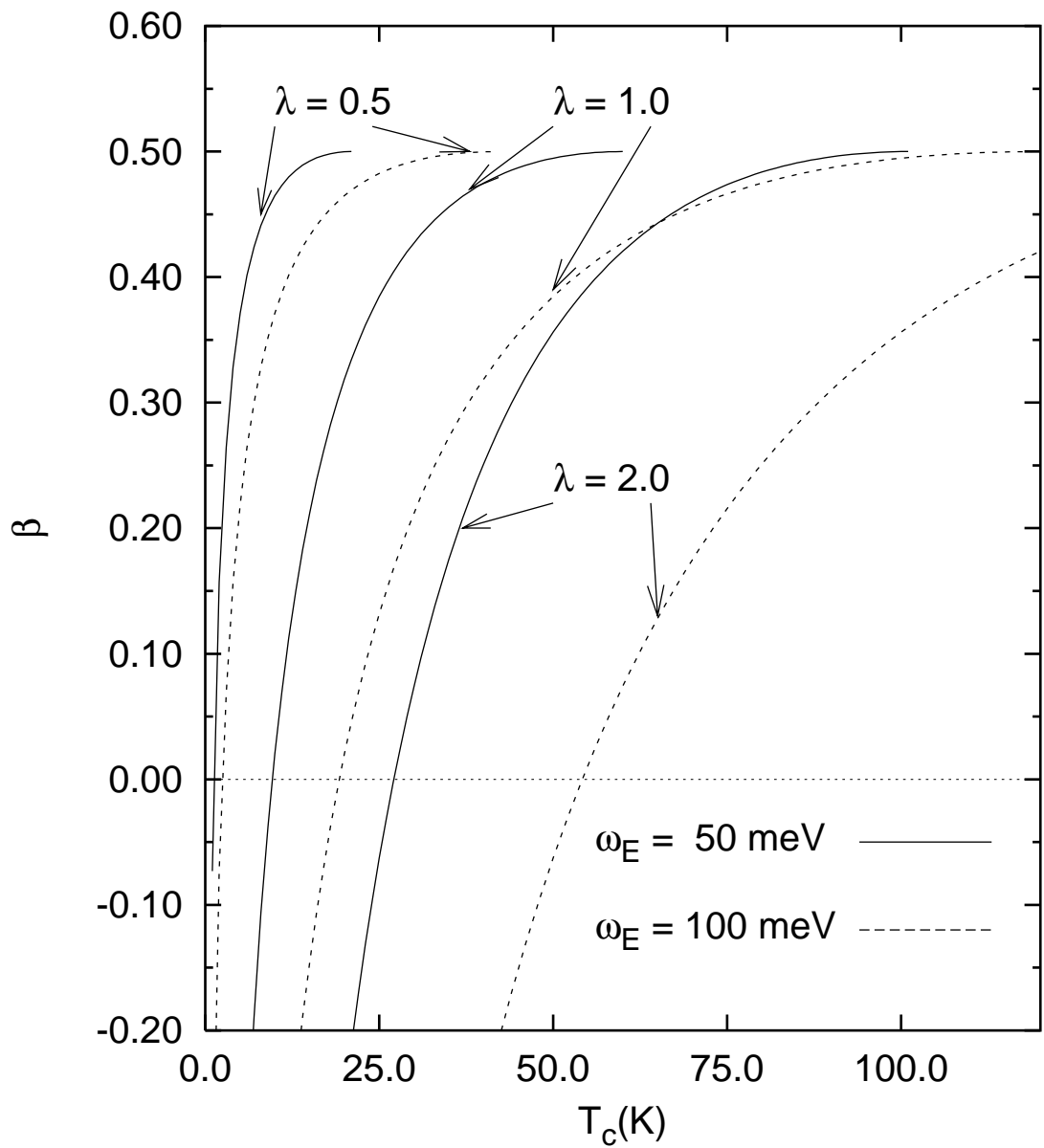}
\caption[]{The isotope coefficient, $\beta$, vs. $T_c$, for various values of
$\lambda$ and $\omega_E$. Along each curve $T_c$ changes because the Coulomb
pseudopotential $\mu^\ast$ is being varied. These results show that a low
value of $\beta$ is difficult to attain with high $T_c$. On the other hand,
for low $T_c$ materials, it is not so difficult.}
\label{ff20}
\end{figure}
                
\begin{figure}
\includegraphics[width=.85\textwidth]{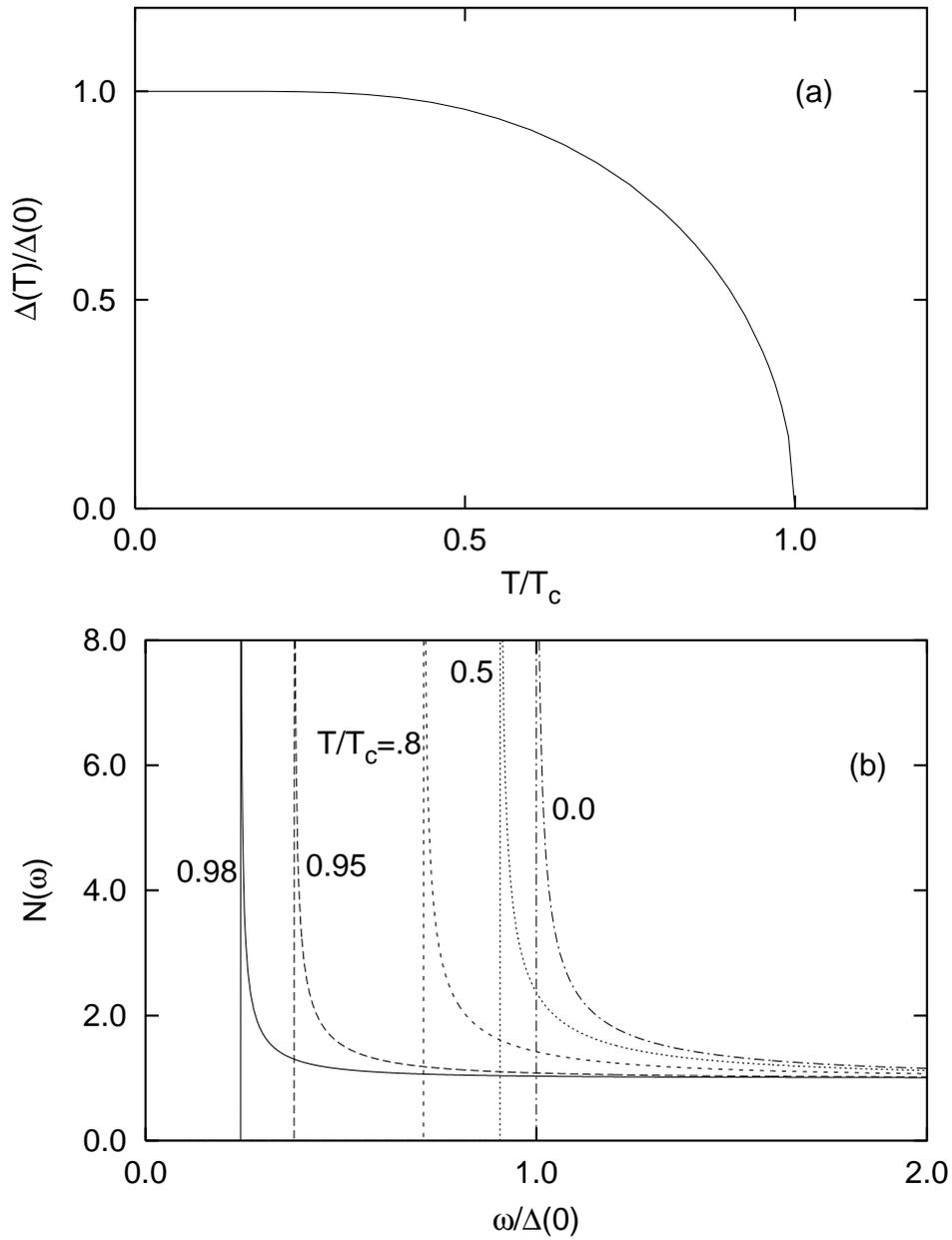}
\caption[]{(a) The temperature dependence of the BCS order parameter, and (b)
the resulting densities of states at various temperatures below $T_c$. The only effect
of finite temperatures on these latter curves is a reduced gap.}
\label{ff21}
\end{figure}
                
\begin{figure}
\includegraphics[width=.8\textwidth]{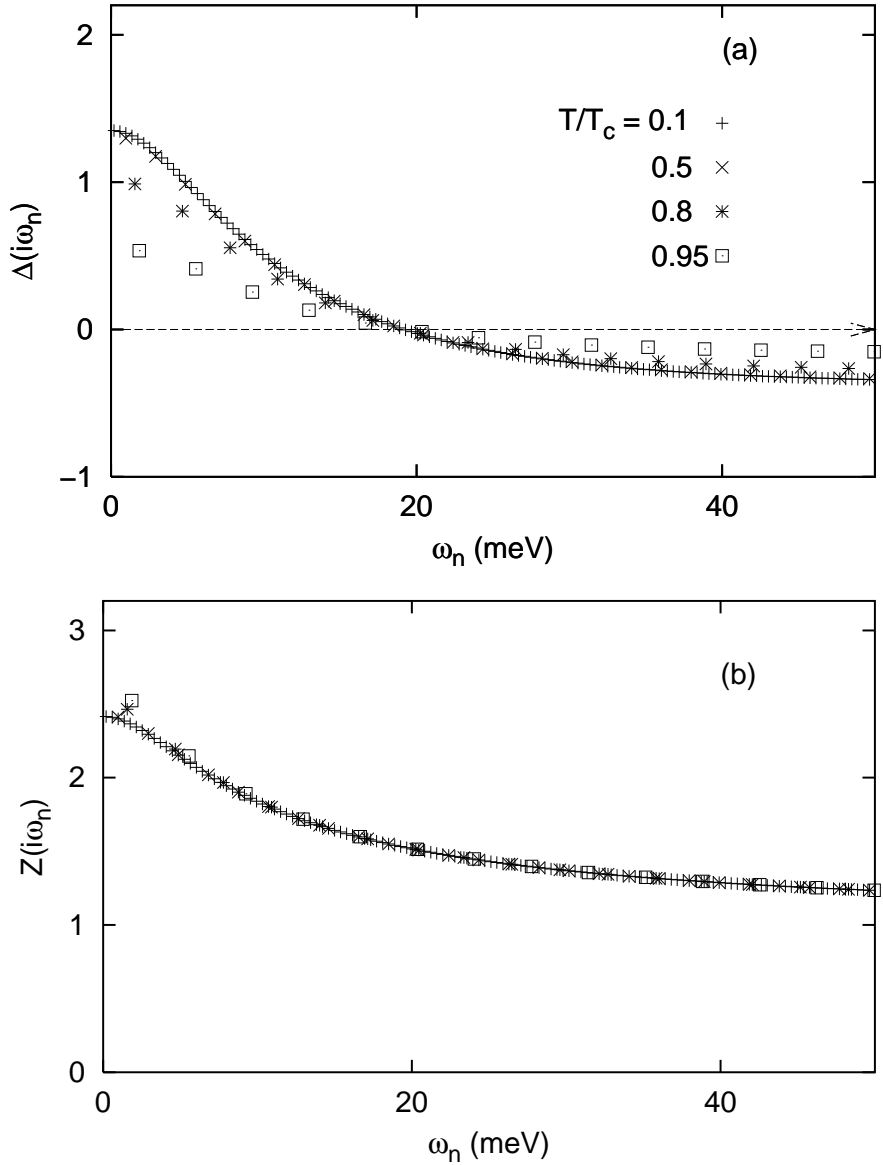}
\caption[]{(a) $\Delta(i\omega_n)$ and $Z(i\omega_n)$ vs $\omega_n$, the fermion Matsubara
frequency, for various temperatures, as indicated. Note that the curves are relatively
smooth and featureless, and at low temperatures little change occurs, except that more
Matsubara frequencies are present. In (a) the units of $\Delta$ are meV. These were produced
for Pb.}
\label{ff22}
\end{figure}
                
\begin{figure}
\includegraphics[width=.8\textwidth]{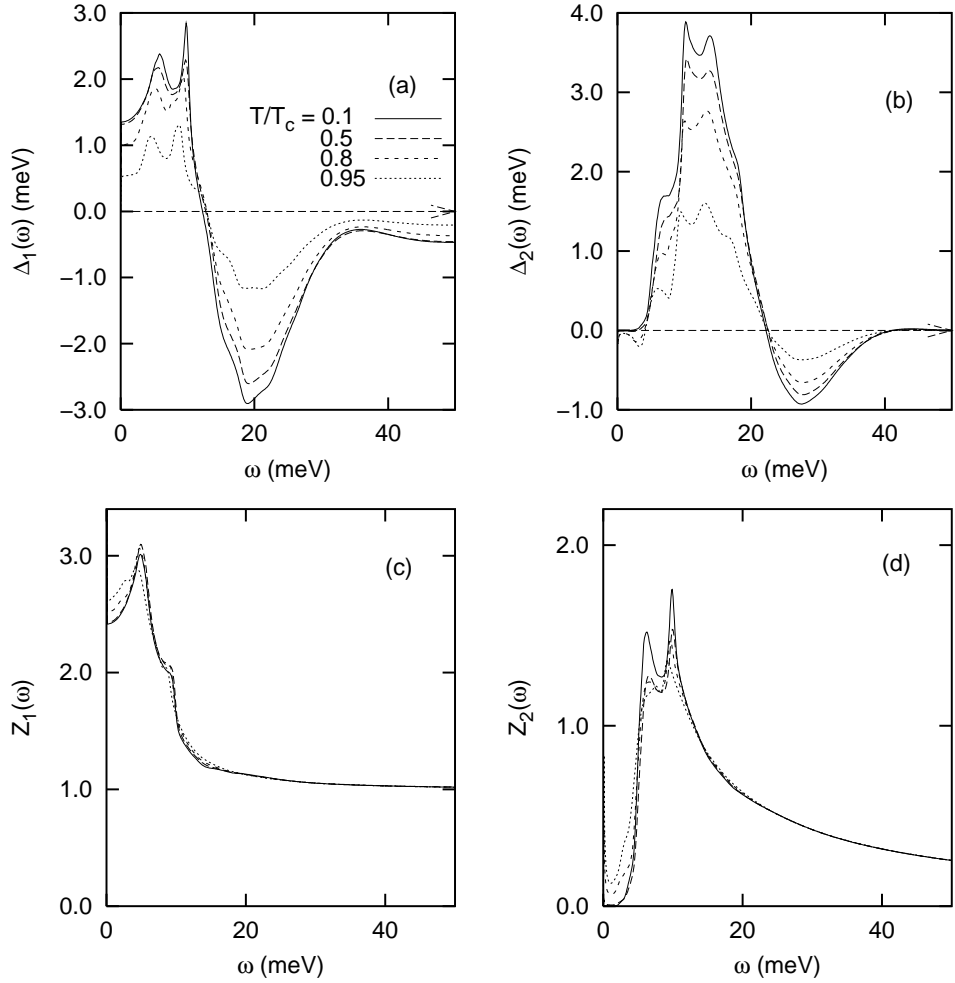}
\caption[]{The (a) real and (b) imaginary parts of the gap function (in meV)
on the real frequency axis, for Pb, for various temperatures, as in the previous figure.
Note the considerable structure present on the real axis. Also shown is the (c)
real and (d) imaginary part of the renormalization function, 
$Z(\omega)$ vs $\omega$.}
\label{ff23}
\end{figure}
                
\begin{figure}
\includegraphics[width=.9\textwidth]{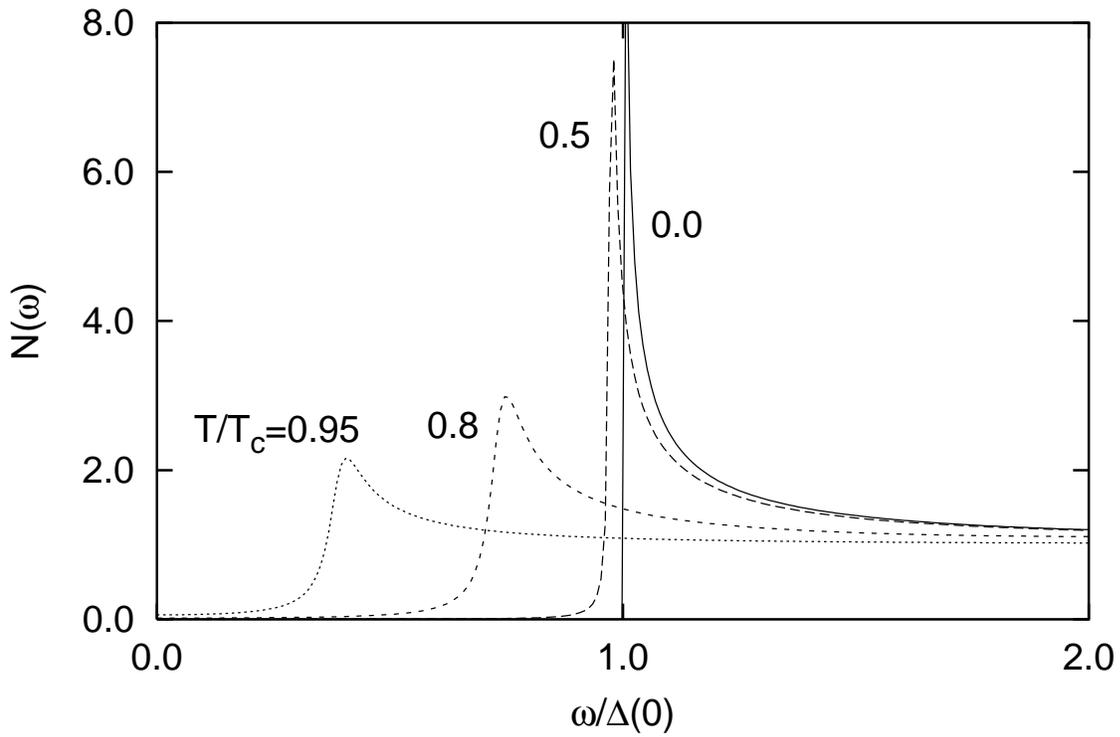}
\caption[]{Calculated densities of states of Pb for various temperatures. In contrast
to the BCS case (Fig. (21b), at high temperatures there is considerable smearing.}
\label{ff24}
\end{figure}
                
\begin{figure}
\includegraphics[width=.74\textwidth]{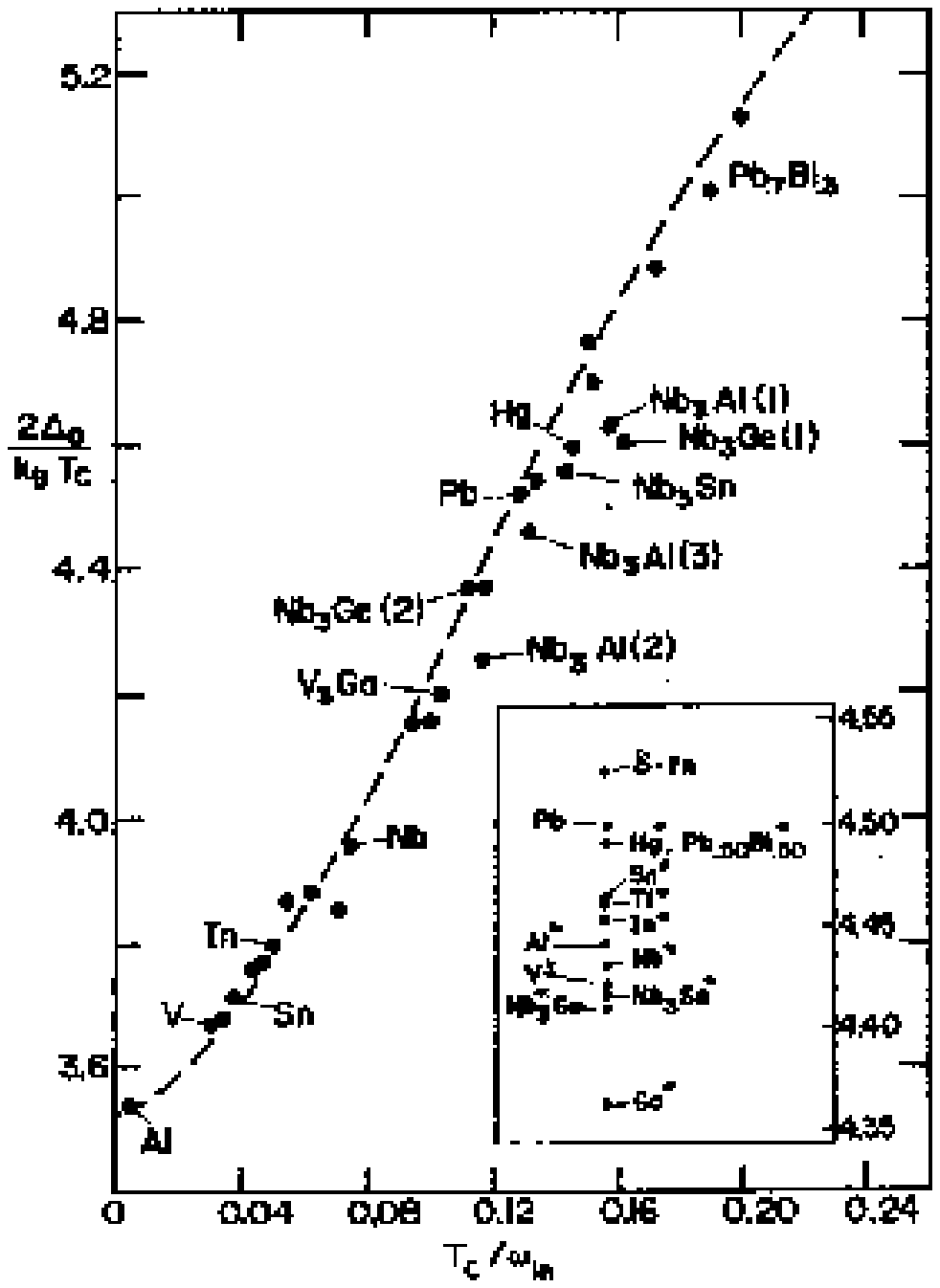}
\caption[]{The ratio $2\Delta _0/k_BT_c$ vs $T_c/\omega _{\ell n}$. 
The solid dots
represent results from the full numerical solutions of the
Eliashberg equations.  Experiment tends to agree to within 10\%.  In increasing
order of $T_c/\omega _{\ell n}$, the dots correspond to the following systems:
$A\ell $, $V$, Ta, Sn, $T\ell $, $T\ell _{0.9}Bi_{0.1}$, In, Nb (Butler), Nb
(Arnold), $V_3Si (1)$, $V_3Si$ (Kihl.), Nb (Rowell), Mo,
$Pb_{0.4}T\ell _{0.6}$, La, $V_3Ga$, $Nb_3A\ell  (2)$, $Nb_3Ge (2)$,
$Pb_{0.6}T\ell _{0.4}$, Pb, $Nb_3A\ell  (3)$, $Pb_{0.8}T\ell _{0.2}$, Hg,
$Nb_3Sn$, $Pb_{0.9}Bi_{0.1}$, $Nb_3A\ell  (1)$, $Nb_3Ge (1)$,
$Pb_{0.8}Bi_{0.2}$, $Pb_{0.7}Bi_{0.3}$, and $Pb_{0.65}Bi_{0.35}$.  The drawn
curve corresponds to $2\Delta _0/k_BT_c=3.53[1+12.5
(T_c/\omega _{\ell n})^2\ell n (\omega _{\ell n}/2T_c)]$.  The insert shows
results for different scaled $\alpha ^2F(\omega )$ spectra.
They all correspond to the same value of $T_c$ and of $\omega _{\ell n}$
as Pb.  They serve to show that some deviation from the general trend is possible.
Reproduced from Ref. \protect\cite{carbotte90}.}
\label{ff25}
\end{figure}
                
\begin{figure}
\includegraphics[width=.9\textwidth]{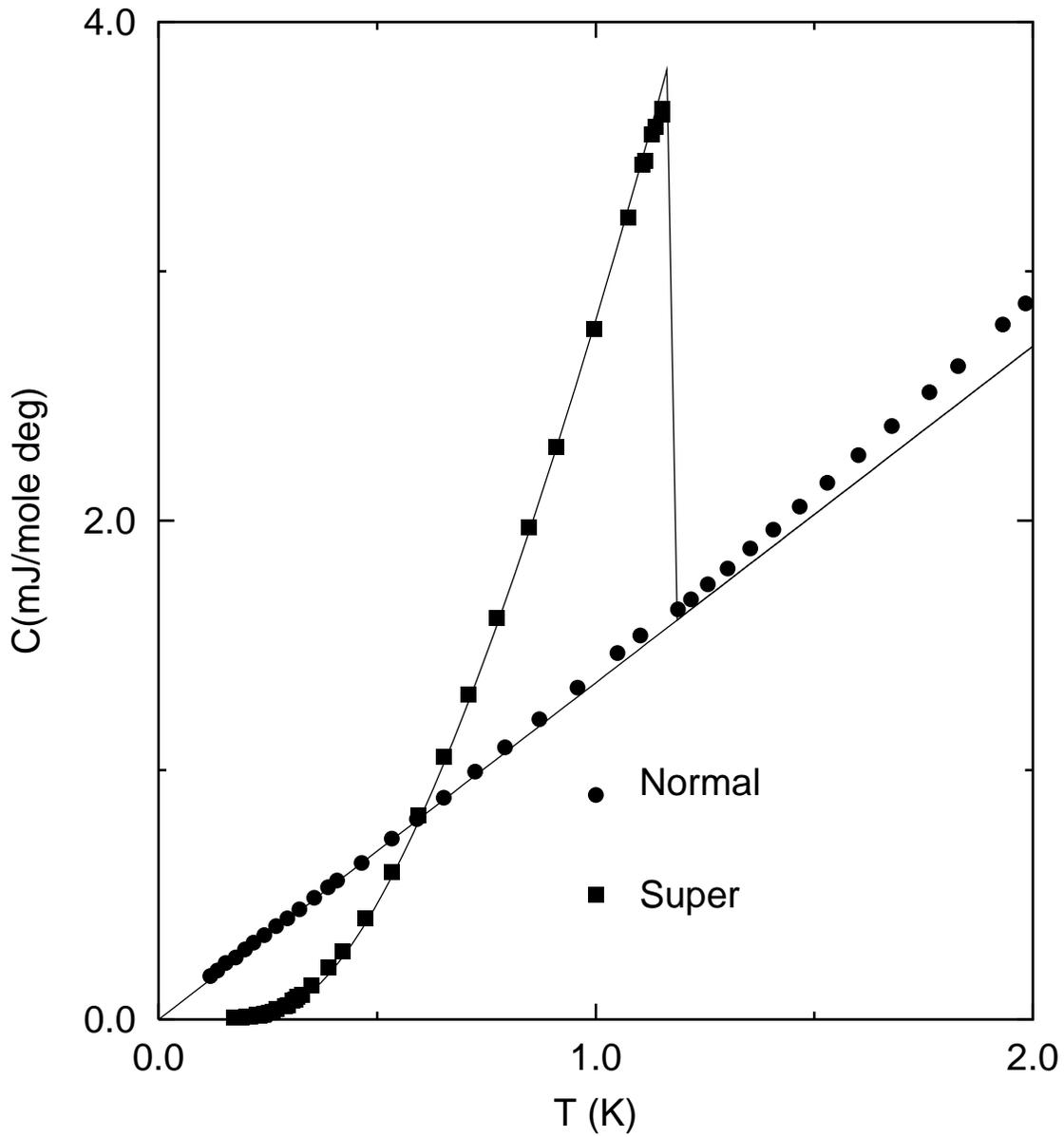}
\caption[]{Specific heat of aluminium as a function of temperature
in the superconducting state and the normal state (applied field of
300 Gauss). Data taken from Ref. \protect\cite{phillips_n59}. 
The BCS prediction,
given the normal state data, is given by the solid curve.}
\label{ff26}
\end{figure}
                
\begin{figure}
\includegraphics[width=.84\textwidth]{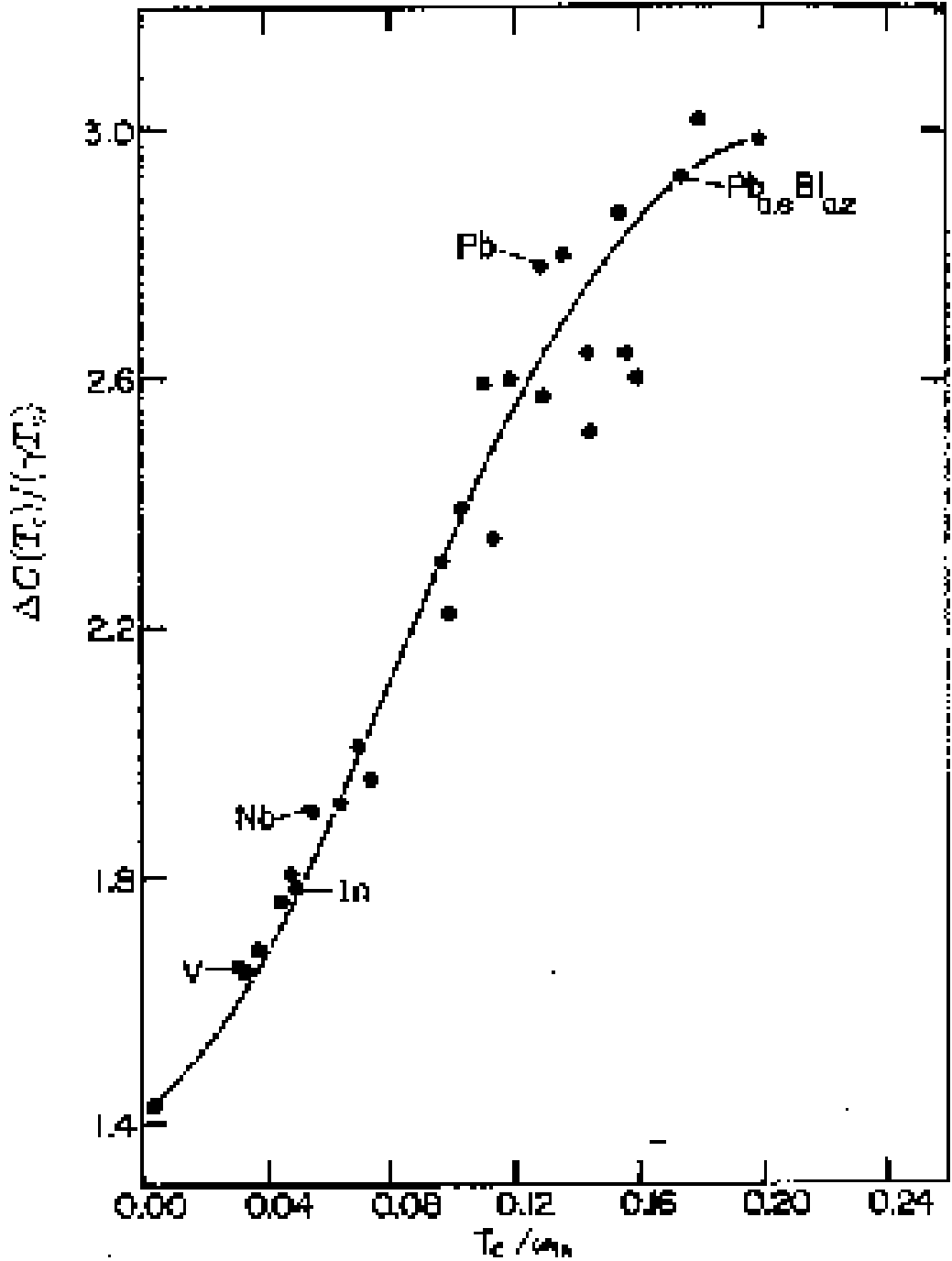}
\caption[]{The specific heat ratio, $\Delta C(T_c)/(\gamma T_c)$ vs
$T_c/\omega _{\ell n}$.  The dots represent results from the full
numerical solutions of the Eliashberg equations.  Experiment tends to agree to
within 10\%.  In increasing order of $T_c/\omega _{\ell n}$, the dots
correspond to the following systems:  $A\ell $, $V$, Ta, Sn, $T\ell $,
$T\ell _{0.9}Bi_{0.1}$, In, Nb (Butler), Nb (Arnold), $V_3Si\ 1$, $V_3Si$
(Kihl.), Nb (Rowell), Mo, $Pb_{0.4}T\ell _{0.6}$, La, $V_3Ga$, $Nb_3A\ell
(2)$, $Nb_3Ge (2)$, $Pb_{0.6}T\ell _{0.4}$, Pb, $Nb_3A\ell  (3)$,
$Pb_{0.8}T\ell _{0.2}$, Hg, $Nb_3Sn$, $Pb_{0.9}Bi_{0.1}$, $Nb_3A\ell  (1)$,
$Nb_3Ge (1)$, $Pb_{0.8}Bi_{0.2}$, $Pb_{0.7}Bi_{0.3}$, and $Pb_{0.65}Bi_{0.35}$.
The drawn curve corresponds to $\Delta C(T_c)/\gamma T_c=1.43(1+53
(T_c/\omega _{\ell n})^2 \ell n (\omega _{\ell n}/3T_c))$. Adapted from Ref.
\protect\cite{marsiglio86}.}
\label{ff27}
\end{figure}
                
\begin{figure}
\includegraphics[width=.83\textwidth]{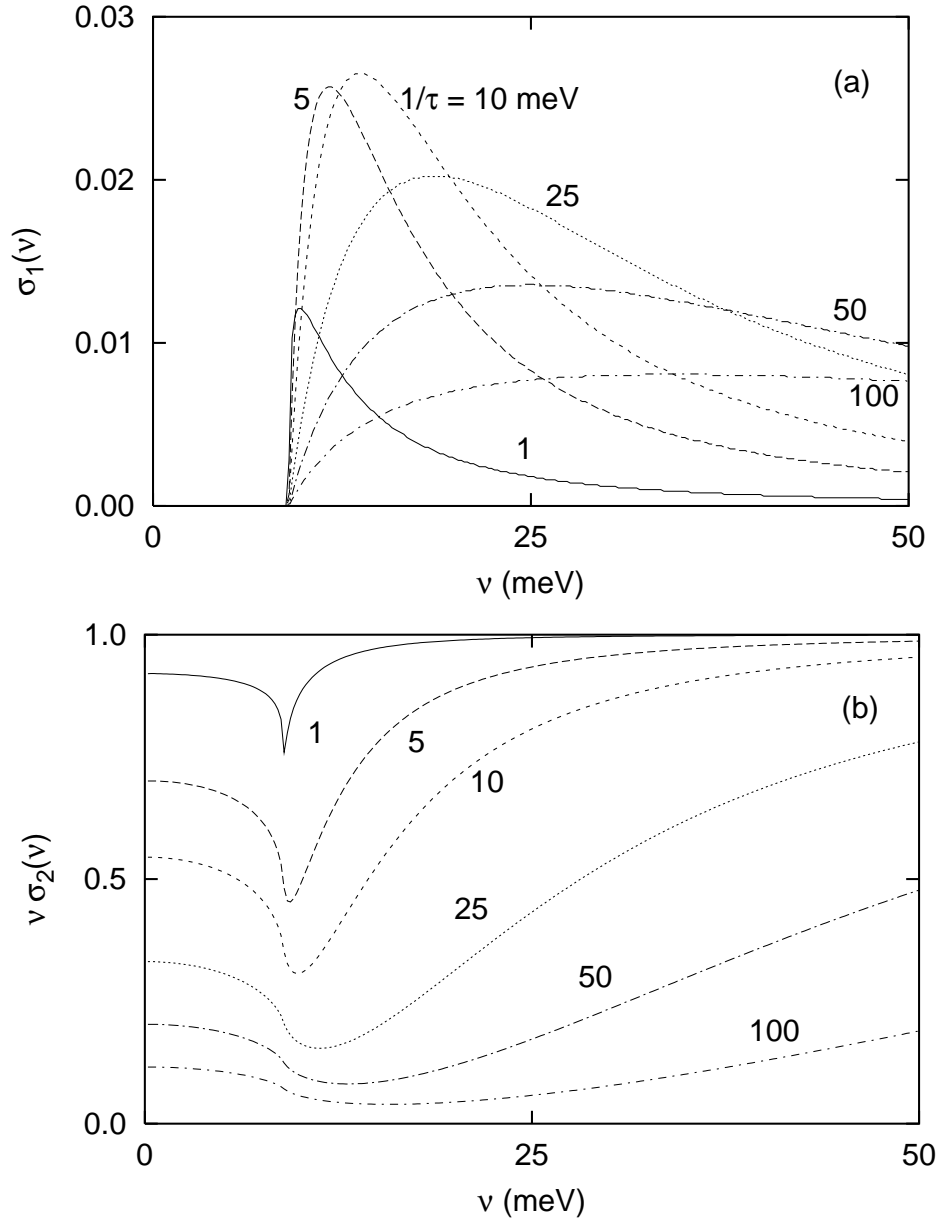}
\caption[]{(a) $\sigma_{1}(\nu)$ vs. $\nu$ in the zero temperature
BCS superconducting state for the various impurity scattering rates indicated.
The absorption onset at $2\Delta(0)$ remains sharp independent of the
scattering rate. A delta-function contribution (not shown) is also
present at the origin.
(b) Same as in (a) except for the frequency times the imaginary part
of the conductivity. The optical gap is a little less evident in the
dirty limit. The conductivity is given in units of $ne^2/m \equiv
\omega_P^2/4\pi$). Taken from Ref. \protect\cite{marsiglio97}.}
\label{ff28}
\end{figure}
                
\begin{figure}
\includegraphics[width=.84\textwidth]{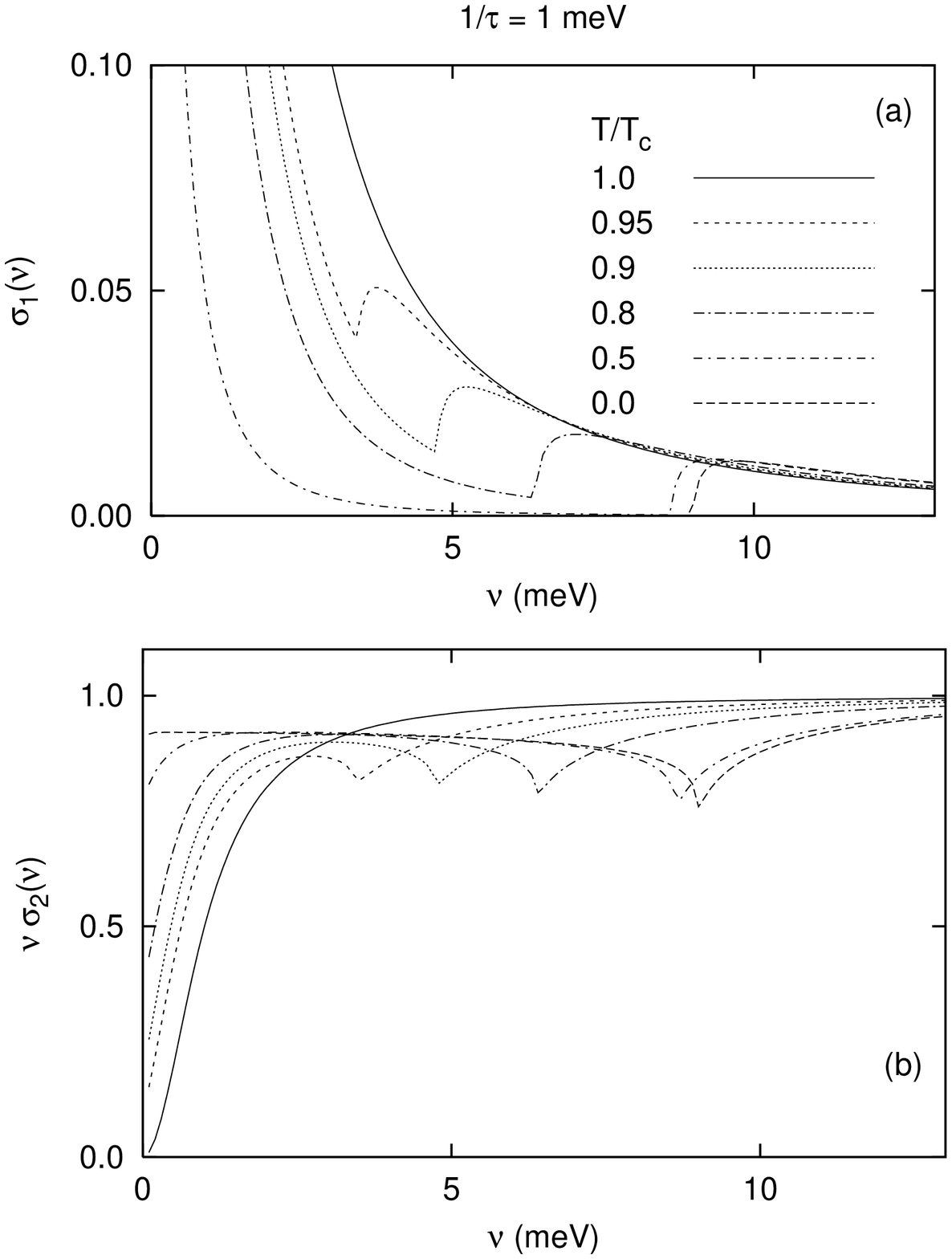}
\caption[]{Frequency dependence of $\sigma_{1}(\nu)$
near the clean limit ($1/\tau = 1$ meV) for various temperatures
in the BCS superconducting state. The appearance of a gap is
evident, even at temperatures close to $T_c$. (b) Same as in (a), but for
$\nu \sigma_{2}(\nu)$. The appearance of a gap is
evident in the imaginary part of the conductivity as well.
The conductivity is given in units of $ne^2/m \equiv \omega_P^2/4\pi$).
Taken from Ref. \protect\cite{marsiglio97}.}
\label{ff29}
\end{figure}
                
\begin{figure}
\includegraphics[width=.84\textwidth]{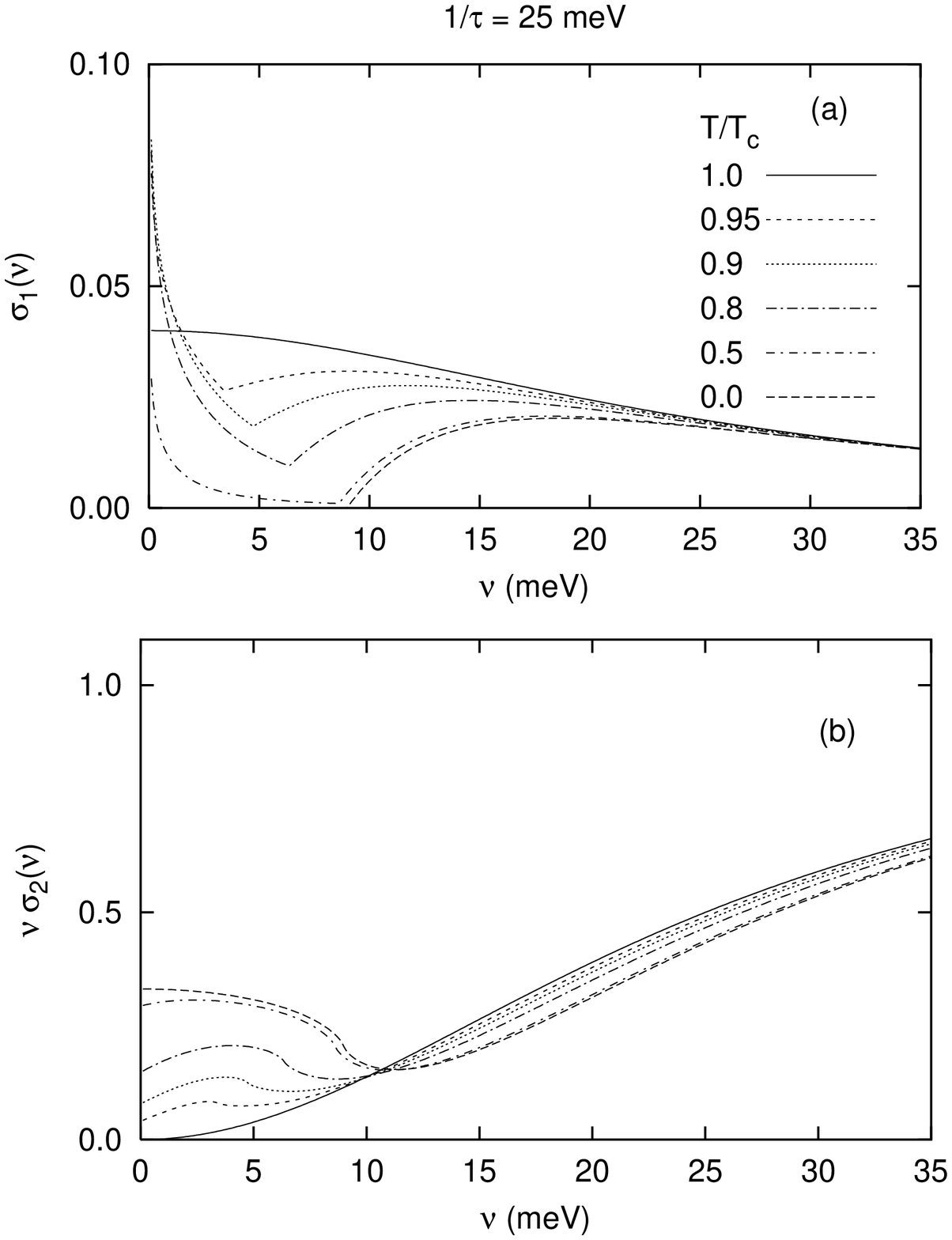}
\caption[]{Frequency dependence of $\sigma_{1}(\nu)$
near the dirty limit ($1/\tau = 25$ meV) for various temperatures
in the BCS superconducting state. The appearance of a gap is
evident, even at temperatures close to $T_c$. (b) Same as in (a), but for
$\nu \sigma_{2}(\nu)$. The appearance of a gap is
evident in the imaginary part of the conductivity as well.
The conductivity is given in units of $ne^2/m \equiv \omega_P^2/4\pi$).} \label{ff30}
\end{figure}
                
\begin{figure}
\includegraphics[width=.9\textwidth]{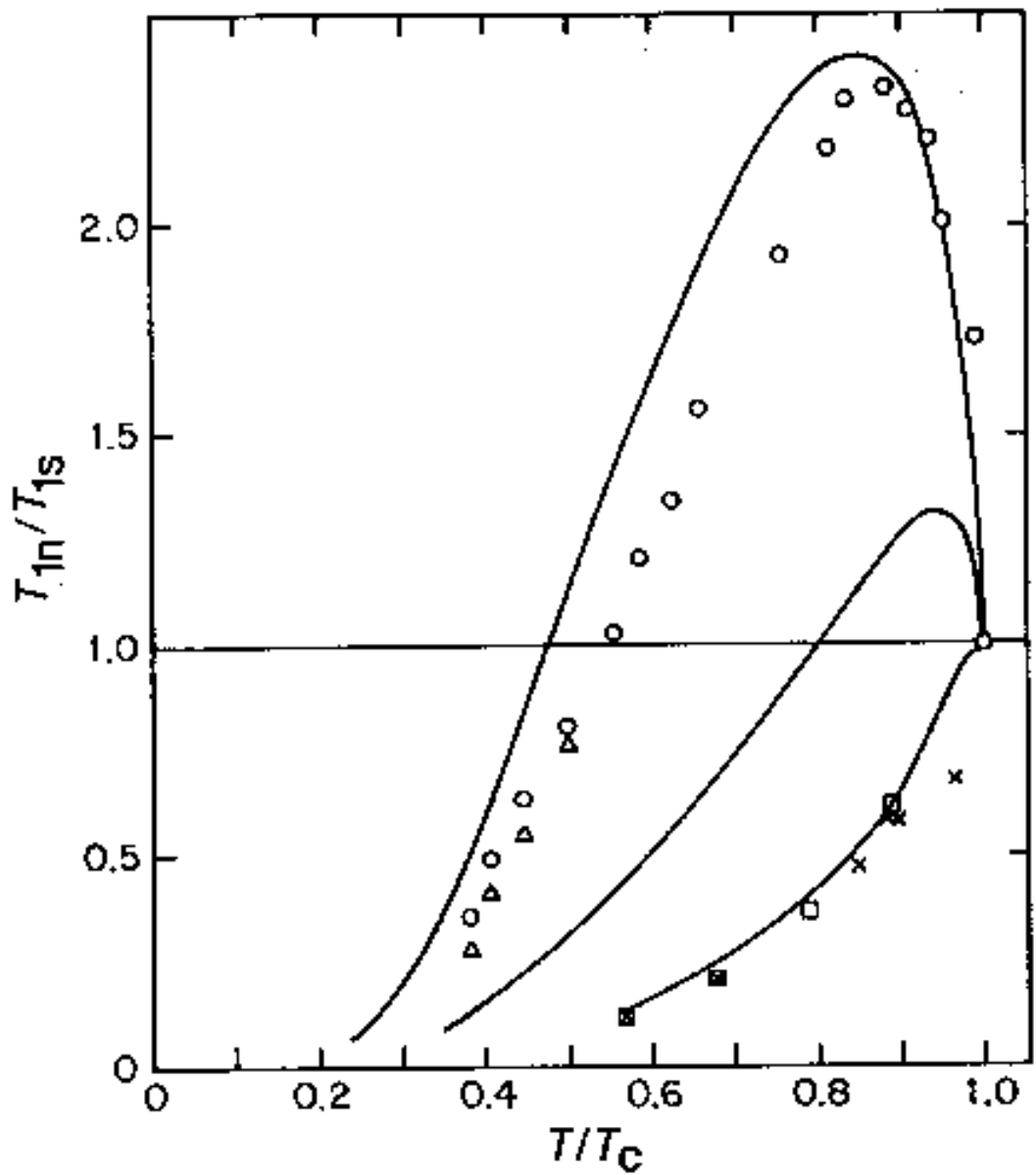}
\caption[]{Nuclear spin relaxation rate vs. reduced temperature. Data points for
Indium are indicated by circles and triangles, while data for YBa$_2$Cu$_3$O$_7$
are indicated by squares and crosses. The solid curves are calculated with Eliashberg
theory for Indium (upper curve) and two model spectra with $\lambda = 1.66$ and
$3.2$ (lowest curve). Agreement is good in the case of Indium and the lowest curve.
Reproduced from \protect\cite{allen91}.}
\label{ff31}
\end{figure}

\clearpage
                
\begin{figure}
\includegraphics[width=0.65\textwidth]{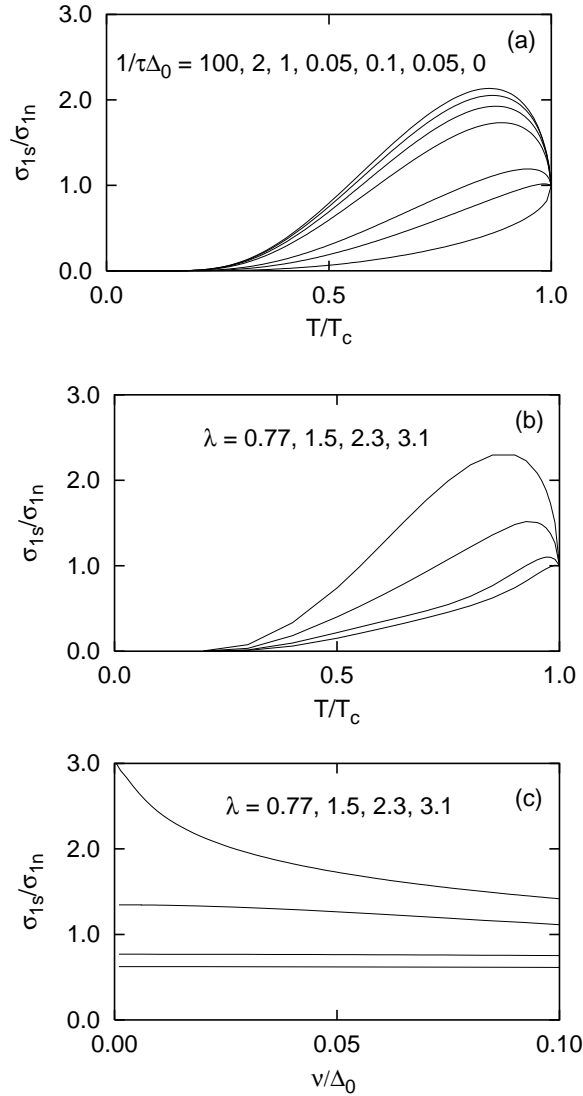}
\caption[]{(a) Conductivity ratio, $\sigma_{1s}/\sigma_{1n}$, versus reduced
temperature, $T/T_c$, in the BCS limit, for various impurity scattering rates.  From
top to bottom the curves are calculated for ${1\over \tau \Delta _0} = 100, 2, 1, 0.5,
0.1, 0.05$, and $0$.  The frequency used was $\nu/\Delta_0 = 0.02.$
In the clean limit
$(1/\tau = 0)$ the coherence peak has disappeared. (b) Same quantity as in (a), but
for different coupling strengths, $\lambda = 0.77, 1.5, 2.3$, and $3.1$.  
(The peak diminishes with increasing coupling
strength).  These were computed in the dirty limit $(1/\tau = 500$ meV) and
for $\nu = 0.05$ meV.  The result for $\lambda = 0.77$ (largest maximum) is nearly identical
with the BCS result. (c)  Conductivity ratio versus frequency normalized to the zero temperature
gap edge, $\nu/\Delta_0$, for the same coupling strengths as in (b).
The curves decrease in magnitude with increasing coupling strength.  The
maximum apparent in (b) for $\lambda = 0.77$ and $1.5$ is also clear here since the
two uppermost curves have magnitude greater than unity.  As the coupling strength
increases the conductivity ratio becomes independent of frequency.  Calculations are in
the dirty limit, with $T/T_c = 0.85$.}
\label{ff32}
\end{figure}
                
\begin{figure}
\includegraphics[width=.9\textwidth]{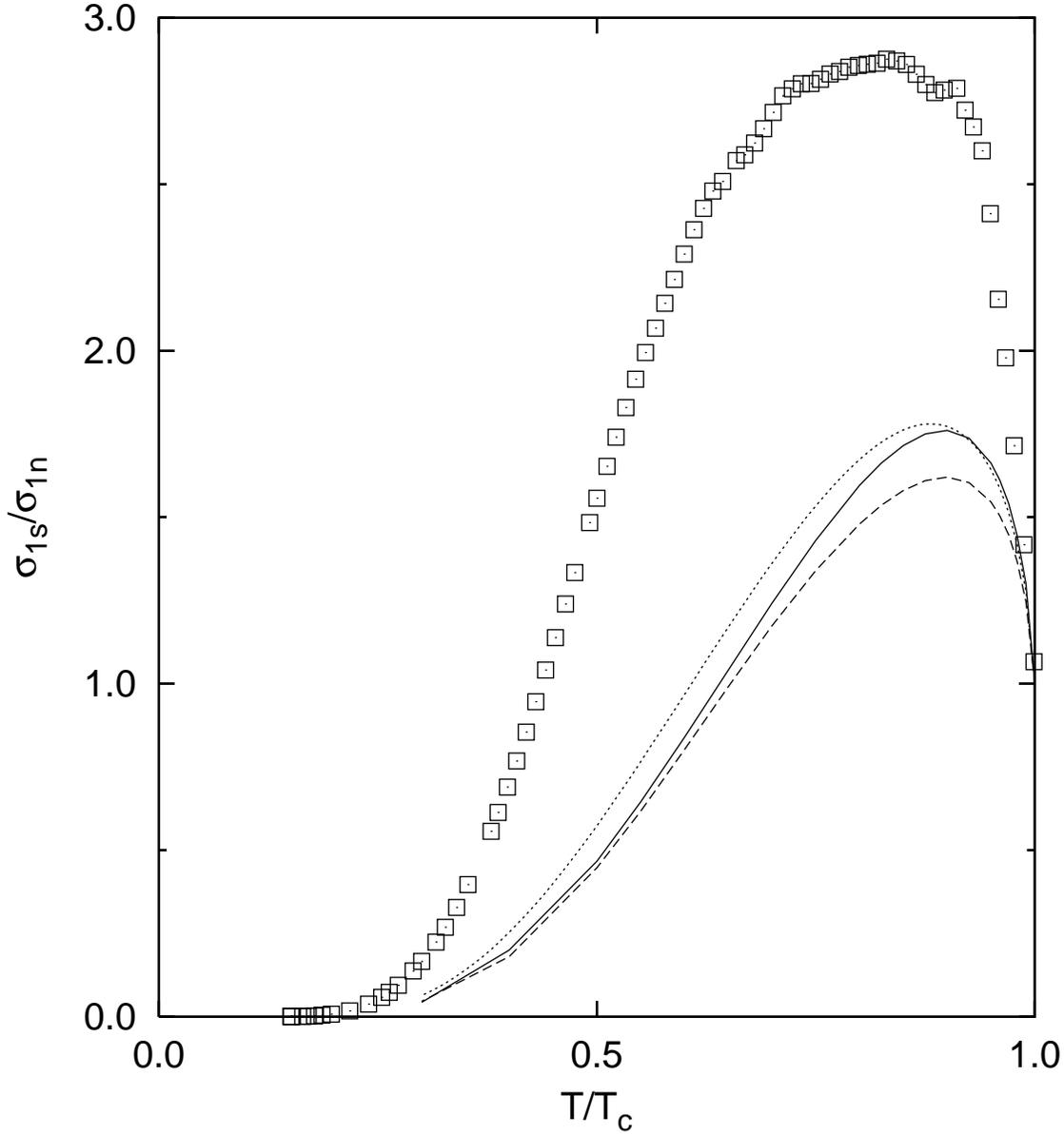}
\caption[]{Microwave conductivity normalized to the normal state,
$\sigma_{1s}/\sigma_{1n}$, as a function of reduced temperature $T/T_c$.
The open squares are the data for Nb.  The
dotted curve is the BCS result with experimental frequency $\omega =
17$ GHz and impurity scattering rate $1/\tau = 100.0$ meV (dirty limit).  The
solid and dashed curves are the results of full Eliashberg calculations with two
different $(\alpha^2 F(\omega))$ spectra. None of the theoretical curves can
reproduce the data.} 
\label{ff33}
\end{figure}
                
\begin{figure}
\includegraphics[width=.8\textwidth]{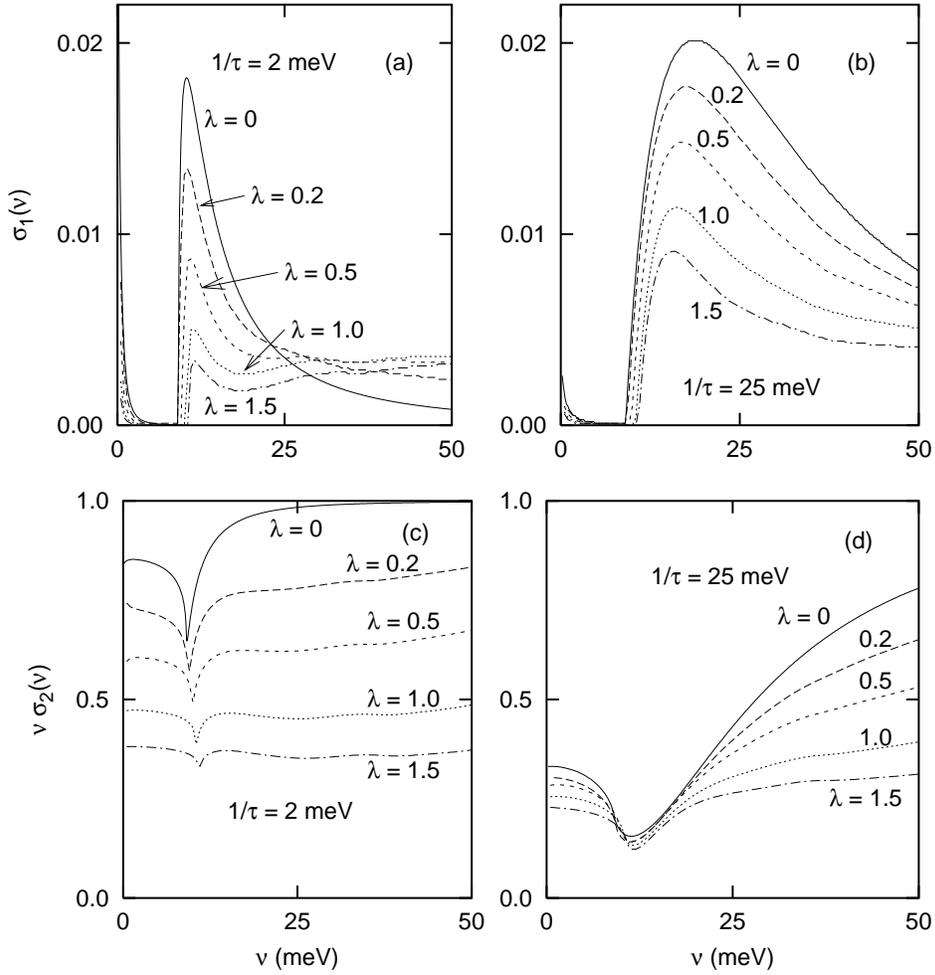}
\caption[]{The real part (a,b) and the imaginary part (c,d) of the
conductivity at essentially zero temperature ($T/T_c = 0.3$)
with $1/\tau = 2$ meV (a,c) and $1/\tau = 25$ meV (b,d). In all cases
we have used the BKBO spectrum scaled to give the designated value of,
$\lambda$, while $T_c$ is held fixed at 29 K by adjusting $\mu^\ast$.
Increased coupling strength suppresses both $\sigma_1(\nu)$ and
$\nu \sigma_2(\nu)$ and broadens the minimum in the latter
at $2\Delta$. Note that $2\Delta$ increases slightly as the coupling
strength is increased.
The conductivity is given in units of $ne^2/m \equiv \omega_P^2/4\pi$).}
\label{ff34}
\end{figure}
                
\begin{figure}
\includegraphics[width=.9\textwidth]{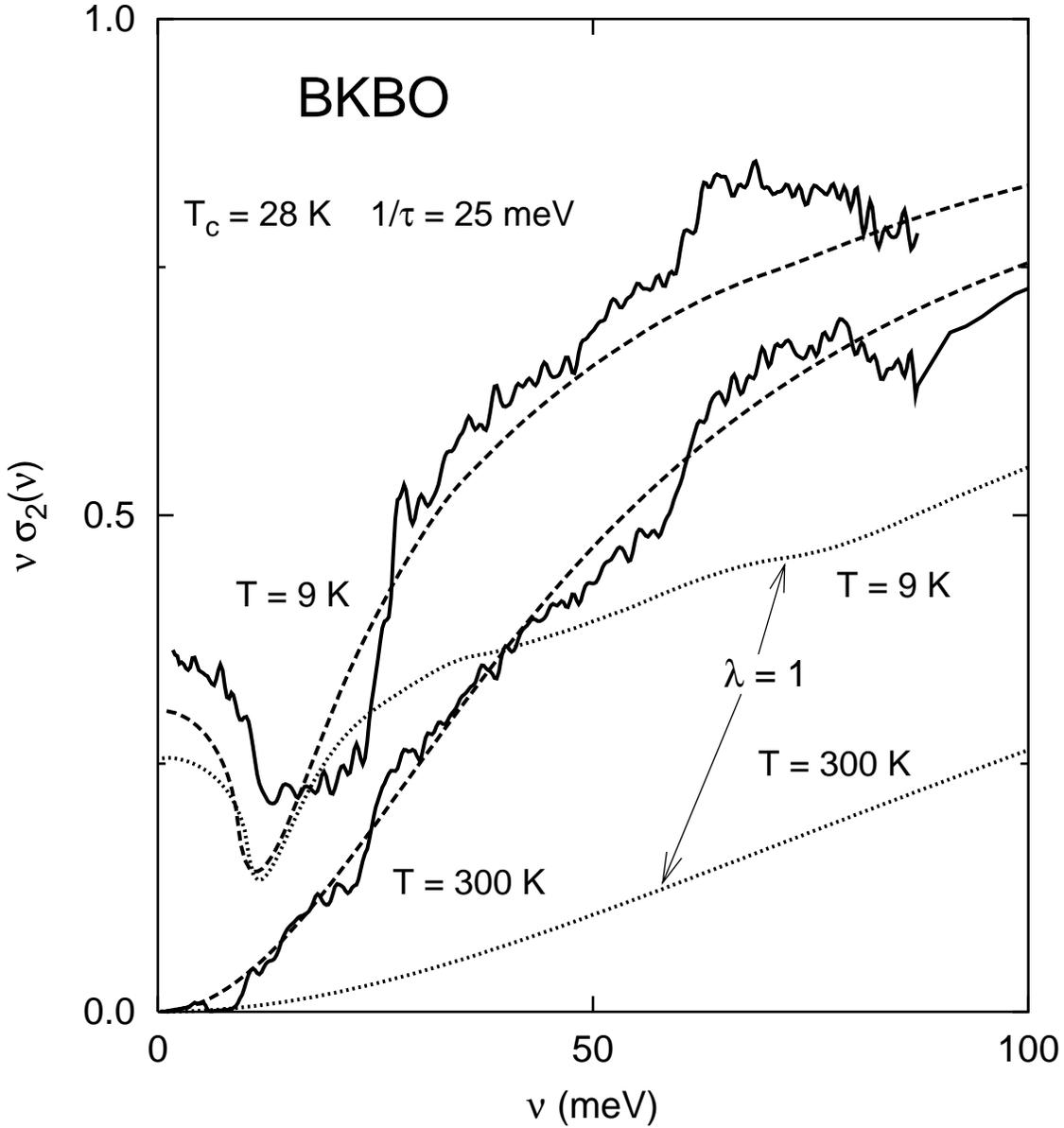}
\caption[]{Measured $\nu \sigma_2(\nu)$ vs. frequency at $T = 9$ K and at
$T = 300$ K (solid curves).
Also shown are the theoretical fits,
using the BKBO spectrum, scaled so that $\lambda = 0.2$ (dashed curves).
$T_c$ is kept fixed to the experimental value with a negative $\mu^\ast$.
Finally, theoretical fits are also shown with $\lambda = 1$ (dotted curves).
The latter curves are clearly incompatible with the experimental results.
Adapted from Ref. \protect\cite{marsiglio96}.}
\label{ff35}
\end{figure}
                
\begin{figure}
\includegraphics[width=.9\textwidth]{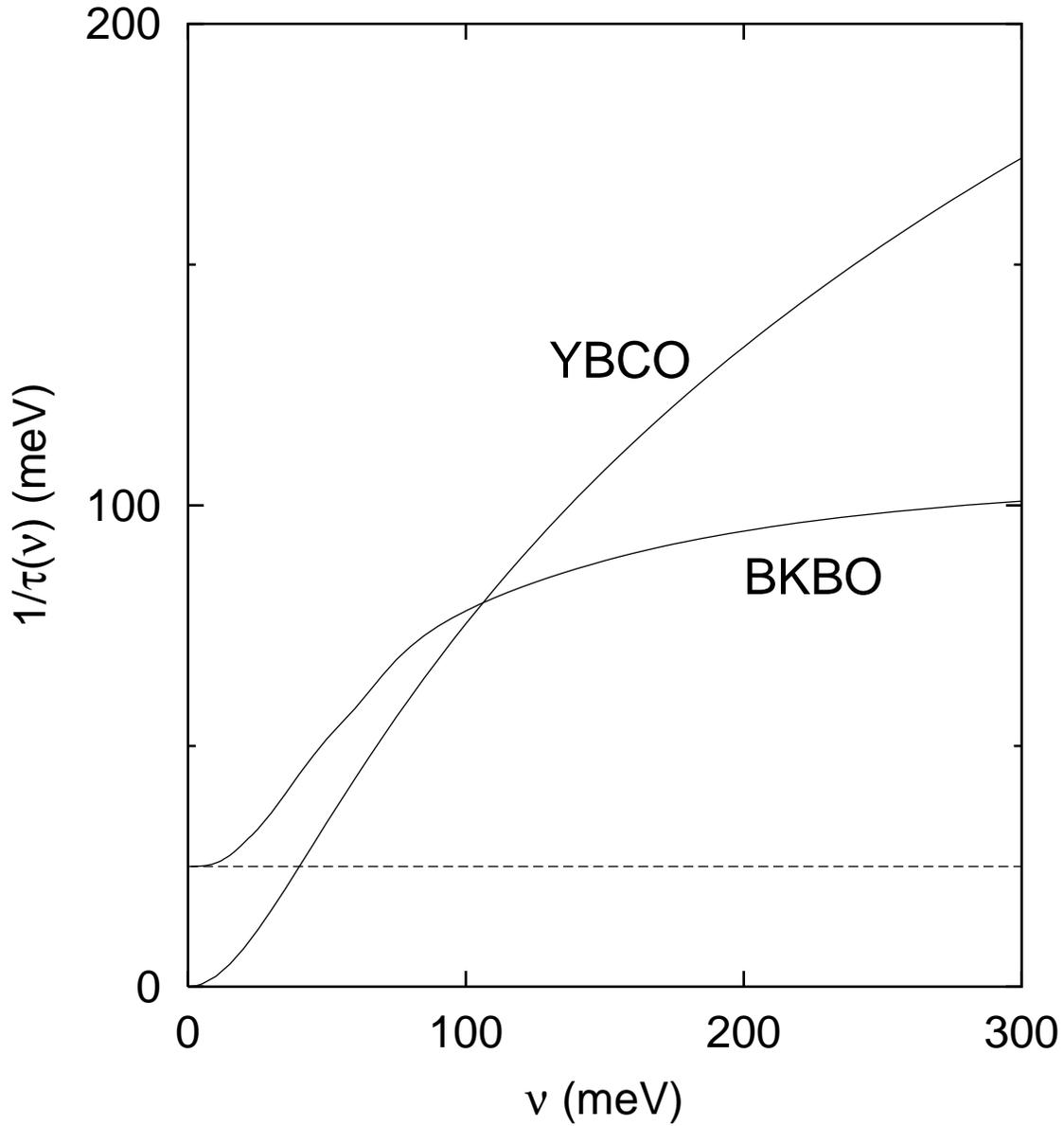}
\caption[]{The conductivity-derived scattering rate, $1/\tau(\nu) \equiv
{\omega_P^2 \over 4\pi }
Re \ \bigl(1/\sigma(\nu) \bigr)$ vs. frequency in the normal state
for pure elastic scattering (dashed line), combined elastic and
inelastic scattering (BKBO spectrum with $\lambda = 1$), and pure
inelastic scattering using a model spin fluctuation spectrum
appropriate to YBCO. Because of the difference in
spectral function frequency scales, the result for YBCO continues to rise with
frequency, even at 300 meV. Reproduced from \protect\cite{marsiglio97}.}
\label{ff36}
\end{figure}
                
\begin{figure}
\includegraphics[width=.8\textwidth]{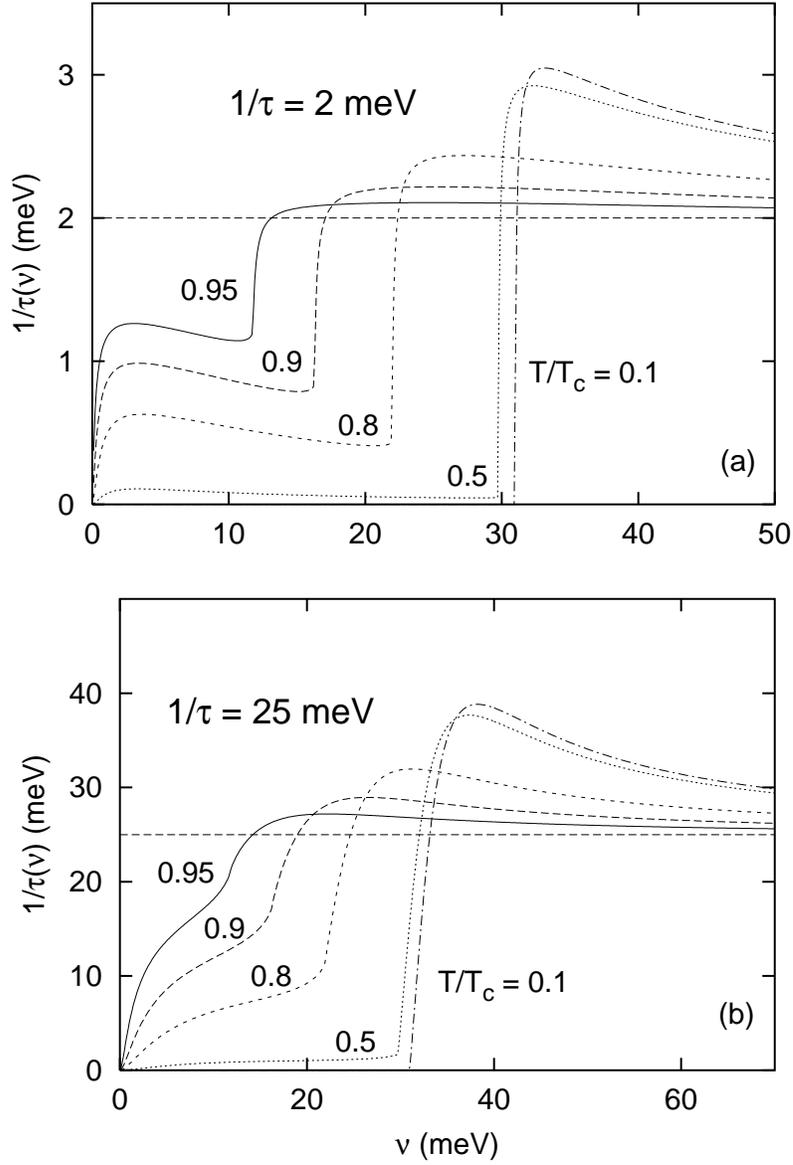}
\caption[]{Conductivity-derived scattering rate, $1/\tau(\nu)$
vs. frequency in the BCS s-wave superconducting state for (a) $1/\tau = 2$
meV and (b) $1/\tau = 25$ meV. An abrupt onset of absorption at the
optical gap at temperatures near $T_c$ is more apparent in (a) than
in (b). The horizontal dashed line indicates the
normal state result. Reproduced from \protect\cite{marsiglio97}.}
\label{ff37}
\end{figure}
                
\begin{figure}
\includegraphics[width=.8\textwidth]{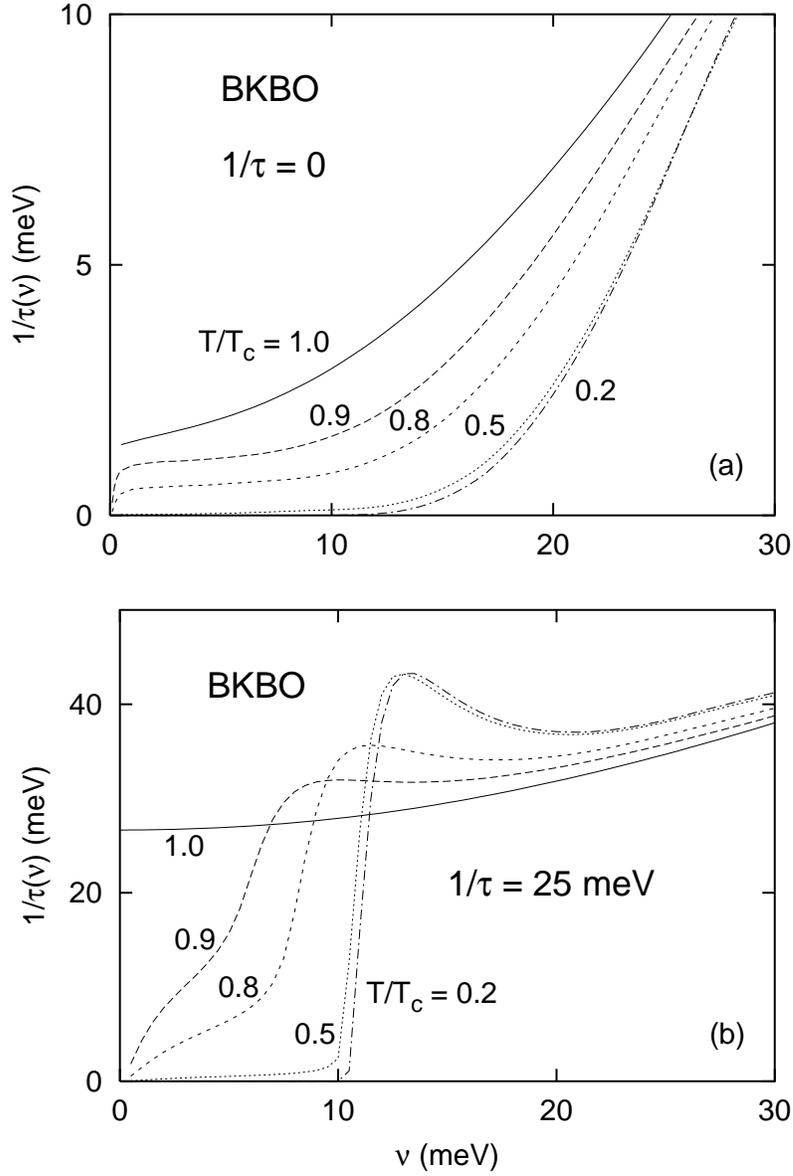}
\caption[]{The conductivity-derived scattering rate, $1/\tau(\nu)$
vs. frequency in the s-wave superconducting state for (a) $1/\tau = 0$
meV and (b) $1/\tau = 25$ meV, for temperatures as indicated.
In both  cases we used the BKBO spectrum with
$\lambda = 1$. In (a) there is no signature for a gap, while one
remains at low temperatures in (b).}
\label{ff38}
\end{figure}
                
\begin{figure}
\includegraphics[width=.9\textwidth]{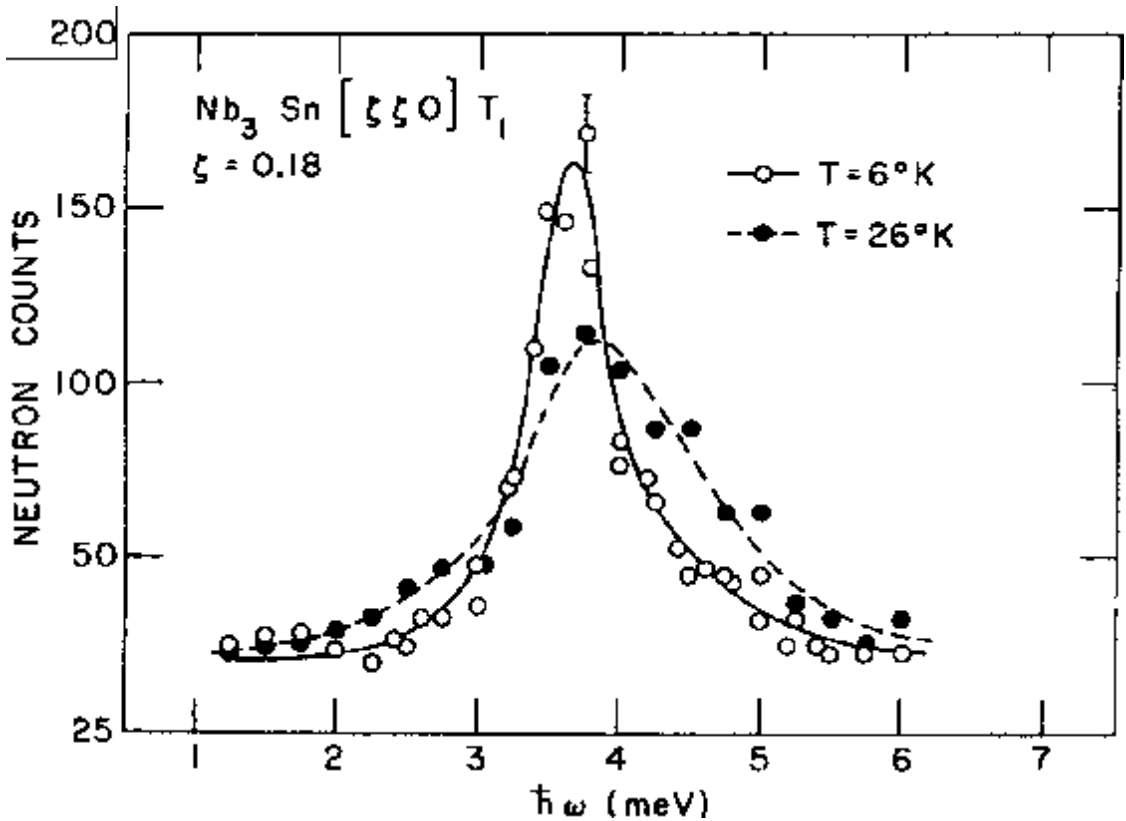}
\caption[]{The widths of low energy $[\zeta \zeta 0]T_1$ acoustic 
phonons broaden
appreciably at temperatures above $T_c$, the superconducting transition
temperature.  This figure shows the same phonon profile above and below
$T_c \approx 18.0\ K$ Reproduced from \protect\cite{axe73}.}
\label{ff39}
\end{figure}
                
\begin{figure}
\includegraphics[width=.8\textwidth]{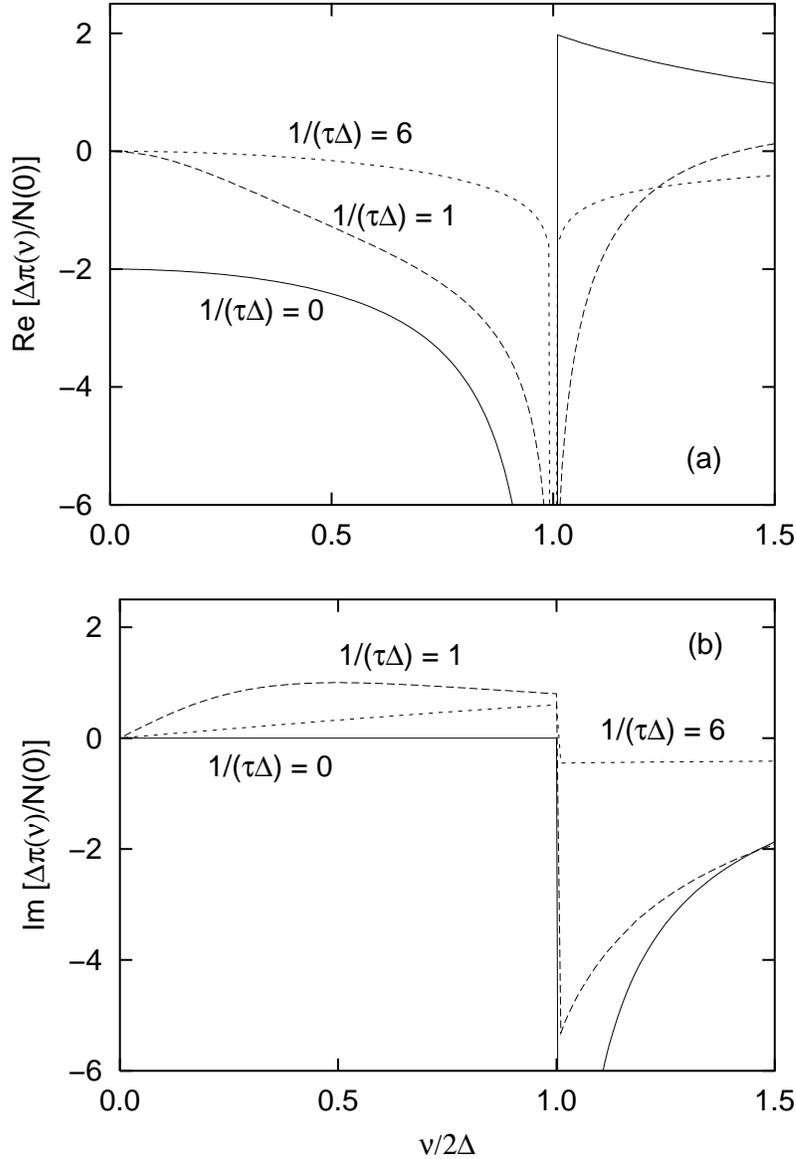}
\caption[]{(a) Real and (b) Imaginary part of $ \Delta \Pi(\nu+i\delta)/N(0)$ vs
$\nu/(2\Delta_0)$ at zero temperature, for various impurity scattering
rates, $1/(\tau\Delta_0) = 0$ (solid), 1 (dotted), and 6 (dashed),
in the weak coupling (BCS) approximation. Below twice the gap edge the phonons
soften; above twice the gap edge they harden in the
clean limit and soften in the dirty limit. Note the narrowing that occurs
below the gap edge in the presence of impurity scattering.}
\label{ff40}
\end{figure}


\begin{thebibliography} {999}
\bibitem{remark2} Indeed, the electron-phonon mechanism has been questioned
many times over the years since BCS, not just on new superconductors, but
on conventional ones as well.
\bibitem{remark4} See Ref. \cite{allen82} for more discussion, but 
note that our notation differs from theirs.
\bibitem{remark8} Of course, physically we know that Eliashberg theory
gets this limit completely wrong; $T_c$ should peak at some point and then
decrease to zero as the coupling strength increases. That an approximation
to Eliashberg theory improves upon this doesn't justify that approximation;
it simply means two errors have undone one another to get closer to the
``right answer''.
\bibitem{remark9} This is not to be taken as a statement regarding the
phase of the order parameter; we have ignored the phase up to this point,
or, equivalently, set it to zero. Rather, as is evident from Eqs. 
(\ref{anal_zz},\ref{anal_phi2}), in Eliashberg theory the quantity
which takes on the role of the order parameter acquires an imaginary
part, whose significance is associated with lifetime effects.
\bibitem{remark10} See, for example, Refs. 
\cite{nam67,lee89,shulga01,bickers90,marsiglio91a,akis91,marsiglio96,marsiglio97,kaufmann98}; 
there have been many others in the last ten years especially.
\bibitem{remark11} Details can be found in Ref. \cite{marsiglio92a}.
\end{thebibliography}
\end{document}